\documentclass[pdflatex,sn-nature]{sn-jnl}

\usepackage{graphicx}%
\usepackage{multirow}%
\usepackage{amsmath,amssymb,amsfonts,mathtools}%
\usepackage{amsthm}%
\usepackage{mathrsfs}%
\usepackage[title]{appendix}%
\usepackage{xcolor}%
\usepackage{textcomp}%
\usepackage{manyfoot}%
\usepackage{booktabs}%
\usepackage{algorithm}%
\usepackage{algorithmicx}%
\usepackage{algpseudocode}%
\usepackage{listings}%
\usepackage{bm}
\usepackage{geometry}
\usepackage{float}
\usepackage{pdfpages}

\theoremstyle{thmstyleone}%
\theoremstyle{thmstyletwo}%

\theoremstyle{thmstylethree}%

\DeclareMathOperator*{\argmin}{arg\,min}
\begin{document}

\title[Article Title]{Generative Latent Diffusion Model for Inverse Modeling and Uncertainty Analysis in Geological Carbon Sequestration}

\author[1,2,3]{\fnm{Zhao} \sur{Feng}}

\author[2]{\fnm{Xin-Yang} \sur{Liu}}

\author[1]{\fnm{Meet Hemant} \sur{Parikh}}

\author[1]{\fnm{Junyi} \sur{Guo}}

\author[2]{\fnm{Pan} \sur{Du}}

\author*[3]{\fnm{Bicheng} \sur{Yan}}\email{bicheng.yan@kaust.edu.sa}

\author*[1,2]{\fnm{Jian-Xun} \sur{Wang}}\email{jw2837@cornell.edu}

\affil*[1]{\orgdiv{Sibley School of Mechanical and Aerospace Engineering}, \orgname{Cornell University}, \orgaddress{\city{Ithaca}, \postcode{14853}, \state{NY}, \country{USA}}}

\affil*[2]{\orgdiv{Department of Aerospace and Mechanical Engineering}, \orgname{University of Notre Dame}, \orgaddress{\city{Notre Dame}, \postcode{46556}, \state{IN}, \country{USA}}}

\affil*[3]{\orgdiv{Physical Science and Engineering Division}, \orgname{King Abdullah University of Science and Technology (KAUST)}, \orgaddress{\city{Thuwal}, \postcode{23955-6900}, \country{Saudi Arabia}}}

\abstract{Geological Carbon Sequestration (GCS) has emerged as a promising strategy for mitigating global warming, yet its effectiveness heavily depends on accurately characterizing subsurface flow dynamics. The inherent geological uncertainty, stemming from limited observations and reservoir heterogeneity, poses significant challenges to predictive modeling. Existing methods for inverse modeling and uncertainty quantification are computationally intensive and lack generalizability, restricting their practical utility. Here, we introduce a Conditional Neural Field Latent Diffusion (CoNFiLD-geo) model, a generative framework for efficient and uncertainty-aware forward and inverse modeling of GCS processes. CoNFiLD-geo synergistically combines conditional neural field encoding with Bayesian conditional latent-space diffusion models, enabling zero-shot conditional generation of geomodels and reservoir responses across complex geometries and grid structures. The model is pretrained unconditionally in a self-supervised manner, followed by a Bayesian posterior sampling process, allowing for data assimilation for unseen/unobserved states without task-specific retraining. Comprehensive validation across synthetic and real-world GCS scenarios demonstrates CoNFiLD-geo's superior efficiency, generalization, scalability, and robustness. By enabling effective data assimilation, uncertainty quantification, and reliable forward modeling, CoNFiLD-geo significantly advances intelligent decision-making in geo-energy systems, supporting the transition toward a sustainable, net-zero carbon future.}

\keywords{Generative modeling, Geological carbon sequestration, Uncertainty quantification, Data assimilation, Bayesian learning}



\maketitle

\section{Introduction}\label{sec_intro}
The escalating threats posed by climate change represent an urgent global imperative for Earth's ecosystems and human civilization, necessitating the deployment of effective decarbonization strategies. Geological Carbon Sequestration (GCS) has emerged as a promising solution for large-scale reduction of atmospheric carbon dioxide ($\mathrm{CO_2}$)~\cite{RN23}. GCS involves injecting captured $\mathrm{CO_2}$, sourced from industrial emissions or direct air capture, into deep subsurface geological formations such as saline aquifers~\cite{huppert_fluid_2014}, depleted hydrocarbon reservoirs~\cite{lyu_role_2021}, and basaltic rocks~\cite{matter_rapid_2016} for permanent storage. This injected $\mathrm{CO_2}$ displaces the in-situ fluids, forming complex multicomponent multiphase flow systems. This process leads to gaseous plume migration and pressure buildup, which together constitute key components of the dynamic reservoir responses. Accurately characterizing these dynamic reservoir responses is crucial for reliable assessment of storage capacity, leakage risks, and operational decisions. However, intrinsic geological uncertainties arising from limited observations and inherent subsurface heterogeneity significantly compromise the predictive reliability.

Inverse modeling, or data assimilation, provides a systematic framework for inferring the uncertain geological parameters from field observational data. Inversion methods broadly fall into deterministic and stochastic categories~\cite{oliver_inverse_2008}. Deterministic approaches identify optimal geomodels by minimizing discrepancies between observed reservoir responses and model predictions through gradient-based~\cite{gao_improved_2006, shirangi_improved_2016} or heuristic  optimization~\cite{lee_field_2019, jeong_theoretical_2019}. However, these methods yield only point estimates, thus limiting their capability to explore the full posterior distribution for uncertainty quantification (UQ). In contrast, stochastic methods leverage Bayesian inference to update prior distributions using observational data, employing sampling-based techniques such as Markov Chain Monte Carlo (MCMC)~\cite{han_surrogate_2024,wang_efficient_2021} or optimization-based ensemble algorithms~\cite{gao2021bi,tang_deep_2022,jiang_history_2024}. Traditional inverse modeling workflows, however, are computationally demanding due to repeated forward model evaluations, exacerbated by the strong nonlinearities, multiphysics coupling, and extensive spatiotemporal scales inherent in GCS.

Recent advancements in machine learning (ML) and deep learning (DL) offer promising pathways to tackle these challenges, introducing computationally efficient surrogate models to bypass costly forward simulations. Convolutional Neural Networks (CNNs), in particular, have gained popularity for subsurface flow prediction on structured grids, framing the tasks as image-to-image regression~\cite{wen_towards_2021, feng_hybrid_2025, seabra_ai_2024, qin_fluid_2025, xu_novel_2023}. Typically employing encoder–decoder architectures, these models encode geological parameters into latent representations and reconstruct flow dynamics through autoregressive rollouts~\cite{mo_deep_2019}, recurrent connections~\cite{tang_deep-learning-based_2022, feng_encoder-decoder_2024}, or explicit temporal conditioning~\cite{yan_physics-constrained_2022, tariq2024transunet}. Graph Neural Networks (GNN) have similarly demonstrated utility in unstructured grid applications~\cite{jiang_graph_2023}, especially for emulating GCS in faulted reservoirs~\cite{ju2024learning}. Additionally, neural operators, such as Fourier Neural Operators (FNOs)~\cite{li_fourier_2021}, Deep Operator Networks (DeepONets)~\cite{lu_learning_2021} and their variants, have been used to learn functional mappings from geological parameters to reservoir responses. Examples include U-Net enhanced FNO (U-FNO)~\cite{wen_u-fnoenhanced_2022}, nested FNO~\cite{wen_real-time_2023}, and hybrid frameworks combining DeepONet and FNO~\cite{jiang_fourier-mionet_2024,lee_efficient_2024} for surrogate GCS modeling. Despite their computational advantages, these surrogate approaches generally require task-specific retraining or repeated inference steps when applied to new observational scenarios, thereby limiting generalizability and practical utility. Moreover, errors introduced by surrogates may propagate into inversion outcomes, particularly beyond training regimes. 

A complementary line of investigation leverages dimension reduction techniques to compactly represent the complex geomodel spatial variability within low-dimensional latent spaces, thereby rendering traditional inversion algorithms computationally tractable. Classic methods such as Principal Component Analysis (PCA)~\cite{oliver_multiple_1996, vo_new_2014, wang_physics-informed_2024}, CNN-PCA~\cite{liu_deep-learning-based_2019,liu_3d_2021}, and Convolutional Autoencoders (CAEs)~\cite{xiao_modelreduced_2022,razak_latent-space_2022} achieve substantial reductions in parameter dimensionality, thus simplifying the subsequent inversion process. More recently, deep generative models, particularly Variational Autoencoders (VAEs)~\cite{laloy_inversion_2017,jiang_deep_2021, bao_variational_2022} and Generative Adversarial Networks (GANs)~\cite{laloy_training-image_2018, fu_deep_2023, ling_improving_2024, zhan_integrated_2022, tetteh_leveraging_2024}, have been adopted to further refine latent representations. These generative architectures often explicitly enforce Gaussian-distributed latent variables, facilitating compatibility with Bayesian inversion techniques, particularly ensemble-based approaches~\cite{feng_deep_2024}. Nevertheless, despite the use of probabilistic generative models, existing workflows predominantly exploit these methods as dimension reduction tools, with posterior exploration still relying on traditional inversion algorithms. Consequently, the intrinsic probabilistic capabilities offered by generative modeling remain largely underutilized. Additionally, reliance on iterative posterior updates and repeated forward simulations continues to constrain the computational efficiency and practical applicability of these methods. Therefore, there remains a need for a unified inversion framework capable of leveraging the full strengths of generative models to directly model the posterior, avoiding iterative retraining or excessive forward simulations.

Diffusion probabilistic models have recently emerged as a powerful paradigm in generative modeling, demonstrating exceptional success in image/video generation to solving and Beyond~\cite{yang2023diffusion}. These models, broadly classified into denoising diffusion probabilistic models (DDPMs)\cite{sohl-dickstein_deep_2015,ho_denoising_2020} and score-based generative models\cite{song_generative_2019}, share a unified theoretical foundation within the stochastic differential equation (SDE) framework~\cite{song_score-based_2021}. Compared to traditional generative methods, diffusion models stand out by offering stable training, high-fidelity sample generation, and particularly effective conditional sampling capabilities~\cite{dhariwal_diffusion_2021}. These properties make them particularly attractive from a Bayesian perspective for inverse problems, allowing for generating posterior samples conditioned on observations~\cite{daras2024survey}. Recent advances have demonstrated the growing potential of diffusion models for inverse problems and data assimilation in computational mechanics~\cite{gao2024bayesian,du_conditional_2024,liu_confild-inlet_2024,li_learning_2024,dasgupta2025conditional,valencia2025learning,fan2025neural}. A particularly notable contribution is the Conditional Neural Field Latent Diffusion (CoNFiLD) framework, which integrates conditional neural fields with latent diffusion modeling~\cite{du_conditional_2024}. CoNFiLD has exhibited strong performance across various challenging tasks, such as turbulent flow generation and reconstruction. In these applications, CoNFiLD has enabled accurate Bayesian conditional sampling, producing physically consistent and observation-consistent flow fields directly from sparse and noisy data without the need for retraining. Studies such as CoNFiLD-inlet and other conditional diffusion-based models~\cite{liu_confild-inlet_2024} have demonstrated the framework's robustness, generalizability, and data-efficiency in modeling highly nonlinear and stochastic spatiotemporal systems.

Despite these promising results, the application of diffusion models to subsurface multiphase flow inversion remains relatively unexplored. Some recent studies have explored this direction, offering initial insights. Zhan et al.~\cite{zhan_toward_2025} employed a Latent Diffusion Model (LDM) to infer aquifer heterogeneity, conditioning the generative process on fully observed flow fields. While the approach demonstrated promise for both unconditional generation and surrogate modeling, it required retraining for each new data assimilation scenario and was constrained by the structured-grid limitations of convolutional VAEs. In another study, Wang et al.~\cite{wang_generative_2025} utilized diffusion models with classifier-free guidance to directly generate geomodels from sparse observations. Although effective in sparse-data settings, the method showed limited generalization to new observational configurations. Moreover, by operating directly in physical space, the model faced scalability challenges in high-dimensional settings, and generated only static geological realizations, thus requiring separate numerical simulations to obtain the associated flow responses.

To address the aforementioned limitations, we propose the Conditional Neural Field Latent Diffusion model for geoscience (CoNFiLD-geo), a unified generative framework that jointly models geological parameters and spatiotemporal reservoir responses without the need for task-specific retraining. CoNFiLD-geo synergistically integrates Conditional Neural Fields (CNFs) with latent diffusion probabilistic models, enabling the learning of joint distributions over geomodels and reservoir responses in a shared, mesh-agnostic latent space. By leveraging CNFs for nonlinear dimension reduction, the framework supports both structured and unstructured grids, allows continuous querying at arbitrary spatial locations, and scales efficiently to high-dimensional settings. CoNFiLD-geo is pretrained in an unconditional, self-supervised manner, during which it learns joint prior distribution of both geological structures and flow dynamics. At inference, observational data are incorporated via a Bayesian posterior sampling strategy that guides the generative process without requiring retraining. This zero-shot conditional generation capability enables robust and adaptive data assimilation across diverse data assimilation scenarios. In contrast to the original CoNFiLD framework~\cite{du_conditional_2024} and most existing literature that focus on solely modeling the distribution of states, CoNFiLD-geo captures the joint distribution of both the parameter (input) and the state (output) spaces simultaneously, which allows the framework to operate as both a forward surrogate and as a generative inverse solver, offering comprehensive data assimilation and UQ for subsurface multiphase flow systems. We demonstrate the efficacy of CoNFiLD-geo across a range of GCS applications, including both synthetic and field-scale case studies, targeting hydrodynamic and stratigraphic inversion tasks. The results highlight the model's ability to generate physically consistent geomodel–response pairs from sparse, noisy, and geometrically complex observational data. While our focus is on GCS, the CoNFiLD-geo framework is broadly applicable to other geo-energy systems involving coupled multiphysics processes. As such, this work presents a generalizable, inference-efficient generative modeling paradigm for geoscience and marks a step toward realizing artificial general intelligence in data-scarce, physics-informed domains.

\section{Results}\label{sec_results}
 In this section, an overview of the proposed CoNFiLD-geo framework is first presented, followed by numerical experiments over a wide range of GCS scenarios, including $\mathrm{CO_2}$ drainage in heterogeneous reservoirs, field-scale $\mathrm{CO_2}$ sequestration at the Sleipner site in Norway, and $\mathrm{CO_2}$ injection coupled with brine production in stratigraphically complex formations. The first two cases aim to quantify uncertainty in hydrodynamic parameters, whereas the final case is dedicated to inferring stratigraphic heterogeneity.
 
\subsection{Overview of the CoNFiLD-geo framework}
\begin{figure}[htb!]
\centering
\includegraphics[width=\textwidth]{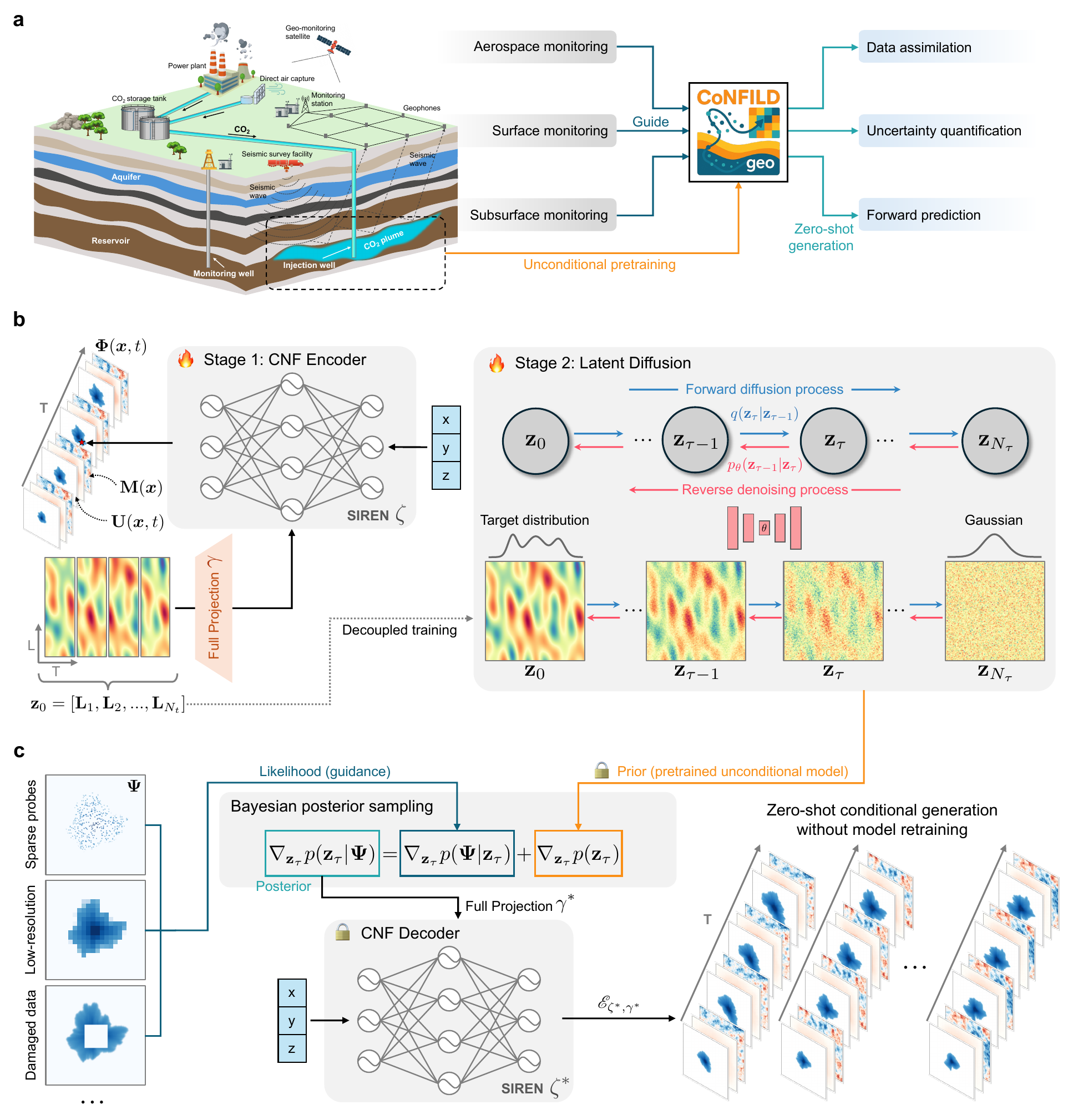}
\caption{Overview of the proposed CoNFiLD-geo framework. (a) Left: schematic of GCS and its monitoring system. Right: schematic of CoNFiLD-geo; the unconditional pretrained model serves as a prior, while multi-source of monitoring data is fused to guide the zero-shot conditional generation. (b) In the pretraining stage, the CNF encodes the spatiotemporal field $\bm\Phi(\bm{x},t)$, which is the concatenation of the geomodel $\bm{M}(\bm{x})$ and the corresponding reservoir responses $\bm{U}(\bm{x},t)$, into a latent variable $\bm{z}_0$. The LDM then learns the joint distribution in the latent through bidirectional diffusion and denoising processes. (c) In the generation stage, arbitrary types of conditional information $\bm\Psi$ (e.g., sparse probed measurements, low resolution fields, and damaged data) are utilized to steer the generation within a Bayesian formulation. The generated latent variable is decoded back into physical space by the trained CNF. No training is involved in the conditional generation process.}\label{fig1}
\end{figure}

CoNFiLD-geo is a probabilistic generative framework, synergistically integrating a Conditional Neural Field (CNF) and a Latent Diffusion Model (LDM) through a modular design. It functions through a two-stage process: an offline pretraining stage and an online generation stage. During pretraining, CoNFiLD-geo learns the joint distribution of geomodels and the corresponding reservoir responses, implicitly capturing the prior physical knowledge across parameter and solution spaces. At deployment, CoNFiLD-geo allows for the real-time integration of diverse observational data sources, such as satellite remote sensing, surface seismic monitoring, and subsurface borehole measurements, to guide the generation process. CoNFiLD-geo can be broadly applied to data assimilation, UQ, and surrogate modeling, providing a unified framework for geoscientific inverse and forward modeling (Fig.~\ref{fig1}a).

A schematic illustration of the pretraining process is shown in Fig.~\ref{fig1}b. Given a realization of geomodel $\bm{M}(\bm{x})\in\mathbb{R}^{N_d\times N_m}$, the corresponding spatiotemporal reservoir responses resulting from engineering perturbations are denoted by $\bm{U}(\bm{x},t)\in\mathbb{R}^{N_d\times N_t\times N_u}$, where $N_d$ represents the spatial dimension,  $N_m$ the set of geological parameters, $N_t$ the temporal length, and $N_u$ the number of state variables. The model parameter-solution pairs are simultaneously encoded by a CNF, which implicitly captures their spatial and temporal correlations in a unified representation. Specifically, the static geomodel is concatenated with the dynamic reservoir responses via temporal broadcasting, forming a joint spatiotemporal field $\bm\Phi(\bm{x},t)\in\mathbb{R}^{N_d\times N_t\times N_\Phi}$, where $N_\Phi = N_m+N_u$ corresponds to the total feature dimension combining both geological parameters and state variables. The CNF $\mathscr{E}_{\zeta,\gamma}(\bm{X},\bm{L})$ is then applied to encode the high-dimensional $\bm{\Phi}(\bm{x},t)$ into a sequence of latents $\bm{z}_0\in\mathbb{R}^{N_l\times N_t}$, where each column $\bm{L}_t\in\mathbb{R}^{N_l}$ corresponds to the low-dimensional latent vector at time step $t$, i.e., $\bm{z}_0=[\bm{L}_1, \bm{L}_2,...,\bm{L}_{N_t}]$. In this work, the CNF encoder $\mathscr{E}_{\zeta,\gamma}(\bm{X},\bm{L})$ is formulated in an auto-decoding fashion \cite{park_deepsdf_2019}, where the learnable condition vector $\bm{L}$ is fed into a hypernetwork parameterized by $\gamma$, which modulates the main network $\zeta$ through full-projection conditioning~\cite{liu_confild-inlet_2024}. The latent variable $\bm{z}_0$ serves as a reduced-dimensional joint representation that bridges the functional spaces of the geological parameters and the corresponding reservoir responses. This property facilitates the LDM to learn the joint distribution in a compact latent space. A diffusion probabilistic model~\cite{ho_denoising_2020} is employed to approximate $p(\bm{z}_0)$ through a Markovian denoising transition kernel starting from a standard Gaussian distribution. This process involves asymptotically adding handcrafted Gaussian noise to the latent variable until the original representation becomes an isotropic Gaussian. Subsequently, a neural network (parameterized by $\theta$) is trained to reverse this diffusion process by denoising the latent variables in a step-wise manner. Once trained, the diffusion model can generate new realizations of $\bm{z}_0$ from randomly sampled white noises by iteratively applying the learned denoising network. To ensure optimization stability, the training of CoNFiLD-geo is decoupled into two stages: the CNF encoder is pretrained, followed by the training of the latent diffusion model.

In the online inference stage, various types of observational data (denoted as $\bm\Psi$) can be utilized on-the-fly to guide the generation of the full field $\bm\Phi$ without retraining (Fig.~\ref{fig1}c). This is accomplished via Bayesian posterior sampling, where the pretrained unconditional CoNFiLD-geo serves as a prior, and the likelihood is evaluated based on the discrepancy between the given observations $\bm\Psi$ and the modeled observables computed from generated latent variable $\bm{z}_\tau$ (see Methods for detailed derivation). The differentiable nature of the entire framework enables continuous refinement of the latents during sampling, thereby ensuring that the conditionally generated results satisfy both observed conditions and follow the prior learned by the pretrained model. The ultimate posterior $\bm{z}_0$ is subsequently decoded into physical space by the trained CNF $\mathscr{E}_{\zeta^*,\gamma^*}$, yielding the joint spatiotemporal field $\bm\Phi(\bm{x},t)$. The synthesized geomodel $\bm{M}(\bm{x})$ and its associated reservoir responses $\bm{U}(\bm{x},t)$ can be retrieved by disentangling $\Phi(\bm{x},t)$ along the feature dimension. The static component $\bm{M}(\bm{x})$ is obtained by temporal averaging, which potentially mitigates numerical artifacts and enhances the consistency of the generated geomodel. An ensemble of inferred geomodel and reservoir responses can be generated in batch mode, facilitating UQ of the flow system in a real-time manner.

\subsection{$\mathrm{CO_2}$ drainage in heterogeneous reservoirs}

As a fundamental physical process in GCS operations, we first consider the $\mathrm{CO_2}$-$\mathrm{H_2O}$ displacement in 2D synthetic reservoirs characterized by heterogeneous permeability. The heterogeneity is represented by spatially correlated structures realized from a geological continuity model, specifically a Gaussian random field. This case simulates $\mathrm{CO_2}$ migration within a 640 m $\times$ 640 m aquifer domain for 500 days, with a constant injection rate of 0.45 kg/s imposed at the left boundary, while a fixed hydraulic pressure is prescribed at the right boundary. We collect the heterogeneous permeability fields $\bm K$ and the corresponding state variables ($\mathrm{CO_2}$ saturation $\bm{S_g}$ and pressure $\bm{P}$) from high-fidelity numerical simulation (see Supplementary Note 2.2 for setup details). CoNFiLD-geo is unconditionally trained to learn the joint distribution of $\bm\Phi=[\bm{K},\bm{S_g},\bm{P}]$, and then tested on unseen observations $\bm\Psi$ via zero-shot conditional generation. We investigate two representative observation scenarios: time-lapse seismic monitoring of $\mathrm{CO_2}$ plume and sparse well measurements.

\begin{figure}[htb!]
\centering
\includegraphics[width=\textwidth]{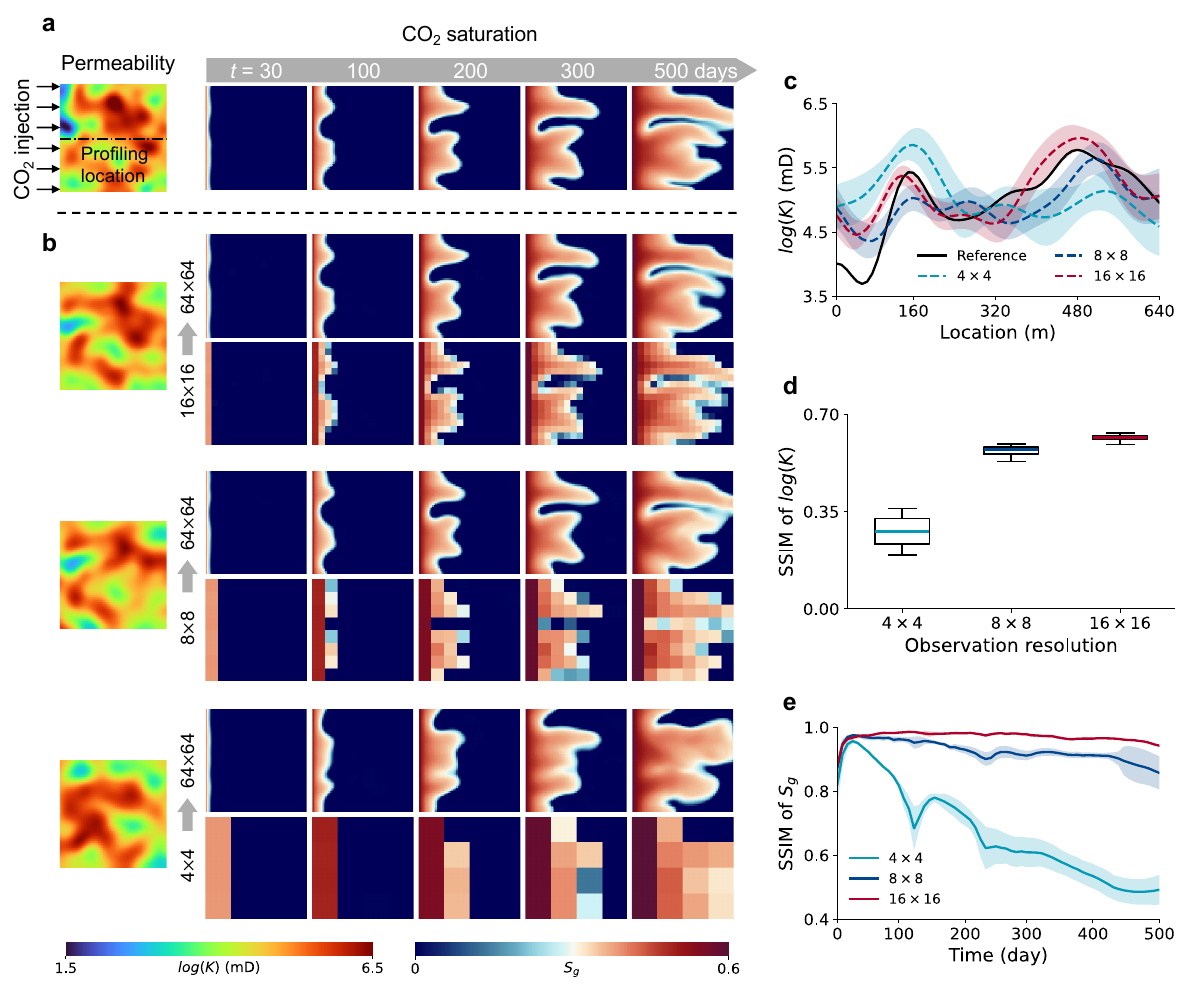}
\caption{Inferring heterogeneous permeability fields (log transformed) and $\mathrm{CO_2}$ saturation from time-lapse seismic monitoring data. 
(a) Reference permeability field and $\mathrm{CO_2}$ saturation trajectory from high-fidelity simulations, with snapshots taken at 30, 100, 200, 300, and 500 days. $\mathrm{CO_2}$ is injected from the left boundary at a constant rate. 
(b) Permeability field and $\mathrm{CO_2}$ saturation snapshots generated by CoNFiLD, conditioned on varying levels of low-resolution $\mathrm{CO_2}$ plume seismic monitoring data (obtained by downsampling the reference saturation fields to resolutions of $4\times 4$, $8\times 8$, and $16\times 16$). 
(c) Comparison between the inferred and reference permeability fields along the profiling location (denoted by the dash-dot line in (a)). Shaded areas indicate the standard deviation (uncertainty) across an ensemble of 10 generated samples. 
(d) Structural Similarity Index Measure (SSIM) between the inferred and reference permeability fields under the three low-resolution conditions. 
(e) Temporal variation of SSIM between the inferred and reference $\mathrm{CO_2}$ saturation fields for three low-resolution settings, with shaded regions indicating standard deviations.}\label{fig2}
\end{figure}

Time-lapse seismic monitoring is an effective and non-intrusive tool for characterizing $\mathrm{CO_2}$ migration in subsurface reservoirs~\cite{reynolds_introduction_nodate, eid_seismic_2015, ajayi_review_2019}. The interpreted $\mathrm{CO_2}$ plume, constrained by observational precision, may be coarsely resolved and potentially noisy. Figure~\ref{fig2} presents the results of inferring the high-resolution fields from different levels of low-resolution seismic monitoring data. Figure~\ref{fig2}a depicts a heterogeneous permeability field and its corresponding snapshots of $\mathrm{CO_2}$ saturation from numerical simulation at 30, 100, 200, 300, and 500 days. This trajectory $\bm{\Phi}\in\mathcal{A}_\mathrm{test}$ is downsampled to various lower resolutions to emulate coarse observation data $\bm\Psi$, which serve as conditional inputs for CoNFiLD-geo to synthesize samples consistent with the available observations. As the observation resolution increases (from bottom to top in Fig.~\ref{fig2}b), the conditionally generated high-resolution $\mathrm{CO_2}$ plume patterns and the inferred permeability field exhibit progressively better alignment with the reference. We refer the reader to Supplementary Note 6.1 for visualizing the contours of pressure. A distinguishing feature of CoNFiLD-geo is its ability to quantify the uncertainties inherent in the generated full fields. Fig.~\ref{fig2}c displays the permeability values along the profiling location (dash-dot line in Fig.~\ref{fig2}a) evaluated across an ensemble of 10 generated samples (ensemble size is 10 throughout this manuscript unless otherwise specified). Complementarily, Fig.~\ref{fig2}d shows the Structural Similarity Index Measure (SSIM) between the generated and reference permeability fields, where values closer to 1 indicate greater similarity to the reference, Together, these results substantiate that CoNFiLD-geo can effectively infer the geological parameters from low-resolution $\mathrm{CO_2}$ plume observations. The uncertainty decreases monotonically as more resolved observational information becomes available, which aligns with the expectations from Bayesian perspective. In addition, CoNFiLD-geo provides accurate predictions of the spatiotemporal evolution of $\mathrm{CO_2}$ saturation, as shown in Fig.~\ref{fig2}e. The generated saturation trajectory converges to the reference one as the observation resolution increases. More generated samples can be found in Supplementary Note 7.1.

\begin{figure}[htb!]
\centering
\includegraphics[width=\textwidth]{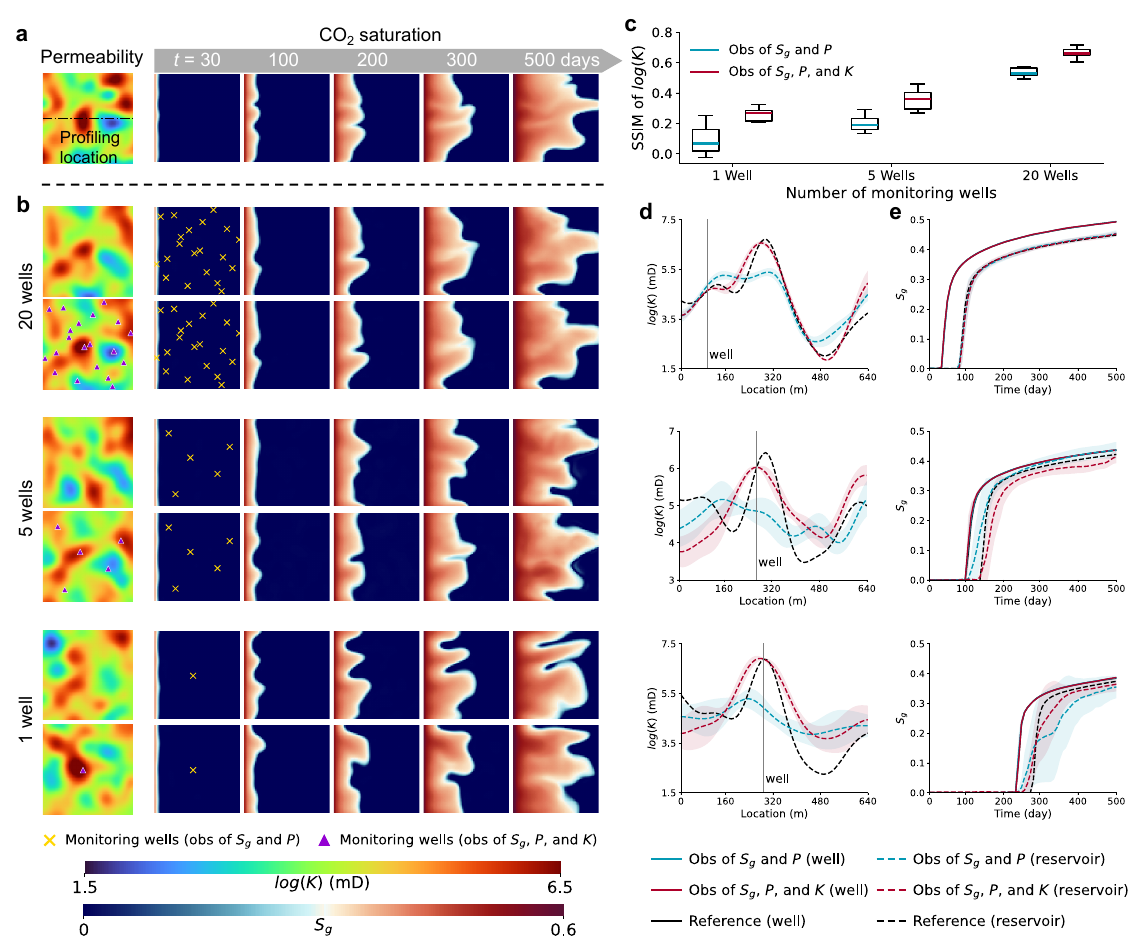}
\caption{Reconstructing heterogeneous permeability fields and $\mathrm{CO_2}$ saturation from sparse well measurements. 
(a) Reference permeability field and $\mathrm{CO_2}$ saturation snapshots at 30, 100, 200, 300, and 500 days. 
(b) Permeability field and $\mathrm{CO_2}$ saturation snapshots generated by CoNFiLD, conditioned on different number of monitoring wells (1, 5, and 20). For each well configuration, the upper row shows generated results from wells with probed reservoir responses (saturation and pressure; wells marked by crosses), while the lower row presents generated results from wells with probed reservoir responses and permeability values (wells marked by triangles). 
(c) SSIM of the inferred permeability under different well configurations and types of probed data. 
(d) From bottom to top, spatial variation of the generated and reference permeability fields along the profiling location (denoted by the dash-dot line in (a)) for 1, 5 and 20 wells. The vertical line indicates the well location. The shaded regions denote the uncertainty.
(e) From bottom to top, temporal variation of the generated and reference $\mathrm{CO_2}$ saturation fields at the well location (solid line) and a nearby reservoir location (dashed line) location for 1, 5 and 20 wells. The reservoir location is offset by 50 m from the corresponding well. The shaded regions denote the uncertainty.}\label{fig3}
\end{figure}

In subsurface engineering, drilling wells provides direct measurements of flow states and reservoir properties~\cite{matter_rapid_2016, massarweh_co2_2024}. Fig.~\ref{fig3} presents the results of generating the full fields conditioned on different number of wells and different type of well measurements. These conditions $\bm\Psi$ inform the generation process by querying the CNF only at specific data points instead of reconstructing the entire field, thereby significantly reducing the computational burden for LDM. The inferred permeability fields and $\mathrm{CO}_2$ saturation spatiotemporal dynamics (Fig.~\ref{fig3}b) increasingly align with the reference (Fig.~\ref{fig3}a) as more monitoring well data become available. Moreover, direct measurements of permeability (such as core sampling and well logging) provide more precise constraints on the permeability distribution (Fig.~\ref{fig3}c). We further quantify the uncertainties of the inferred results. When permeability is directly probed at the wells, the inferred values at those locations exactly match the reference due to the provided ground truth. As the distance from the wells increases, both the discrepancies and associated uncertainties grow due to diminishing conditional constraints. In contrast, when permeability is not directly observed, the inferred fields tend to be overly smoothed and exhibit broader uncertainty bands (Fig.~\ref{fig3}d). For saturation, the generated trajectories consistently align with the reference at well locations, while noticeable uncertainties emerge in unobserved regions of the reservoir (Fig.~\ref{fig3}e). Introducing additional monitoring wells effectively reduces the uncertainties in both the geomodel and the associated reservoir responses. These results underscore the ability of CoNFiLD-geo to achieve accurate inverse modeling even under highly sparse observational conditions, as demonstrated by the scenario with 20 wells probing both state variables and reservoir parameters, which accounts for only 0.5\% of the total data points. We have also compared the generation performance under regularly placed monitoring wells in Supplementary Note 7.1. In addition to scenarios involving $\mathrm{CO_2}$ seismic monitoring data and sparse well measurements, we also consider settings where observations of either permeability or saturation are incomplete. Notably, CoNFiLD-geo enables data restoration under such conditions, which is particularly valuable in the presence of corrupted or missing data, thereby ensuring continuity and reliability in GCS monitoring (see Supplementary Note 3 for details).

\subsection{Field-scale $\mathrm{CO_2}$ sequestration at the Sleipner site}
\begin{figure}[htb!]
\centering
\includegraphics[width=0.88\textwidth]{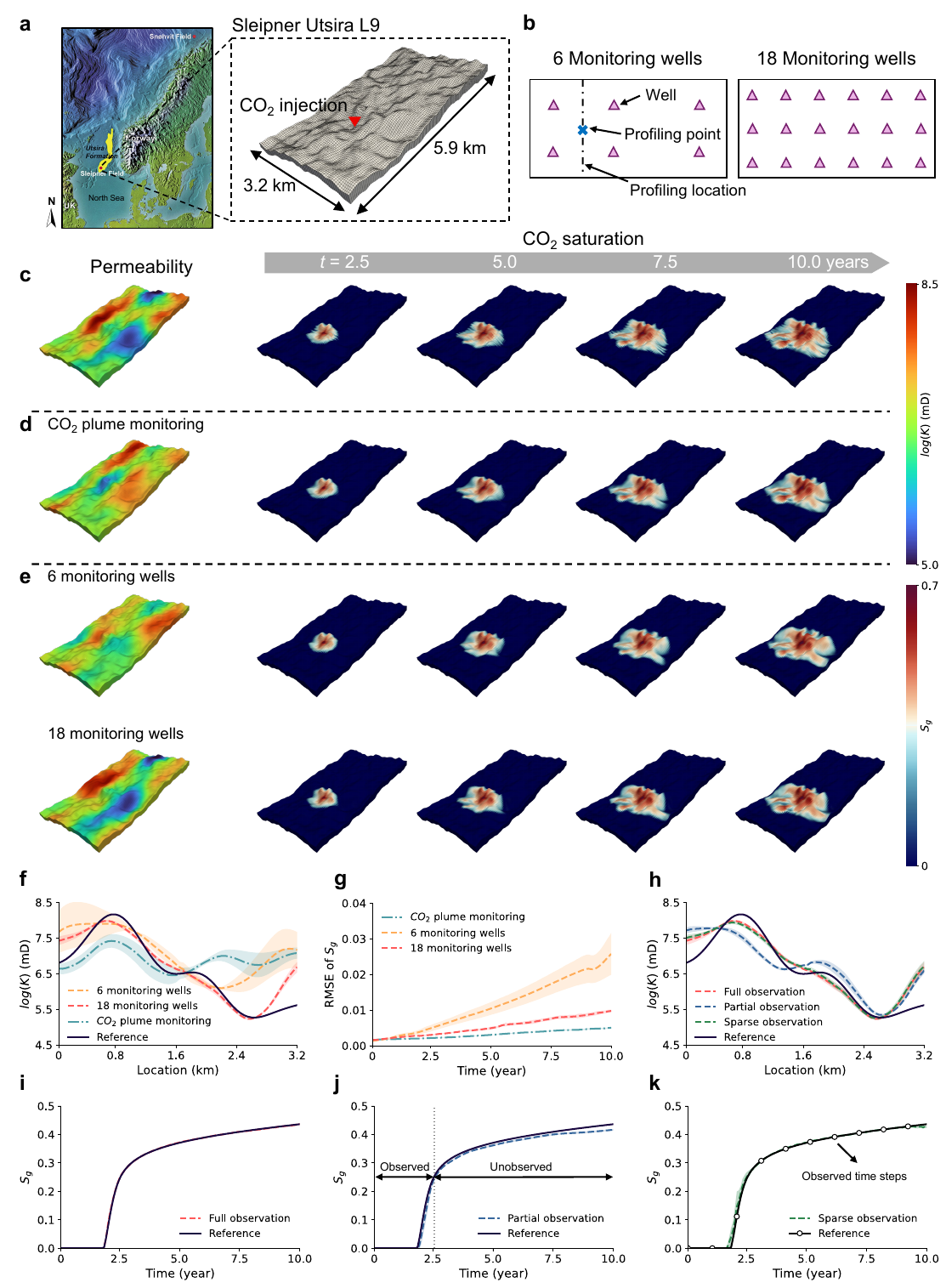}
\caption{Inferring heterogeneous permeability fields and $\mathrm{CO_2}$ saturation at the Sleipner site in Norway. 
(a) Map of the Sleipner GCS project. The zoom-in box shows the geometric configuration of the Utsira L9 layer. 
(b) Schematic illustration of 6 and 18 monitoring well configurations (well marked by triangles).
(c) Reference permeability field and $\mathrm{CO_2}$ saturation snapshots at 2.5, 5, 7.5 and 10 years. 
(d) Generated permeability field and $\mathrm{CO_2}$ saturation snapshots conditioned on $\mathrm{CO_2}$ plume monitoring data. 
(e) Upper row: generated permeability field and $\mathrm{CO_2}$ saturation snapshots under 6 monitoring wells with probed permeability, saturation and pressure. Lower row: results under 18 monitoring wells.
(f) Spatial profiles of the generated and reference permeability fields along the dash-dot profiling location (denoted in (b)) for the three monitoring strategies. Shaded regions indicate uncertainty. 
(g) Root Mean Square Error (RMSE) of generated $\mathrm{CO_2}$ saturation over time for the three monitoring strategies, with shaded regions indicating standard deviation.
(h) Spatial profiles of the generated and reference permeability fields under different levels of temporal sparsity in the 18 monitoring strategy. Shaded regions indicate uncertainty.
(i-k) Temporal evolution of the generated and reference $\mathrm{CO_2}$ saturation fields under different levels of temporal sparsity: (i) full observations at all time steps, (j) partial observations in the initial 2.5 years (k) annual observations.}\label{fig4}
\end{figure}
The Sleipner project is the world's first industrial-scale GCS initiative. Since 1996, separated $\mathrm{CO_2}$ from a nearby gas field has been injected into the Utsira Sand formation, a saline aquifer located at a depth of $1012\ \mathrm{m}$ beneath the North Sea, offshore Norway (see Fig.~\ref{fig4}a for the location). The complete numerical model of the Sleipner site is described in Supplementary Note 2.2. In this study, we focus on a specific sublayer, Utsira L9, which is bounded by a low-permeability shale interlayer and an overlying caprock, forming a relatively independent hydrosystem. We adopt the realistic stratigraphic structure of the reservoir as determined by field-based geological surveys~\cite{sleipner2019}. The reservoir extends over an area of $3.2\ \mathrm{km} \times 5.9\ \mathrm{km}$, exhibiting spatial variability in both depth and thickness arising from natural sedimentary processes. The domain is discretized using a lateral resolution of $50\times 50\ \mathrm{m}$ and a single vertical layer, as the simulation primarily targets the lateral migration of the $\mathrm{CO_2}$ plume. Given the large horizontal extent of the reservoir relative to its modest thickness of approximately $50\ \mathrm{m}$, vertical flow is considered to play a limited role in governing plume dynamics. All boundaries are modeled as open-flow conditions, except for the impermeable top and bottom. $\mathrm{CO_2}$ is injected through a single injection well operating at a constant rate of $32\ \mathrm{kg/s}$. The numerical model setup is illustrated in Fig.~\ref{fig4}a and detailed in Supplementary Note 2.2.

To explore a diverse range of geological scenarios, we generate the heterogeneous permeability fields using a Gaussian covariance kernel. In combination with the corresponding simulated state variables, our objective is to infer the full fields $\bm\Phi$ (Fig.~\ref{fig4}c) conditioned on specified observations $\bm\Psi$. We first consider the availability of $\mathrm{CO_2}$ seismic data. In this scenario, the generated $\mathrm{CO_2}$ saturation closely resembles the reference enabled by access to direct observations. However, the inferred permeability deviates notably from the reference one (Fig.~\ref{fig4}d). We further investigate scenarios involving sparse well measurements with probed state variables and reservoir parameters. Results indicate that using 18 monitoring wells yields markedly improved inversion accuracy for both permeability and saturation, relative to the scenario with only 6 wells (Fig.~\ref{fig4}e). The permeability profiles along the cross-sectional line (dash-dotted in Fig.~\ref{fig4}b) further substantiate that deploying 18 monitoring wells significantly reduces uncertainty and more accurately captures the underlying geological heterogeneity (Fig.~\ref{fig4}f). While seismic plume data offer the highest precision in inferring $\mathrm{CO_2}$ saturation (Fig.~\ref{fig4}g), the 18-well configuration provides the best overall performance by jointly improving the reconstruction of both permeability and saturation fields. These results showcase that CoNFiLD-geo provides a remarkable solution for inverse modeling in realistic GCS projects, while simultaneously enabling UQ without the need for task-specific retraining.

Considering the long operational lifespan of GCS projects (e.g., 10 years in this case), continuous access to monitoring well data may be limited due to equipment maintenance, calibration periods, or cost-saving considerations. Accordingly, we assess the robustness of CoNFiLD-geo in performing inverse modeling when well measurements are temporally sparse or incomplete. Overall, as shown in Fig.~\ref{fig4}h, CoNFiLD-geo is capable of recovering permeability variations with only minor discrepancies near the periphery, even when observations are limited to a quarter of the temporal sequence (25\% of total time steps) or collected exclusively on an annual basis (7.8\% of total time steps). Compared to the fully observed monitoring data (Fig.~\ref{fig4}i), CoNFiLD-geo can still generate reliable saturation dynamics with minimal uncertainty. A slight increase in uncertainty is observed at time steps farther from the observation points due to the temporal dilution of conditional information (Fig.~\ref{fig4}j,k). See Supplementary Note 7.2 for additional conditional generation results and Supplementary Note 6.2 for pressure fields. These findings underscore CoNFiLD-geo’s scalability to real-world applications, demonstrating its flexibility in accommodating diverse monitoring data and its robustness to spatial and temporal sparsity.

\subsection{$\mathrm{CO_2}$ injection and brine production in stratigraphically complex reservoirs}
\begin{figure}[htb!]
\centering
\includegraphics[width=0.9\textwidth]{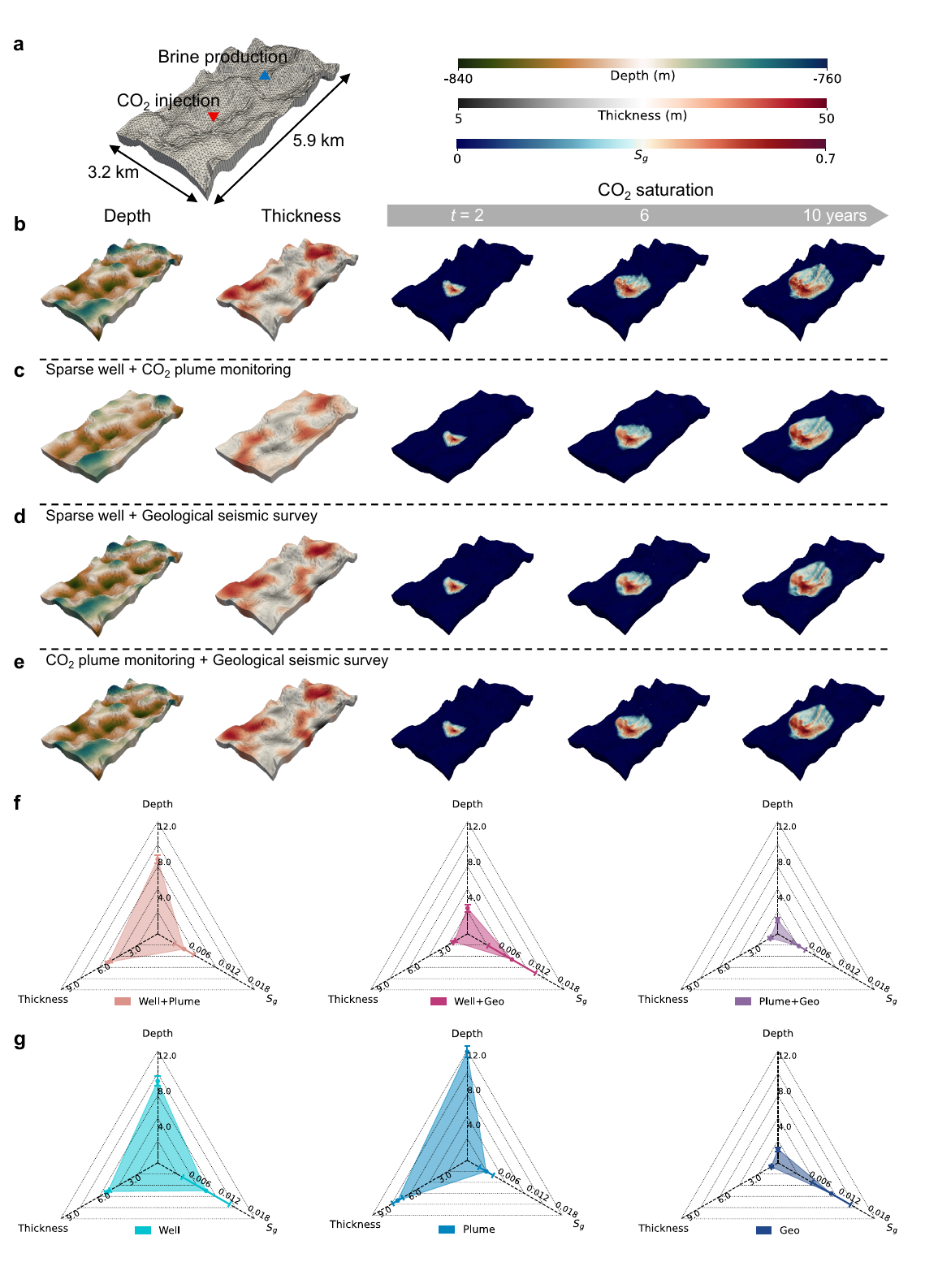}
\caption{Inverse modeling of reservoir geometry (depth and thickness) and $\mathrm{CO_2}$ saturation from multi-source monitoring data. 
(a) Geometric configuration of a stratigraphically complex reservoir. The domain is discretized using unstructured triangular grids. 
(b) Reference reservoir geometry and $\mathrm{CO_2}$ saturation snapshots at 2, 6, and 10 years. 
(c) Generated reservoir geometry and $\mathrm{CO_2}$ saturation conditioned on sparse well measurements and $\mathrm{CO_2}$ plume monitoring data. The sparse well configuration is identical to the 18-well setup in the Sleipner case. 
(d) Generated reservoir geometry and $\mathrm{CO_2}$ saturation conditioned on sparse well measurements and geological seismic data. 
(e) Generated reservoir geometry and $\mathrm{CO_2}$ saturation conditioned on $\mathrm{CO_2}$ plume monitoring data and geological seismic data. 
(f) Performance of the three multi-source monitoring strategies evaluated by RMSE for generated reservoir depth, thickness and $\mathrm{CO_2}$ saturation. Error bars along the radial directions represent standard deviations.
(g) Performance of the three single-source monitoring strategies evaluated by RMSE for generated reservoir depth, reservoir thickness and $\mathrm{CO_2}$ saturation.}\label{fig5}
\end{figure}
The Sleipner case reveals that reservoir geometry, particularly depth and thickness, exerts a substantial influence on the spatiotemporal dynamics of the state variables. Previous studies have predominantly focused on the inverse modeling of seepage-related parameters~\cite{laloy_inversion_2017, jiang_history_2024, han_surrogate_2024, tang_deep-learning-based_2020, zhu_bayesian_2018}, leaving the stratigraphic structure of the reservoir largely overlooked, despite its critical role in flow dynamics. Therefore, in this case, we assess the capability of CoNFiLD-geo to jointly reconstruct reservoir geometry and the corresponding flow responses ($\bm\Phi$) based on integrated multi-source monitoring data ($\bm\Psi$). The operation is configured as $\mathrm{CO_2}$ injection coupled with brine production within a closed hydrosystem (Fig.~\ref{fig5}a), aiming to introduce additional physical complexity, thereby enabling a rigorous evaluation of CoNFiLD-geo’s ability to accommodate varied operational conditions. The lateral extent is retained from the Sleipner case, while the reservoir depth and thickness are constructed using Gaussian random fields to span a broad range of structural variability. To facilitate a high-fidelity representation of complex reservoir geometry, the domain is discretized using an unstructured triangular mesh (Fig.~\ref{fig5}a). This gridding scheme is natively compatible with CoNFiLD-geo, as the CNF operates as a mesh-agnostic method. In contrast, traditional CNN-based dimension reduction techniques are typically limited to structured grids and therefore lack the capacity to resolve such geometric complexity. Details of the numerical model setup are provided in Supplementary Note 2.3.

We consider the integration of different monitoring data modalities as conditional information for inverse modeling. In general, monitoring approaches can be categorized into either intrusive methods, such as monitoring wells, or non-intrusive methods, such as seismic surveys. Although non-intrusive techniques are often more cost-effective, they are typically subject to a certain level of noise. In this study, non-intrusively acquired data — such as the $\mathrm{CO_2}$ plume and geological seismic surveys — are assumed to contain 5\% noise, whereas the intrusive well measurements are considered noise-free. Under conditions of sparse well data and noisy plume observations, CoNFiLD-geo produces saturation fields that are consistent with the reference (Fig.~\ref{fig5}b), despite the inferred depth and thickness fields being marginally smoother (Fig.~\ref{fig5}c). When geological seismic surveys, which provide global but noisy estimates of depth and thickness, are combined with sparse well measurements, the reconstruction of reservoir geometry becomes accurate, whereas minor discrepancies appear in the $\mathrm{CO_2}$ saturation near the plume front (Fig.~\ref{fig5}d). The generated field $\bm\Phi$ aligns most closely with the reference when both $\mathrm{CO_2}$ plume data and geological seismic information are available (Fig.~\ref{fig5}e). This is further corroborated by the radar plot in Fig.~\ref{fig5}g, which shows that combining the two non-intrusive monitoring modalities yields the most accurate reconstruction results. The results for single-source monitoring are presented in Fig.~\ref{fig5}g and Supplementary Note 7.3. Overall, relying on a single data source turns out to be a suboptimal strategy. Together with the pressure generation results in Supplementary Note 6.3, the findings underscore that CoNFiLD-geo is capable of accurately reconstructing the geomodel and associated reservoir responses, even under noisy observational conditions. Its ability to flexibly fuse multi-source monitoring data further highlights its versatility as an inverse modeling framework for GCS applications under realistic and challenging settings.

\section{Discussion}\label{sec_discussion}

We have presented CoNFiLD-geo, a generative framework for zero-shot conditional reconstruction of geomodels and corresponding reservoir responses from diverse types of observational data, enabling real-time inversion with UQ in realistic GCS projects. In CoNFiLD-geo, a CNF-based dimension reduction module is first employed to compress high-dimensional geodata into a compact latent space, followed by the LDM to learn the joint distribution of both input parameters and output states. Once pretrained, CoNFiLD-geo can conditionally generate state samples consistent with observations via Bayesian posterior sampling, alleviating the need for task-specific retraining. The capabilities of the framework have been demonstrated on three representative GCS scenarios of different complexity: the $\mathrm{CO_2}$ drainage in 2D heterogeneous reservoirs, field-scale $\mathrm{CO_2}$ sequestration in the Sleipner site of Norway, and $\mathrm{CO_2}$ injection coupled with brine production in stratigraphically complex reservoirs. The results substantiate the predictive fidelity, generalizability, and robustness of CoNFiLD-geo in inverse modeling of subsurface heterogeneity and the resulting spatiotemporal dynamics of reservoir responses, even under sparse and noisy observational conditions. Moreover, as a unified framework, CoNFiLD-geo also enables efficient surrogate modeling in the forward direction. It outperforms the deterministic U-FNO in scenarios where input parameters are low-resolution and sparse, a situation commonly encountered in subsurface environments (Supplementary Note 4).

\begin{figure}[htb!]
\centering
\includegraphics[width=\textwidth]{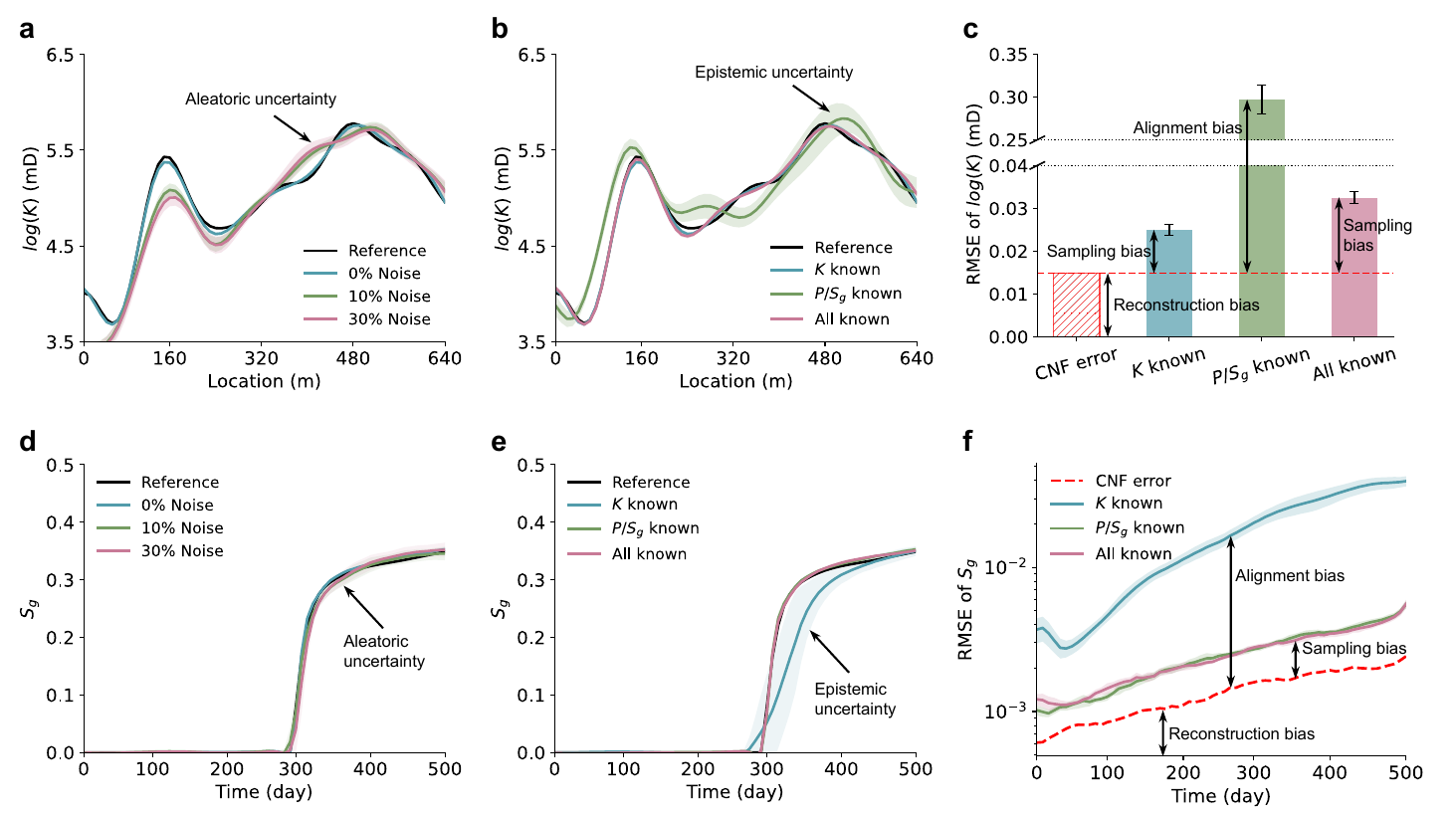}
\caption{Uncertainty and bias analysis in CoNFiLD-geo, based on the first 2D case.
(a) Spatial variation of the reference and inferred permeability fields under varying noise levels (0\%, 10\% and 30\%). The conditional input is fully observed permeability field perturbed by different levels of Gaussian noise. The shaded regions indicate the aleatoric uncertainty. 
(b) Spatial variation of the reference and inferred permeability fields under different types of conditioning inputs drawn from distinct functional spaces: either the parameter space (permeability), the solution space (pressure and saturation), or both. All conditioning inputs are fully observed and noise-free. The shaded regions indicate the epistemic uncertainty. 
(c) Comparison of RMSE in permeability fields for CNF decoding (serving as a reconstruction baseline) and for different conditioning scenarios. 
(d) Temporal variation of the reference and inferred saturation fields under varying noise levels. The conditional inputs are fully observed saturation fields perturbed by different levels of Gaussian noise. 
(e) Temporal variation of the reference and inferred saturation fields under different types of conditioning inputs drawn from distinct functional spaces. 
(f) Comparison of RMSE in saturation fields for CNF decoding and for different conditioning scenarios. }\label{fig6}
\end{figure}

CoNFiLD-geo can also function as a fast numerical emulator through unconditional generation, facilitating preliminary screening of reservoir responses (Supplementary Note 8). The computational costs of both conditional and unconditional generation, relative to numerical simulation, are reported in Supplementary Note 9, revealing that CoNFiLD-geo achieves a substantial speedup over conventional simulation methods. We further analyze the sources of uncertainties and biases within the framework. The generated field exhibits aleatoric uncertainty when the conditioning input has measurement uncertainties (Fig.~\ref{fig6}a,d). Even when the observation data is noise-free, the generated field still shows minor discrepancies. In this setting, the total generation error comprises two components: the reconstruction bias introduced by the CNF decoder and the sampling bias arising from the stochastic nature of the gradient-based sampling process (Fig.~\ref{fig6}c,f). While the reconstruction bias can be reduced by further tuning the CNF, the sampling bias is inherently irreducible. CoNFiLD-geo models the joint distribution between parameter and solution Spaces. Generating one field conditioned on complete knowledge of the other, without direct observations in the target space, introduces epistemic uncertainty (Fig.~\ref{fig6}b,e). This discrepancy is referred to as alignment bias, originating from a physical misalignment between the two function spaces (Fig.~\ref{fig6}c,f).

As a modular framework, CoNFiLD-geo allows for the integration of alternative dimensionality reduction techniques. In Supplementary Note 5, we evaluate the performance of the well-established Proper Orthogonal Decomposition (POD) as a substitute. However, compared to the expressive CNF, POD remains a suboptimal choice for dimensionality reduction due to its inferior reconstruction fidelity and limited compression capability. In addition, the generative modeling component stands to benefit from ongoing advances in generative AI. In the future, we aim to further accelerate generation by adopting expedited sampling methods such as those used in flow matching models~\cite{lipman_flow_2023,hemant2025conditional}. Moreover, although the current model implicitly learns physical priors through data-intensive training, the incorporation of domain knowledge offers a promising pathway to reduce such dependence~\cite{bastek_physics-informed_2025}. This is particularly critical in subsurface engineering, where high-quality data are costly to acquire. One promising strategy is to introduce physics-based constraints during the sampling stage, which enforce the generation of physically consistent parameter–solution pairs and help mitigate alignment bias.

\section{Methods}\label{sec_methods}
\subsection{Subsurface multiphase flow governing equations}
The dynamics of $\mathrm{CO_2}$ migration in reservoirs are governed by the mass conservation of subsurface multiphase flow in the porous media \cite{pruess_tough2_1999},
\begin{equation}
    \frac{\partial}{\partial t}\left(\varphi\sum_\xi \bm{S}_\xi \rho_\xi \chi_\xi^\nu\right)
     - \nabla\cdot\left(\bm{K}\sum_\xi \chi_\xi^\nu \frac{k_{r,\xi}\rho_\xi}{\mu_\xi}(\nabla\bm{P}_\xi-\rho_\xi \bm{g}) \right)
     -\sum_\xi\rho_\xi\chi_\xi^\nu q^\nu
     = 0.
\end{equation}
Here, the subscript $\xi$ denotes the fluid phase, the superscript $\nu$ represents the fluid component. The reservoir responses of interest are characterized by the spatiotemporal saturation $\bm{S}_\xi(\bm{x},t)$ and pressure $\bm{P}_\xi(\bm{x},t)$. The permeability field $\bm{K}(\bm{x})$ is one of the key hydrodynamic quantities to be inferred through inversion. Other constitutive parameters in the above equation include the porosity $\varphi$, the density $\rho_\xi$, the viscosity $\mu_\xi$, the mass fraction $\chi_\xi^\nu$, the relative permeability coefficient $k_{r,\xi}$,  the gravitational acceleration $\bm{g}$, and the well volumetric flow rate $q^\nu$. The constitutive relatioships are detailed in Supplementary Note 2.1.

\subsection{Conditional neural field with full-projection}
Neural field is a class of coordinate-based neural networks that parameterize a continuous field of interest by learning a mapping from spatial coordinates to field values, i.e., $\bm{X}\in\mathbb{R}^{N_d} \mapsto \bm{\Phi}(\bm{x}, t_i)$. A Conditional Neural Field (CNF) extends this formulation by incorporating a set of latent vectors $\bm{L}$ to modulate the field across different snapshots in time, resulting in a conditional mapping $(\bm{X}, \bm{L}) \mapsto \bm{\Phi}(\bm{x}, t)$. Once trained, the CNF encodes the field $\bm{\Phi}(\bm{x},t)$ as a neural implicit representation defined over space and time, i.e., $\bm{\Phi}(\bm{x},t)\approx\mathscr{E}_{\zeta^*,\gamma^*}(\bm{X},\bm{L})$. The encoding process is implemented in an auto-decoding fashion by solving the following optimization problem,
\begin{equation}
    \bm{L}^*,\zeta^*,\gamma^* = 
    \argmin_{\bm{L},\zeta,\gamma}\sum_{i}^{N_t}\sum_{j}^{N_d}\left\|\bm{\Phi}(\bm{X}_j,t_i)-\mathscr{E}(\bm{X}_j,\bm{L}_i;\zeta,\gamma)\right\|^2,
\end{equation}
where $\bm{L}^*$ denotes the optimized latent vectors, $\zeta^*$ and $\gamma^*$ are the optimal parameters of the main network and the modulation network, respectively. In this work, the main network is constructed using the sinusoidal representation network (SIREN) \cite{sitzmann_implicit_2020},
\begin{align}
    \mathscr{E}_\zeta\big(\bm{x})&=\bm{W}_p(\sigma_{p-1}\circ\sigma_{p-2}\circ...\circ\sigma_0(\bm{x})\big)+\bm{b}_p,\nonumber\\
    \sigma_i(\bm\eta_i)&=\sin{\big(\omega_0(\bm{W}_i\bm\eta_i+\bm{b}_i)\big)},
    \ \ i = 0,...,p-1,
\end{align}
where $\bm\eta_0=\bm{x}$ and $(\bm\eta_i)_{i\ge 1}$ are the hidden outputs throughout the network. $\omega_0\in\mathbb{R}_+$ is a hyperparameter that controls the frequency spectrum of the network, $\bm{W}_i$ and $\bm{b}_i$ are the trainable weights and biases of the $i$-th layer. SIREN requires a specialized initialization scheme, wherein the weights of the first layer are initialized as $\bm{W}_0\sim\mathcal{U}(-1/d,1/d)$ and the subsequent layers as $(\bm{W}_i)_{i\ge 1}\sim\mathcal{U}(-\sqrt{\dfrac{6}{\omega_0^2d}},\sqrt{\dfrac{6}{\omega_0^2d}})$, where $d$ denotes the input dimensionality of the corresponding layer.. This ensures that the inputs to the sinusoidal activation functions are approximately normally distributed with unit variance, thereby stabilizing gradient propagation.

The SIREN is modulated by full-projected conditioning \cite{liu_confild-inlet_2024} for its consistent and expressive representation,
\begin{align}
    \mathscr{E}_{\zeta,\gamma}(\bm{x},\bm{L})&=\bm{W}_p(\sigma'_{p-1}\circ\sigma'_{p-2}\circ...\circ\sigma'_0(\bm{x})\big)+\bm{b}_p,\nonumber\\
    \sigma'_i(\bm\eta_i,\delta\bm{W}_i,\delta\bm{b}_i)&=\sin{\big((\bm{W}_i+\delta\bm{W}_i)\bm\eta_i + \bm{b}_i + \delta\bm{b}_i\big)},\ \ i=0,...,p-1,
\end{align}
where $\delta\bm{W}_i$ and $\delta\bm{b}_i$ are regressed from the condition $\bm{L}$ as,
\begin{align}
    \delta\bm{W}_i(\bm{L})&=\bm{W}_i^w\bm{L}+\bm{b}_i^w,\nonumber\\
    \delta\bm{b}_i(\bm{L})&=\bm{W}_i^b\bm{L}+\bm{b}_i^b.
\end{align}
Here, $\zeta=\{\bm{W}_i,\bm{b}_i\}_{i=0}^p$ denotes the shared parameters of SIREN and $\gamma=\{\bm{W}_i^w,\bm{b}_i^w,\bm{W}_i^b,\bm{b}_i^b\}_{i=0}^{p-1}$ represents the instance associated parameters of the modulation network. The CNF in CoNFiLD-geo serves a dual role as both encoder and decoder. During the offline training stage, the optimized latent vector $\bm{L}^*$ functions as a concise encoding of the physical field $\bm\Phi\in\mathcal{A}_\mathrm{train}$ at a specific time frame. A whole trajectory can then be represented as a 2-D latent image $\bm{z}_0$ by concatenating $\bm{L}^*$ along the temporal dimension. It is noteworthy that CoNFiLD-geo, in contrast to the orginal CoNFiLD, simultaneously encodes the geomodel and its corresponding reservoir responses, thereby providing a joint implicit representation of both the parameter space and the solution space. This enables the generative model to capture the joint distribution, facilitating both inverse and forward modeling. The auto-decoding design works synergistically with the latent generative model, which serves as a “latent optimizer” that circumvents the need to retrain the CNF. Therefore, during the online inference stage, the physical field $\bm\Phi\in\mathcal{A}_\mathrm{test}$ can be retrieved by simply feeding the generated latent vectors and spatial coordinates into the trained CNF. 

\subsection{Latent diffusion probabilistic model}
 A diffusion probabilistic model is employed to approximate the underlying distribution of the CNF-encoded latent variable $\bm{z}_0$ by a neural network $p_\theta(\bm{z}_0)$. This is achieved through learning a Markovian transition kernel, which iteratively transforms a tractable Gaussian prior $p(\bm{z}_{N_\tau})=\mathcal{N}(\bm{z}_\tau;\bm{0},\bm{I})$ to the target distribution over $N_\tau$ steps,
 \begin{equation}\label{diffusion1}
    p_\theta(\bm{z}_{0:N_\tau})=p(\bm{z}_{N_\tau})\prod_{\tau=1}^{N_\tau}p_\theta(\bm{z}_{\tau-1}|\bm{z}_{\tau}),
 \end{equation}
 where the transition kernel at each diffusion step $\tau$ is formulated as a Gaussian conditional density, with its mean and variance parameterized by a neural network with learnable weights $\theta$,
 \begin{equation}\label{diffusion2}
     p_\theta(\bm{z}_{\tau-1}|\bm{z}_{\tau})=\mathcal{N}(\bm{z}_{\tau-1};\bm{\mu}_\theta(\bm{z}_\tau,\tau),\Sigma_\theta(\bm{z}_\tau,\tau)\bm{I}).
 \end{equation}
The above process is referred to as the reverse denoising process. To train the denoising network, a forward diffusion process is defined to progressively corrupt the latent variable $\bm{z}_0$ through a Markov chain,
\begin{equation}
    q(\bm{z}_{1:N_\tau}|\bm{z}_0)=\prod_{\tau=1}^{N_\tau}q(\bm{z}_\tau|\bm{z}_{\tau-1}),
\end{equation}
where the transition kernel is a Gaussian conditional density with fixed variance schedule $\beta_\tau$,
\begin{equation}\label{diffusion4}
    q(\bm{z}_\tau|\bm{z}_{\tau-1})=\mathcal{N}(\bm{z}_\tau;\sqrt{1-\beta_\tau}\bm{z}_{\tau-1},\beta_\tau\bm{I}).
\end{equation}
With sufficient perturbation steps, the marginalized distribution $q(\bm{z}_{N_\tau}|\bm{z}_0)$ converges to a zero-mean Gaussian distribution $p(\bm{z}_{N_\tau})=\mathcal{N}(\bm{z}_\tau;\bm{0},\bm{I})$, which aligns with the prior used in the reverse process and is trival to sample. It is noteworthy that the forward process admits a closed-form expression at any noise level $\tau$ conditioned on the initial latent $\bm{z}_0$ using the re-parameterization trick \cite{kingma_auto-encoding_2022},
\begin{align}
    q(\bm{z}_\tau|\bm{z}_0)&=\mathcal{N}\left(\bm{z}_\tau;\sqrt{\bar{\alpha}_\tau}\bm{z}_0,(1-\bar{\alpha}_\tau)\bm{I}\right), \label{diffusion5}\\
    \bm{z}_\tau&=\sqrt{\bar{\alpha}_\tau}\bm{z}_0+\sqrt{1-\bar{\alpha}_\tau}\bm{\epsilon},\label{diffusion6}
\end{align}
where $\bm{\epsilon}\sim\mathcal{N}(\bm{0}, \bm{I})$ is the isotropic Gaussian noise, $\alpha_\tau=1-\beta_\tau$, and $\bar{\alpha}_\tau=\prod_{s=1}^\tau\alpha_s$.

The neural network is trained by optimizing the variational lower bound ($L_\mathrm{vlb}$) on the negative log likelihood (derivation can be found in \cite{sohl-dickstein_deep_2015}),
\begin{align}
     \min_\theta \mathbb{E}[-\log{p_\theta(\bm{z}_0)}] \leq L_{\mathrm{vlb}} &= \mathbb{E}_q\left[-\log{\frac{p_\theta(\bm{z}_{0:N_\tau})}{q(\bm{z}_{1:N_\tau}|\bm{z}_0)}}\right] \nonumber \\
    & = \mathbb{E}_q\bigg[D_{\mathrm{KL}}(q(\bm{z}_{N_\tau}|\bm{z}_0) \parallel p(\bm{z}_{N_\tau})) \nonumber \\& + \sum_{\tau=2}^{N_\tau}D_{\mathrm{KL}}(q(\bm{z}_{\tau-1}|\bm{z}_\tau,\bm{z}_0) \parallel p_\theta(\bm{z}_{\tau-1}|\bm{z}_\tau)) - \log{p_\theta(\bm{z}_0|\bm{z}_1)}\bigg],
\end{align}
where $D_{\mathrm{KL}}(\cdot\parallel\cdot)$ denotes the Kullback-Leibler (KL) divergence operation. The first term of $L_{\mathrm{vlb}}$ contains no trainable parameters and therefore can be omitted. The forward posterior $q(\bm{z}_{\tau-1}|\bm{z}_\tau,\bm{z}_0)$ is tractable when conditioned on $\bm{z}_0$ by using Bayes' rule and Eqs. (\ref{diffusion4}-\ref{diffusion6}),
\begin{equation}
    q(\bm{z}_{\tau-1}|\bm{z}_\tau,\bm{z}_0) = \frac{q(\bm{z}_{\tau}|\bm{z}_{\tau-1})q(\bm{z}_{\tau-1}|\bm{z}_0)}{q(\bm{z}_\tau|\bm{z}_0)} 
    = \mathcal{N}(\bm{z}_{\tau-1};\tilde{\bm{\mu}}_\tau(\bm{z}_\tau,\bm{\epsilon}),\tilde{\beta}_\tau\bm{I}),
\end{equation}
where
\begin{align}
    \tilde{\bm{\mu}}_\tau(\bm{z}_\tau,\bm{\epsilon})&=\frac{1}{\sqrt{\alpha_\tau}}\left(\bm{z}_\tau-\frac{1-\alpha_\tau}{\sqrt{1-\bar\alpha_\tau}}\bm{\epsilon}\right), \\
    \tilde{\beta}_\tau&=\frac{1-\bar{\alpha}_{\tau-1}}{1-\bar{\alpha}_\tau}\beta_\tau.
\end{align}
Since the two distributions in the second term of $L_\mathrm{vlb}$ are Gaussian, the KL divergence can be evaluated in an analytical form. Ho et al. \cite{ho_denoising_2020} set the variance $\Sigma_\theta(\bm{z}_\tau,t)=\beta_\tau$ as non-trainable constant, and parametrized the mean as follows,
\begin{equation}
    \bm{\mu}_\theta(\bm{z}_\tau,\tau)=\frac{1}{\sqrt{\alpha_\tau}}\left(\bm{z}_\tau-\frac{1-\alpha_\tau}{\sqrt{1-\bar\alpha_\tau}}\bm{\epsilon}_\theta(\bm{z}_\tau,\tau)\right),
\end{equation}
where the noise function $\bm{\epsilon}_\theta(\bm{z}_\tau,\tau)$ is approximated by a U-Net variant with residual blocks, self-attention, and diffusion step embedding. The loss function can then be simplified as,
\begin{equation}
    L_{\mathrm{simple}}=\mathbb{E}_{\tau,\bm{z}_0,\bm{\epsilon}}\left[\left\|\bm\epsilon-\bm\epsilon_\theta(\sqrt{\bar\alpha_\tau}\bm{z}_0+\sqrt{1-\bar\alpha_\tau}\bm\epsilon,\tau)\right\|^2\right],
\end{equation}
where $\tau$ is randomly sampled from a discrete uniform distribution $\mathcal{U}(1,N_\tau)$ and $\bm\epsilon$ is randomly sampled from a standard Gaussian distribution $\mathcal{N}(\bm{0},\bm{I})$. The $t=1$ case corresponds to the last term of $L_\mathrm{vlb}$, while the $t>1$ cases correspond to the second term. However, $L_\mathrm{simple}$ ignores the effect of the reverse kernel's covariance, which potentially undermines the learning efficacy at the initial diffusion steps \cite{nichol_improved_2021}. To this end, a hybrid form of loss function is adopted in this work to account for both mean and variance approximation, resulting in the following optimization objective,
\begin{equation}
    \theta^*=\argmin_\theta \sum_{\mathscr{E}_{\zeta^*,\gamma^*}(\bm{z}_0)\in\mathcal{A}_\mathrm{train}}\bigg(L_\mathrm{simple}+\lambda L_\mathrm{vlb}\bigg),
\end{equation}
where $\lambda$ is a weighting coefficient and $L_\mathrm{vlb}$ is incorporated to guide the network to learn $\Sigma_\theta(\bm{z}_\tau,t)$. Specifically, in additional to the noise function, the network also outputs a vector $\bm{v}$, which functions as an interpolation coefficient between $\beta_\tau$ and $\tilde{\beta}_\tau$, i.e., $\Sigma_\theta(\bm{z}_\tau,t)=\exp(\bm{v}\log{\beta_\tau} + (1-\bm{v})\log{\tilde{\beta}_\tau})$.

\subsection{Bayesian posterior sampling}
Once trained, the diffusion model can generate new realizations of $\bm{z}_0$ by sampling a white noise $\bm{z}_{N_\tau}$ from an isotropic multivariant Gaussian distribution $\mathcal{N}(\bm{0},\bm{I})$, and progressively denoising it through the learned reverse transition kernel $p_{\theta^*}(\bm{z}_{\tau-1}|\bm{z}_{\tau})=\mathcal{N}(\bm{z}_{\tau-1};\bm{\mu}_{\theta^*}(\bm{z}_\tau,\tau),\Sigma_{\theta^*}(\bm{z}_\tau,\tau)\bm{I})$ as,
\begin{equation}
    \bm{z}_{\tau-1}=\frac{1}{\sqrt{\alpha_\tau}}\left(\bm{z}_\tau-\frac{1-\alpha_\tau}{\sqrt{1-\bar\alpha_\tau}}\bm{\epsilon}_{\theta^*}(\bm{z}_\tau,\tau)\right) + \sqrt{\Sigma_{\theta^*}(\bm{z}_\tau,\tau)}\bm{\epsilon},\ \bm\epsilon\sim\mathcal{N}(\bm{0},\bm{I}), \ \tau=N_\tau,...,1.
\end{equation}
This unconditional generation process aligns with the Langevin dynamics sampling in score-based generative models by introducing the Stein score function \cite{song_generative_2019},
\begin{equation}\label{stein}
    \bm{s}_\theta(\bm{z}_\tau,\tau) \coloneq \nabla_{\bm{z}_\tau}\log p_\theta(\bm{z}_\tau) = -\frac{1}{\sqrt{1-\bar\alpha_\tau}}\bm{\epsilon}_\theta(\bm{z}_\tau,\tau),
\end{equation}
and the denoising function can be rewritten as,
\begin{equation}\label{score_sampling}
    \bm{z}_{\tau-1}=\frac{1}{\sqrt{\alpha_\tau}}\bigg(\bm{z}_\tau+(1-\alpha_\tau)\bm{s}_{\theta^*}(\bm{z}_\tau,\tau)\bigg) + \sqrt{\Sigma_{\theta^*}(\bm{z}_\tau,\tau)}\bm{\epsilon},\ \bm\epsilon\sim\mathcal{N}(\bm{0},\bm{I}), \ \tau=N_\tau,...,1.
\end{equation}
The generated latent $\bm{z}_0$ is then decoded back to the physical space via the trained CNF decoder, $\bm{\Phi}=\mathscr{E}_{\zeta^*,\gamma^*}(\bm{x},\bm{z}_0)$. The unconditional generation process can yield synthetic geological parameters $\bm{M}$ along with the corresponding reservoir responses $\bm{U}$ simultaneously, as the joint distribution of $\bm{M}$ and $\bm{U}$ is learned by CoNFiLD-geo after the unconditional training.

Compared to unconditional generation, conditional generation offers deeper insight in geoscientific modeling by leveraging available field observations to quantify the uncertainty inherent in geomodels, thereby facilitating robust decision-making in practical field applications. In the context of subsurface multiphase flow, observations may arise from sparse measurements, low-resolution seismic monitoring, well logging data, or any other accessible field information pertaining to the geomodel and its associated reservoir responses. Let $\bm{\Psi}\in\mathbb{R}^{N_{\bm\Psi}}$ be a condition vector from either the geomodel $\bm{M}$, the reservoir responses $\bm{U}$, or both. The choice of $\bm\Psi$ depends on the type of observations in the specific application. The conditional generation can be conceptualized as a Bayesian inverse problem, which involves sampling from the posterior distribution,
\begin{equation}\label{dps-bayes_formula}
    p(\bm{\Phi}|\bm\Psi)\propto p(\bm\Psi | \bm\Phi)p(\bm\Phi),
\end{equation}
where the prior $p(\bm\Phi)\approx p(\bm\Phi;
\theta^*,\zeta^*,\gamma^*)$ has been learned by CoNFiLD-geo through uncontional training. The relationship between $\bm\Psi$ and $\bm\Phi$ can be generally formulated as \cite{oliver_inverse_2008},
\begin{equation}
    \bm\Psi=\mathcal{F}(\bm\Phi) + \bm{\epsilon}_c = \mathcal{F}\bigg(\mathscr{E}_{\zeta^*,\gamma^*}(\bm{z}_0;\zeta^*,\gamma^*)\bigg) + \bm\epsilon_c,
\end{equation}
where $\mathcal{F}:\ \mathbb{R}^{N_d\times N_t}\rightarrow\mathbb{R}^{N_{\bm\Psi}}$ is a nonlinear mapping from the full field $\bm\Phi$ to the partial observations $\bm\Psi$, $\bm{\epsilon}_c$ denotes the observational error arising from  monitoring uncertainty, typically modeled as  $\bm{\epsilon}_c\sim\mathcal{N}(\bm{0},\sigma_c^2\bm{I})$. Concretely, $\mathcal{F}$ can take various forms depending on the specific application. For instance, $\mathcal{F}$ may act as a downsampling operator when reconstructing fields from low-resolution observations, or as a masking operator that selects specific coordinates when inferring complete fields from sparse or noisy measurements. The likelihood function in Eq. (\ref{dps-bayes_formula}) then takes the form,
\begin{equation}\label{likelihood}
    p(\bm\Psi | \bm\Phi)=\frac{1}{\sqrt{(2\pi)^{N_{\bm\Psi}}\sigma_c^{2N_{\bm\Psi}}}}\exp{\left[-\frac{\|\bm\Psi-\mathcal{F}(\bm\Phi)\|^2}{2\sigma_c^2}\right]}.
\end{equation}
In the latent space, conditional generation corresponds to modifying the score function as $\nabla_{\bm{z}_\tau}p(\bm{z}_\tau|\bm\Psi)$, which can be further decomposed by Bayes' rule,
\begin{equation}\label{dps-zong}
    \nabla_{\bm{z}_\tau}p(\bm{z}_\tau|\bm\Psi)=\nabla_{\bm{z}_\tau}p(\bm{z}_\tau) + \nabla_{\bm{z}_\tau}p(\bm\Psi|\bm{z}_\tau),
\end{equation}
where the first term can be viewed as the pre-trained unconditional generative model $\bm{s}_{\theta^*}(\bm{z}_\tau,\tau)$, and the second term serves as a guidance that steers the generated samples towards satisfying the condition $\bm\Psi$. To deduce the analytical form of the guidance term, we first factorize it by,
\begin{equation}
    p(\bm\Psi|\bm{z}_\tau)
    =\int p(\bm\Psi|\bm{z}_0,\bm{z}_\tau)p(\bm{z}_0|\bm{z}_\tau)d\bm{z}_0
    =\int p(\bm\Psi|\bm{z}_0)p(\bm{z}_0|\bm{z}_\tau)d\bm{z}_0,
\end{equation}
which can be approximated by Jensen's inequality,
\begin{equation}
    p(\bm\Psi|\bm{z}_\tau) = 
    \mathbb{E}_{\bm{z}_0\sim p(\bm{z}_0|\bm{z}_\tau)}[p(\bm\Psi|\bm{z}_0)]
    \simeq p(\bm\Psi|\mathbb{E}_{\bm{z}_0\sim p(\bm{z}_0|\bm{z}_\tau)}[\bm{z}_0]).
\end{equation}
Recalling the forward diffusion function in Eq. (\ref{diffusion6}) and the relation in Eq.(\ref{stein}), the unique mean of $p(\bm{z}_0|\bm{z}_\tau)$ can be derived by Tweedie's formula~\cite{chung_diffusion_2024, efron_tweedies_2011},
\begin{equation}
    \hat{\bm{z}}_0 \coloneq \mathbb{E}[\bm{z}_0|\bm{z}_\tau]=\mathbb{E}_{\bm{z}_0\sim p(\bm{z}_0|\bm{z}_\tau)}[\bm{z}_0]=\frac{1}{\sqrt{\bar{\alpha}_\tau}}\bigg(\bm{z}_\tau + (1-\bar\alpha_\tau)\nabla_{\bm{z}_\tau}\log p(\bm{z}_\tau)\bigg),
\end{equation}
where the unconditional score function $\nabla_{\bm{z}_\tau}\log p(\bm{z}_\tau)$ can be approximated by $\bm{s}_{\theta^*}(\bm{z}_\tau,\tau)$. Now, since $p(\bm\Psi|\bm{z}_\tau)\simeq p(\bm\Psi|\hat{\bm{z}}_0)$, and recalling the formulation in Eq.~(\ref{likelihood}), we obtain the likelihood function in the latent space based on the nested probabilistic relationship,
\begin{align}\label{chainrule_eq}
    \nabla_{\bm{z}_\tau}\log p(\bm\Psi|\bm{z}_\tau)&\simeq
    \nabla_{\bm{z}_\tau}\log p_{\theta^*,\zeta^*,\gamma^*}(\bm\Psi|\bm{z}_\tau)\nonumber\\&=
    -\frac{1}{\sigma_c^2}\nabla_{\bm{z}_\tau}\bigg\|\bm\Psi-\mathcal{F}\bigg(\mathscr{E}_{\zeta^*,\gamma^*}\big(\hat{\bm{z}}_0^*(\bm{z}_\tau,\tau;\theta^*);\zeta^*,\gamma^*\big)\bigg)\bigg\|^2,
\end{align}
where $\hat{\bm{z}}_0^*(\bm{z}_\tau,\tau;\theta^*)$ is the approximated clean latent provided by the pre-trained unconditional generative model. Eq. (\ref{chainrule_eq}) can be computed via the chain rule,
\begin{equation}\label{AD}
    \nabla_{\bm{z}_\tau}\log p_{\theta^*,\zeta^*,\gamma^*}(\bm\Psi|\bm{z}_\tau)=
    -\frac{2}{\sigma_c^2}(\bm\Psi-\mathcal{F}(\mathscr{E}_{\zeta^*,\gamma^*}))\frac{\partial\mathcal{F}(\mathscr{E}_{\zeta^*,\gamma^*})}{\partial{\mathscr{E}_{\zeta^*,\gamma^*}}}\frac{\partial{\mathscr{E}_{\zeta^*,\gamma^*}}(\hat{\bm{z}}_0^*)}{\partial\hat{\bm{z}}_0^*}\frac{\partial\hat{\bm{z}}_0^*(\bm{z}_\tau,\tau;\theta^*)}{\partial\bm{z}_\tau},
\end{equation}
through automatic differentiation (AD) as long as $\mathcal{F}$ is fully differentiable. This can be seamlessly implemented in modern differentiable programming frameworks such as PyTorch \cite{paszke_pytorch_2019}.

Therefore, the conditional score function in Eq. (\ref{dps-zong}) now comes at,
\begin{align}
    \nabla_{\bm{z}_\tau}p(\bm{z}_\tau|\bm\Psi)&\simeq\bm{s}_{\theta^*}(\bm{z}_\tau,\tau) + \nabla_{\bm{z}_\tau}\log p_{\theta^*,\zeta^*,\gamma^*}(\bm\Psi|\bm{z}_\tau)=s^{\mathrm{guide}}_{\theta^*,\zeta^*,\gamma^*}(\bm\Psi,\bm{z}_\tau,\tau),
\end{align}
where $s^{\mathrm{guide}}_{\theta^*,\zeta^*,\gamma^*}(\bm\Psi,\bm{z}_\tau,\tau)$ is the guided score function for the conditional generation. Replacing $\bm{s}_{\theta^*}(\bm{z}_\tau,\tau)$ in Eq. (\ref{score_sampling}) with $s^{\mathrm{guide}}_{\theta^*,\zeta^*,\gamma^*}(\bm\Psi,\bm{z}_\tau,\tau)$, we can generate $\bm\Phi$ given conditions on $\bm\Psi$ without retraining the latent diffusion model. This enables CoNFiLD-geo to synthesize diverse geological models $\bm{M}$ and their associated reservoir responses $\bm{U}$, conditioned on arbitrary unseen observations.  At inference time, only a pre-trained unconditional CoNFiLD-geo model is required, and conditional generation can then be performed in a plug-and-play manner. This distinctive capability paves the way for real-time UQ in practical GCS scenarios.

\section*{Acknowledgment}
The authors would like to acknowledge the startup funds from the College of Engineering at Cornell University.

\section*{Data availability}
All the data and codes needed to evaluate the conclusions in the paper will be made publicly available upon acception.

\section*{Compliance with ethical standards}
Conflict of Interest: The authors declare that they have no conflict of interest.

\bibliography{sn-bibliography}

\clearpage
\appendix


\section*{Supplementary material}

# Supplementary Notes

# Contents

# 1 Neural network architectures

The CoNFiLD-geo framework is constructed through a synergistic integration of two complementary neural networks: the Conditional Neural Field (CNF) and the Latent Diffusion Model (LDM). The architectural and training details of both components are elaborated in the following subsections

## 1.1 Conditional neural field architecture

The schematic of the neural field with full-projection conditioning is illustrated in Fig. S1. The sinusoidal representation network (SIREN; lower box) [1] is modulated in a layer-wise manner by a dedicated modulation network. The SIREN receives spatial coordinates $\boldsymbol{X}$ as input and predicts the physical field $\boldsymbol{\Phi}$ at the corresponding spatial locations. The modulation network (upper box) takes as input a learnable latent vector $\boldsymbol{L}$, which serves as a concise encoding of the underlying field. It consists of a series of linear layers that project the latent vector to each hidden layer of the SIREN, enabling consistent and expressive modulation throughout the network.

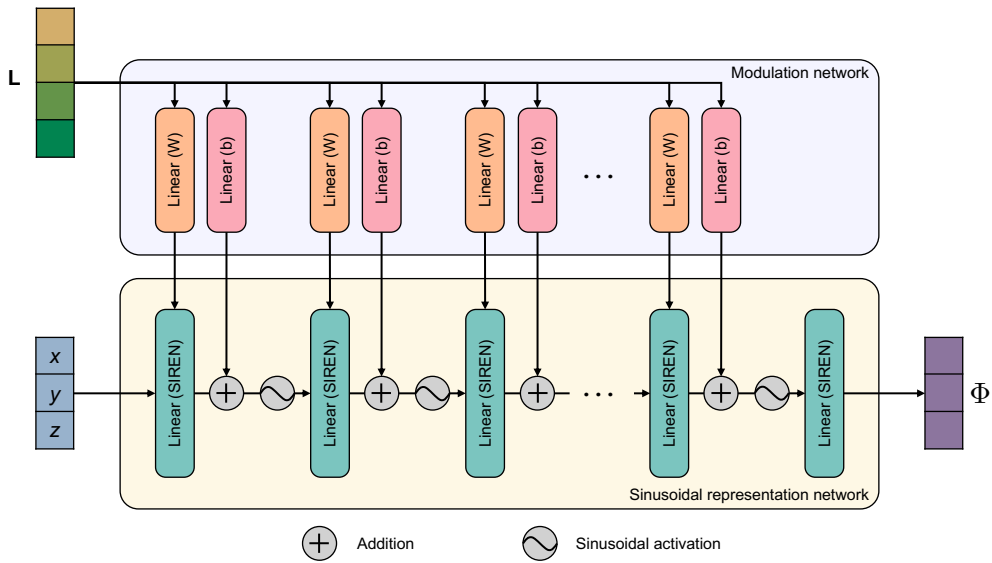

**Fig. S1** Conditional neural field architecture. The upper box is the modulation network. The lower box shows the SIREN base network.

## 1.2 Latent diffusion model architecture

We implement the improved U-Net architecture [2] as the denoising backbone in our diffusion model. It predicts the corresponding noise component $\boldsymbol{\epsilon}_\tau$ given a noisy latent image $\boldsymbol{z}_\tau$ at diffusion timestep $\tau$. As illustrated in Fig. S2, the U-Net is composed of hierarchical residual blocks interconnected via skip concatenations. The residual block contains two convolutional layers with group normalization and SiLU activation. Within each block, the intermediate tensor is modulated by the embedded diffusion timestep, enabling noise level conditioning throughout the network. Multi-head spatial attention blocks are incorporated at the lower-resolution stages to enhance the learning of fine-scale representations.

## 1.3 CoNFiLD-geo hyperparameters

The neural network and training hyperparameters of CoNFiLD-geo for the three cases are summarized in Table S1. The CNF is trained in a distributed manner across five NVIDIA RTX 4090 GPUs, while the LDM is trained on a single NVIDIA RTX 4090. Note that the hyperparameter $\omega_0$, which controls the frequency bandwidth of the SIREN, varies across the three cases. Its value is selected via empirical tuning to accommodate the spatiotemporal complexity of each scenario. We adopt an alternating training strategy for the CNF, in which the latent vectors are updated on a per-batch basis with the CNF parameters held fixed, followed by updating the CNF parameters while keeping the latent vectors temporarily frozen. Moreover, we assign different learning rates to the CNF and the latent vectors, with the latter set to be ten times larger. This setting helps mitigate overfitting in the CNF and promotes stable convergence during training [3].

**Fig. S2** U-Net architecture of the LDM. PE denotes positional embedding. MLP stands for multi-layer perceptron. Down and Up refer to downsampling and upsampling operations, respectively. The residual block and the multi-head spatial attention block are depicted in the lower left and lower right panels, respectively.

**Table S1** Hyperparameters of CoNFiLD-geo. The U-Net channel configuration specifies the number of feature channels in the residual blocks across six stages of the network.

| Case | | 2D synthetic reservoir | 3D Sleipner reservoir | 3D stratigraphically complex reservoir |
|---|---|---|---|---|
| CNF | Latent size ($N_l$) | 256 | 256 | 384 |
| | Hidden layer size | | 128 | |
| | Number of hidden layers | | 5 | |
| | SIREN $\omega_0$ | 5 | 15 | 20 |
| | Batch size | 600 | 300 | 250 |
| | Training epochs | | 5000 | |
| | Latent learning rate | | $1 \times 10^{-3}$ | |
| | Network learning rate | | $1 \times 10^{-4}$ | |
| LDM | U-Net channels | | [64,64,128,128,256,256] | |
| | Latent image size ($N_t \times N_l$) | $64 \times 256$ | $128 \times 256$ | $128 \times 384$ |
| | Batch size | 8 | 32 | 16 |
| | Noise schedule | | Cosine | |
| | Diffusion steps | | 1000 | |
| | Learning rate | | $5 \times 10^{-5}$ | |

3

## 2 Numerical model setup

In this section, we present the auxiliary constitutive relationships that complete the closed form of the governing equations. Detailed configurations of the numerical experiments are also provided. All simulations are performed using the ECO2N module of the finite-volume simulator TOUGH3 [4] on an Intel i9-13900K CPU workstation.

### 2.1 Constitutive relationships of subsurface multiphase flow

In a multiphase flow system, the capillary pressure $P_c(\leq 0)$ measures the pressure difference between two fluid phases that arises due to the interfacial tension between the fluid phases,

$$P_c = P_w - P_{nw}, \tag{1}$$

where $P_w$ is the pressure of the wetting phase (e.g., the aqueous brine phase) and $P_{nw}$ is the pressure of the non-wetting phase (e.g., the gaseous $CO_2$ phase). The van Genuchten [5] capillary pressure relationship is used in this work,

$$P_c = -P_0\big((S^*)^{-\frac{1}{\lambda}} - 1\big)^{1-\lambda}, \tag{2}$$

subject to the restriction,

$$-P_{max} \leq P_c \leq 0, \tag{3}$$

where $S^* = (S_l - S_{lr})/(1 - S_{lr})$ is the effective saturation, $S_{lr}$ is the irreducible water saturation, $P_{max}$ is the maximum capillary pressure, $P_0$ is the entry capillary pressure, and $\lambda$ is the exponential constant.

Relative permeability quantifies the effective permeability of a given fluid phase in a multiphase porous medium, normalized by its permeability under single-phase flow conditions. The Corey's [6] relationship is chosen to characterize the $CO_2$-brine system,

$$k_{r,l} = \hat{S}^4, \tag{4}$$

$$k_{r,g} = (1 - \hat{S})^2(1 - \hat{S}^2), \tag{5}$$

where $\hat{S} = (S_l - S_{lr})/(1 - S_{lr} - S_{gr})$ is the effective saturation and $S_{gr}$ is the residual gaseous saturation.

The above constitutive relationships, together with the governing mass conservation law, form a closed system of partial differential equations (PDEs) that can be solved numerically using methods such as the finite volume method. The specific values of the constitutive parameters for the three cases are detailed in Table S2.

**Table S2** Constitutive parameter values.

| Capillary pressure parameters | Value |
|---|---|
| $\lambda$ | 0.254 |
| $S_{lr}$ | 0.1 |
| $P_0$ | 19,600 Pa |
| $P_{max}$ | 12,500 Pa |
| Relative permeability parameters | Value |
| $S_{lr}$ | 0.1 |
| $S_{gr}$ | 0.01 |

### 2.2 Case 1: $CO_2$ drainage in heterogeneous reservoirs

Case 1 simulates the migration of $CO_2$ in a synthetic heterogeneous 2D reservoir. A horizontal saline aquifer spanning $640 \times 640$ m$^2$ is discretized using a $64 \times 64$ uniform grid. The initial in-situ pore pressure is 12 MPa with a constant temperature of 45 °C. The salinity of the aquifer is assumed to be 15% by weight. The porosity has a constant value of 0.2. No-flux boundary conditions are imposed at the top and bottom boundaries. A constant $CO_2$ injection rate of 0.45 kg/s is prescribed at the left boundary, while the right boundary is set to maintain constant hydraulic pressure and saturation. The simulation spans 500 days and consists of 64 time steps. Detailed model setup is also list in Table S3.

**Table S3** Numerical setup for Case 1.

| Domain size | Number of mesh | Simulation time | Number of time steps | $CO_2$ injection rate |
|---|---|---|---|---|
| 640 m × 640 m | 64 × 64 | 500 days | 64 | 0.45 kg/s |

The reservoir is characterized by random log-normal permeability fields to represent a wide range of geological heterogeneity. The distribution has a mean of ln(100) (corresponding to a geometric mean of 100 mD) and a standard deviation of 1.0. An isotropic correlation length of 80 m is assumed along the horizontal directions. A total of 2,000 permeability field realizations are obtained using the Gaussian covariance model of the open-source geostatistical Python package GSTools [7].

## 2.3 Case 2: field-scale $CO_2$ sequestration at the Sleipner site

In Case 2, we present a realistic GCS scenario based on an actual field-scale project to demonstrate the capability of CoNFiLD-geo in addressing practical geo-engineering problems. This case study serves as a representative testbed for evaluating the framework's performance under realistic geological complexity and operational constraints. This subsection provides an overview of the Sleipner GCS project in Norway and details the setup of the numerical model.

Since October 1996, Statoil and its Sleipner partners have been injecting $CO_2$ — separated from natural gas extracted at the Sleipner field — into the Utsira Formation, a saline aquifer located at a depth of 1012 m below sea level [8]. The Utsira Formation primarily consists of highly porous, very permeable, and weakly consolidated sandstones, located at depths ranging from approximately 800 m to 1100 m. Around the injection site, the stratigraphy comprises nine sandstone units, designated Utsira L1 (the lowermost) to Utsira L9 (the uppermost), each separated by low-permeability shale layers (Fig. S3). Shale layers separating L1 through L8 are relatively thin ($\sim 1$ m), while the shale between L8 and L9 is significantly thicker ($\sim 7$ m). Overlying L9 is a thick shale formation with very low permeability, serving effectively as a caprock. Therefore, L9 forms a relatively isolated hydrological system from surrounding sandstone units, making it a favorable candidate for studying $CO_2$ migration within the reservoir.

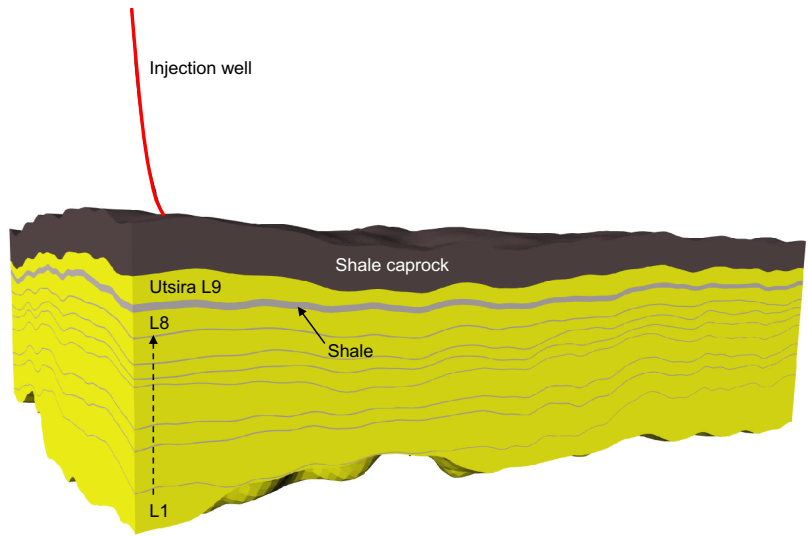

**Fig. S3** The Sleipner simulation model comprising all nine sandstone layers. The red line indicates the $CO_2$ injection well. The target layer used in this study, Utsira L9, is vertically isolated by low-permeability shale layers. Modified from [9].

We select Utsira L9 from the Sleipner 2019 Benchmark Model [9] as the target reservoir for simulation (Fig. S3). In the horizontal plane, the reservoir extends 3.2 km in the $x$-direction and 5.9 km in the $y$-direction. The reservoir topography (i.e., thickness) is interpreted from geophysical survey data, with an average thickness of approximately 50 m. The reservoir depth ranges from $-854$ m to $-770$ m. The lateral resolution of the model grid is set to $50 \times 50$ m, balancing the need to resolve fine-scale characteristics of

the $CO_2$ plume against the computational tractability. As the horizontal extent of the domain is significantly larger than its vertical thickness, vertical $CO_2$ plume dynamics are expected to have minimal impact. Accordingly, the model employs a single-layer grid in the vertical direction to reduce computational complexity while retaining the essential features of lateral migration. Despite employing only a single grid layer in the vertical direction, the injected $CO_2$ exhibits complex migration behavior as gravity is accounted for in the simulation. This complexity results from the interplay between buoyancy and lateral driving forces (e.g., viscous and capillary forces), compounded by the structural undulations of the reservoir. The perforated injection well is consistent with the actual project configuration. The injection rate is set to a constant value of 32 kg/s, equivalent to a storage capacity of 1 Mt/year, closely matching the target capacity of the reference site. Impermeable boundary conditions are applied at the top and bottom, reflecting the sealing effect of overlying and underlying shale formations. Constant hydraulic boundary conditions are imposed on the lateral sides to emulate connectivity with an infinite-acting aquifer. An injection period of 10 years is considered, which is discretized into 128 time steps. Detailed model setup is also summarized in Table S4.

**Table S4** Numerical setup for Case 2.

| Lateral domain size | Number of mesh | Simulation time | Number of time steps | $CO_2$ injection rate |
|---|---|---|---|---|
| 3.2 km × 5.9 km | 64 × 118 × 1 | 10 years | 128 | 32 kg/s |

An in-situ pressure of 9 MPa is prescribed based on the reservoir depth, and pressure equilibrium is established through a preliminary simulation phase prior to injection. The reservoir temperature is constant at 32 °C. The salinity of the pore water is 33500 ppm [10]. The Utsira sandstone has a high porosity of 0.36 [9]. Permeability is modeled as a spatial random field based on the Gaussian covariance model. Permeability values are sampled from a log-normal distribution with a logarithmic mean of $\ln 1000$ and a standard deviation of 0.4. The correlation lengths are set to 500 m and 1000 m in the $x$ and $y$ directions, respectively

## 2.4 Case 3: $CO_2$ injection and brine production in stratigraphically complex reservoirs

Case 3 represents an extension of Case 2 toward increased geological and operational complexity, with stochastically generated reservoir structures and the simultaneous modeling of $CO_2$ injection and brine production. While the reservoir shares the same lateral extent as Case 2, its vertical depth and thickness are stochastically defined using geostatistical tools to account for structural uncertainty. Specifically, the reservoir depth is modeled as a Gaussian random field with a mean of $-800$ m and a standard deviation of 200 m. Similarly, the thickness is modeled as a Gaussian random field with a mean of 30 m and a standard deviation of 60 m. The isotropic correlation lengths for depth and thickness are set to 400 m and 500 m, respectively. The relatively large correlation length for thickness ensures that no negative values occur, even though the standard deviation exceeds the mean. To better characterize the structural complexity of stratigraphically heterogeneous reservoirs, the model domain is discretized using an unstructured triangular mesh, allowing enhanced fidelity in representing complex reservoir geometry. Note that the unstructured mesh can be naturally handled by CNF, whereas traditional convolution-based methods are limited to structured grids. The domain is discretized into 7,720 triangular elements, which offer adequate spatial resolution to resolve the dynamic behavior of $CO_2$ in the reservoir. In Case 3, a closed hydrological system is assumed. To maintain safe reservoir pressure, an additional brine production well is introduced. The bottomhole pressure (BHP) of the production well is controlled at 9.5 MPa for effective pressure management. The model setup is also listed in Table S5. Porosity and permeability are assumed to be constant at 0.36 and 2000 mD, respectively. All other in-situ conditions are identical to those in Case 2.

**Table S5** Numerical setup for Case 3.

| Lateral domain size | Number of mesh | Simulation time | Number of time steps | $CO_2$ injection rate | Production BHP |
|---|---|---|---|---|---|
| 3.2 km × 5.9 km | 7720 | 10 years | 128 | 32 kg/s | 9.5 MPa |

# 3 Data restoration from incomplete observations

In this section, we demonstrate the capability of CoNFiLD-geo to perform data restoration, that is, to recover the complete field of interest from incomplete or missing observations. We consider two scenarios: (i) spatially occluded $CO_2$ plume monitoring data, and (ii) permeability fields with missing regions, as detailed in the following subsections. For simplicity, we adopt Case 1 as a representative test case. More complex scenarios can be seamlessly accommodated within the CoNFiLD-geo framework without requiring any structural modifications.

## 3.1 Spatially occluded $CO_2$ plume monitoring data

$CO_2$ plume monitoring data obtained from time-lapse seismic surveys may suffer from spatial incompleteness due to subsurface occlusions or acquisition-related issues such as equipment malfunction [11]. Such data loss can hinder accurate interpretation and prediction, necessitating reliable data restoration techniques. We showcase the novel application of CoNFiLD-geo to simultaneously infer the unknown permeability field and restore the complete $CO_2$ plume dynamics from spatially occluded plume monitoring data.

With the reference data shown in Fig.S4a, the damaged monitoring data is defined as removing a central subregion of the $CO_2$ saturation field (Fig.S4b). This degradation is mathematically formulated as a spatiotemporal masking operation. Fig. S4c presents three representative examples of the generated permeability fields along with their corresponding $CO_2$ saturation snapshots. CoNFiLD-geo can consistently restore the complex spatiotemporal $CO_2$ migration patterns within the occluded region, integrating smoothly with the surrounding observed data without noticeable discrepancies at the interfaces (Fig.S4c). The overall reconstructed $CO_2$ saturation fields appear physically plausible and closely match the reference, as further corroborated by the Structural Similarity Index Measure (SSIM) evaluation metric (Fig.S4e). Along the profiling line (dash-dot lines indicated in Fig.S4b), the variation in reconstructed saturation generally follows the reference trend, with slight discrepancies and elevated uncertainties within the occluded region (Fig.S4f,g). Moreover, CoNFiLD-geo is capable of inversely inferring the underlying permeability fields from the incomplete $CO_2$ saturation monitoring data. The high-permeability pathways are reasonably well captured by the model (Fig.S4c), showing a general agreement with the reference subsurface structure (Fig.S4d).

## 3.2 Permeability fields with missing region

For field-scale GCS projects, petrophysical properties, such as permeability, may be entirely unknown in certain regions due to the absence of geological surveys or the loss of historical subsurface data. This lack of information is common in areas with limited exploration, inaccessible terrains, or incomplete data archives, posing significant challenges for accurate reservoir characterization and predictive modeling. In such contexts, CoNFiLD-geo offers a promising solution for inferring missing information based on incomplete observations.

Take the same parameter-solution pair from testing set as reference data (Fig.S5a), we deliberately remove a central subregion of the permeability field and treat it as completely unknown (Fig.S5b). This spatially masked region serves as the conditioning input for CoNFiLD-geo, which aims to infer the missing permeability values and generate the corresponding spatiotemporal evolution of the $CO_2$ saturation field. The restored full permeability fields closely resemble the reference (Fig.S5c), with the mean probabilistic density function (PDF) of the generated samples aligning well with that of the reference (Fig.S5d). In addition, the permeability values along the profiling lines (dash-dot lines indicated in Fig. S5b) exhibit a generally consistent trend with the reference, suggesting that the model reasonably captures the spatial variation within the unknown region (Fig.S5f,g). The increase in uncertainty from the periphery to the center of the masked region (Fig.S5f,g) arises from stronger spatial covariance with the surrounding known geological information. Consequently, the uncertainty is lower near the boundaries and becomes more pronounced toward the center, where the influence of known data diminishes. The simultaneously predicted $CO_2$ plumes share similar overall spatiotemporal patterns with the reference, particularly in capturing the major fingering structures along the highly permeable pathway (Fig.S5a,c). The SSIM metric is initially high and gradually decreases over time as the $CO_2$ plume evolves and becomes more complex. Nevertheless, it remains consistently above 0.6 throughout the simulation, indicating reliable reconstruction performance (Fig.S5e).

In summary, the proposed CoNFiLD-geo framework enables flexible data restoration from either parameter or solution space via zero-shot conditional generation, with the ability to quantify associated uncertainties. This capability paves the way for broader applications in real-world subsurface modeling and monitoring tasks, particularly in scenarios involving damaged or missing data.

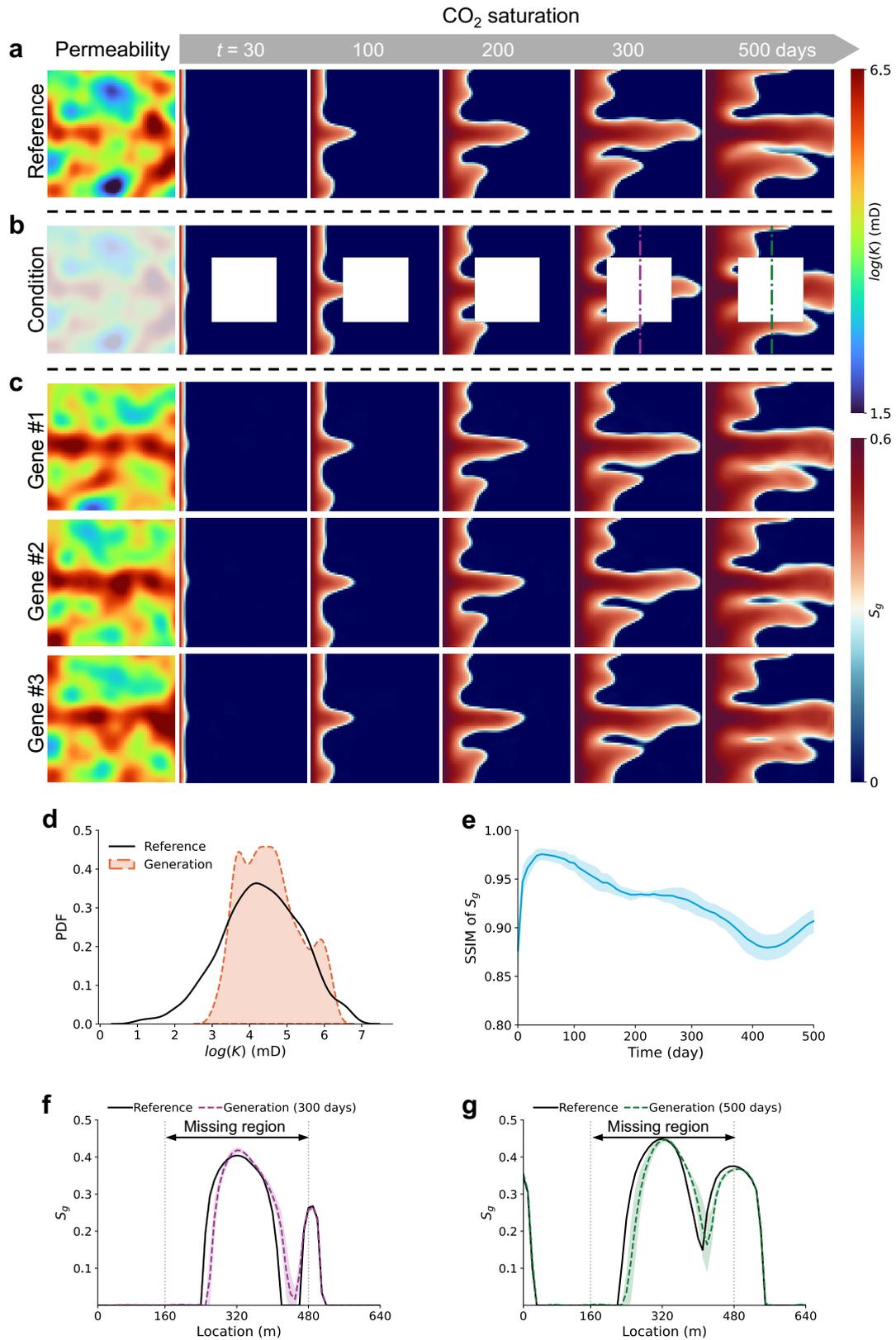

**Fig. S4** Data restoration under spatially occluded $CO_2$ plume monitoring data. (a) Reference permeability field and corresponding $CO_2$ saturation trajectory with snapshots at 30, 100, 200, 300, and 500 days. (b) Spatially occluded $CO_2$ plume observations used as conditional inputs for CoNFiLD-geo. (c) Three representative generated permeability fields and their corresponding $CO_2$ saturation dynamics. (d) Probabilistic density function (PDF) plot of the reference permeability field the sample-mean generated permeability fields. (e) Temporal variation of SSIM for $CO_2$ saturation, with shaded regions indicating standard deviation. (f) $CO_2$ saturation profile along the purple dash-dot line at 300 days (see panel b), with shaded regions denoting uncertainty. (g) Same as (f) but at 500 days.

8

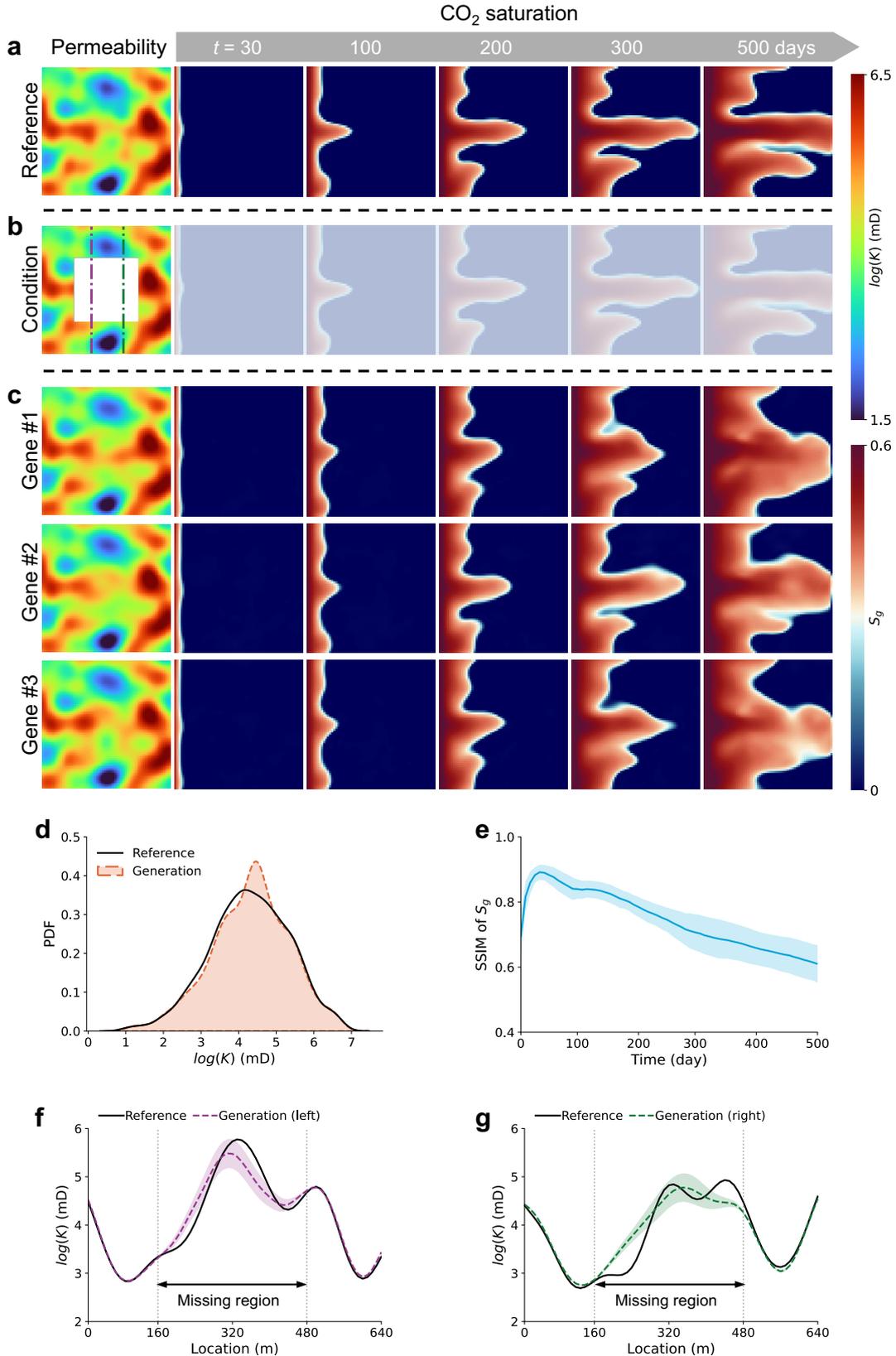

**Fig. S5** Data restoration under permeability with missing region. (a) Reference permeability field and corresponding $CO_2$ saturation trajectory with snapshots at 30, 100, 200, 300, and 500 days. (b) Permeability field with a missing central region used as conditional inputs for CoNFiLD-geo. (c) Three representative generated permeability fields and their corresponding $CO_2$ saturation dynamics. (d) PDF plot of the reference permeability field the sample-mean generated permeability fields. (e) Temporal variation of SSIM for $CO_2$ saturation, with shaded regions indicating standard deviation. (f) Permeability profile along the purple dash-dot line on the left-hand side (see panel b), with shaded regions denoting uncertainty. (g) Same as (f) but on the right-hand side.

9

# 4 Comparison with deterministic forward models

CoNFiLD-geo, by virtue of its probabilistic nature, serves as a generative modeling framework that approximates the conditional distribution of the target field given partial observations or inputs. In contrast, deterministic models yield only single-point predictions and inherently lack the capacity to characterize uncertainty. In the context of GCS, numerous studies have explored deep learning-based surrogate models as deterministic forward solvers to replace computationally expensive numerical simulations. In this section, we compare CoNFiLD-geo with the well-established U-FNO [12] surrogate model to demonstrate CoNFiLD-geo's capability for forward modeling with quantified uncertainty. This feature is particularly important in practical scenarios, where full knowledge of input parameters is often unavailable or subject to significant uncertainty.

## 4.1 Baseline U-FNO model

We adopt U-FNO [12], a U-Net enhanced Fourier Neural Operator (FNO) network, as our baseline deterministic model. U-FNO integrates the global modeling capabilities of neural operators with the local feature extraction strengths of Convolutional Neural Networks (CNNs), and has demonstrated superior performance over vanilla FNO [13] and CNNs in geological carbon sequestration tasks [12, 14].

The deterministic model learns a functional mapping from the parameter space to the solution space, i.e., $\mathcal{G}_\theta : \boldsymbol{M} \in \mathbb{R}^{N_d \times N_m} \mapsto \boldsymbol{U} \in \mathbb{R}^{N_d \times N_t \times N_u}$. A U-FNO network $\mathcal{G}_\theta$ with $I$ Fourier layers and $J$ U-Fourier layers is formulated as,

$$\mathcal{G}_\theta = \mathcal{Q} \circ \underbrace{\sigma(\mathcal{K}_J + \mathcal{U}_J + \mathcal{W}_J) \circ ... \circ \sigma(\mathcal{K}_1 + \mathcal{U}_1 + \mathcal{W}_1)}_{\text{U}-\text{Fourier layers}} \circ \underbrace{\sigma(\mathcal{K}_I + \mathcal{W}_I) \circ ... \circ \sigma(\mathcal{K}_1 + \mathcal{W}_1)}_{\text{Fourier layers}} \circ \mathcal{P}, \qquad (6)$$

where $\mathcal{P}$ and $\mathcal{Q}$ denote the linear lifting and projecting operator, respectively. In each intermediate layer, $\mathcal{K}$ denotes the kernel integral operator, $\mathcal{U}$ represents a U-Net CNN operator, $\mathcal{W}$ is a linear operator, and $\sigma$ is a nonlinear activation function. The kernel integral operator applied on layer input function $v_l$ is defined by,

$$\big(\mathcal{K}(v_l)\big)(x) \coloneqq \int \kappa(x, y) v_l(y) dy, \qquad (7)$$

and is parameterized in the Fourier space [13],

$$\big(\mathcal{K}(v_l)\big)(x) = \mathcal{F}^{-1}\big(\mathcal{R} \cdot \mathcal{F}(v_l)\big)(x), \qquad (8)$$

where $\mathcal{F}$ and $\mathcal{F}^{-1}$ represents the Fourier transformation and its inverse, respectively, and $\mathcal{R}$ is the Fourier transform of a periodic function $\kappa$.

**Table S6** U-FNO model architecture. The Padding layer is used to accommodate the non-periodic boundaries. The first two channels of the output tensor denote the spatial grids along the $x$ and $y$ directions, respectively, the third channel denotes time dimension, the last channel denotes the feature dimension. The input features include the permeability field and the time step information.

| Name | Layer | Output shape |
| --- | --- | --- |
| Input | - | (64, 64, 64, 2) |
| Padding | Padding | (72, 72, 72, 2) |
| Lifting | Linear | (72, 72, 72, 36) |
| Fourier 1 | Fourier3d/Conv1d/Add/ReLu | (72, 72, 72, 36) |
| Fourier 2 | Fourier3d/Conv1d/Add/ReLu | (72, 72, 72, 36) |
| Fourier 3 | Fourier3d/Conv1d/Add/ReLu | (72, 72, 72, 36) |
| U-Fourier 1 | Fourier3d/Conv1d/UNet3d/Add/ReLu | (72, 72, 72, 36) |
| U-Fourier 2 | Fourier3d/Conv1d/UNet3d/Add/ReLu | (72, 72, 72, 36) |
| U-Fourier 3 | Fourier3d/Conv1d/UNet3d/Add/ReLu | (72, 72, 72, 36) |
| Projection 1 | Linear | (72, 72, 72, 128) |
| Projection 2 | Linear | (72, 72, 72, 1) |
| De-padding | - | (64, 64, 64, 1) |

In this section, we utilize U-FNO as a data-driven surrogate model to approximate the spatiotemporal reservoir responses using the corresponding geological parameters. The input to U-FNO consists of the geological parameters concatenated with broadcasted time step scalars. The target output comprises $N_u$ state variables at $N_t$ time steps. In practice, we train $N_u$ identical neural networks from scratch for each state variable, as proposed in Wen et al [12]. Since U-FNO is constrained to structured data formats due to its reliance on convolution operations, we train it on the first 2D case. The detailed architecture of U-FNO is illustrated in Table S6.

U-FNO is trained as a forward deterministic mapping using full observations of $M$. Following the approach of Huang et al [15], during the testing stage, U-FNO is tasked with predicting $U$ given either a low-resolution $M$ or sparse measurements of $M$, in a manner consistent with CoNFiLD-geo. We present the comparison results in the following two subsections.

## 4.2 Forward modeling with low-resolution permeability

We begin with predicting the $CO_2$ saturation based on low-resolution permeability. Results for pressure are expected to exhibit analogous trends and are therefore omitted for brevity. This synthetic test case reflects practical scenarios where permeability information is only coarsely available — for example, as interpreted from sparse monitoring wells or derived through spatial averaging. As illustrated in Fig. S6a and b, CoNFiLD-geo consistently produces results that closely match the reference solutions, whereas U-FNO exhibits significant performance degradation as input resolution decreases, occasionally generating artifacts even at early time steps. It is noteworthy that, beyond predicting $CO_2$ saturation, CoNFiLD-geo is also capable of generating high-resolution permeability fields inferred from low-resolution inputs. This enhanced flexibility stems from the unified treatment of inverse and forward modeling within a single generative framework. The generated high-resolution permeability fields more closely resemble the reference fields as the input resolution increases, as expected (Fig S6c). CoNFiLD-geo can fully recover the reference field using only a $16 \times 16$ resolution input, highlighting its robustness to severely under-resolved inputs (Fig S6c). The SSIM and root mean square error (RMSE) metrics for $CO_2$ saturation over time are presented in Fig. S6d and e. While U-FNO yields only single-point predictions, CoNFiLD-geo produces ensembles that not only align more closely with reference data but also enable explicit uncertainty quantification. CoNFiLD-geo outperforms U-FNO at most time steps, and notably achieves superior performance across the entire time horizon when the input resolution is as low as $16 \times 16$. Overall, CoNFiLD-geo demonstrates clear advantages over U-FNO in terms of robustness to coarse inputs, unified forward and inverse modeling capability, and the ability to quantify predictive uncertainty.

## 4.3 Forward modeling with sparse measurements of permeability

We further evaluate the forward prediction performance of both models under sparse observations of the permeability field. In Fig. S7, we present the results for cases where only a small fraction of the permeability values are known, specifically 0.05%, 1%, and 3% of the total number of grid cells. CoNFiLD-geo can simultaneously recover the full permeability field as well as the spatiotemporal dynamics of the corresponding $CO_2$ saturation (Fig. S7a,b), offering a flexible forward and inverse modeling capability in a unified framework. Nevertheless, U-FNO fails to generate accurate predictions under sparse observations, yielding overly smoothed and physically implausible plume migration patterns (Fig. S7a,b). The PDF plots demonstrate that CoNFiLD-geo can produce decent estimation of the full permeability field using only 3% probed data, highlighting its robustness under extremely sparse observation conditions (Fig. S7c). The error metrics further showcase the superior forward prediction accuracy of CoNFiLD-geo compared to U-FNO (Fig. S7d,e). These results underscore the advantage of probabilistic modeling over deterministic approaches in handling realistic GCS scenarios, as it enables uncertainty quantification and robust inference under sparse or low-resolution observations, which are inherent capabilities of the proposed CoNFiLD-geo framework.

## 4.4 Forward modeling with fully observed permeability

We also include results under the condition of full permeability field availability, although such a setting is rarely attainable in realistic GCS scenarios. It is observed that CoNFiLD-geo is capable of generating decent permeability fields and consistent accurate $CO_2$ saturation dynamics compared to the reference (Fig. S8a,b). In this fully observed setting, the performance of CoNFiLD-geo and U-FNO is largely comparable (Fig. S8b,c). Based on the error metrics (Fig. S8d,e), CoNFiLD-geo shows slightly lower accuracy than U-FNO, particularly at later time steps. We argue that this trade-off is acceptable, as CoNFiLD-geo offers additional capabilities beyond deterministic prediction — most notably, the ability to handle partial

observations and to quantify uncertainties. These advantages are critical in realistic GCS scenarios where full knowledge of input parameters is rarely available.

It should be noted that the CoNFiLD-geo framework could be further enhanced by incorporating physics-informed sampling strategies [15–17]. Importantly, the incorporation of PDE constraints is confined to the sampling stage, preserving the zero-shot conditional generation capability of CoNFiLD-geo without requiring additional training. In such a case, the generated parameter–solution pairs would inherently satisfy the constraints imposed by the governing PDEs, thereby enabling the model to surpass deterministic approaches in terms of physical consistency and realism. This direction will be the focus of our future research.

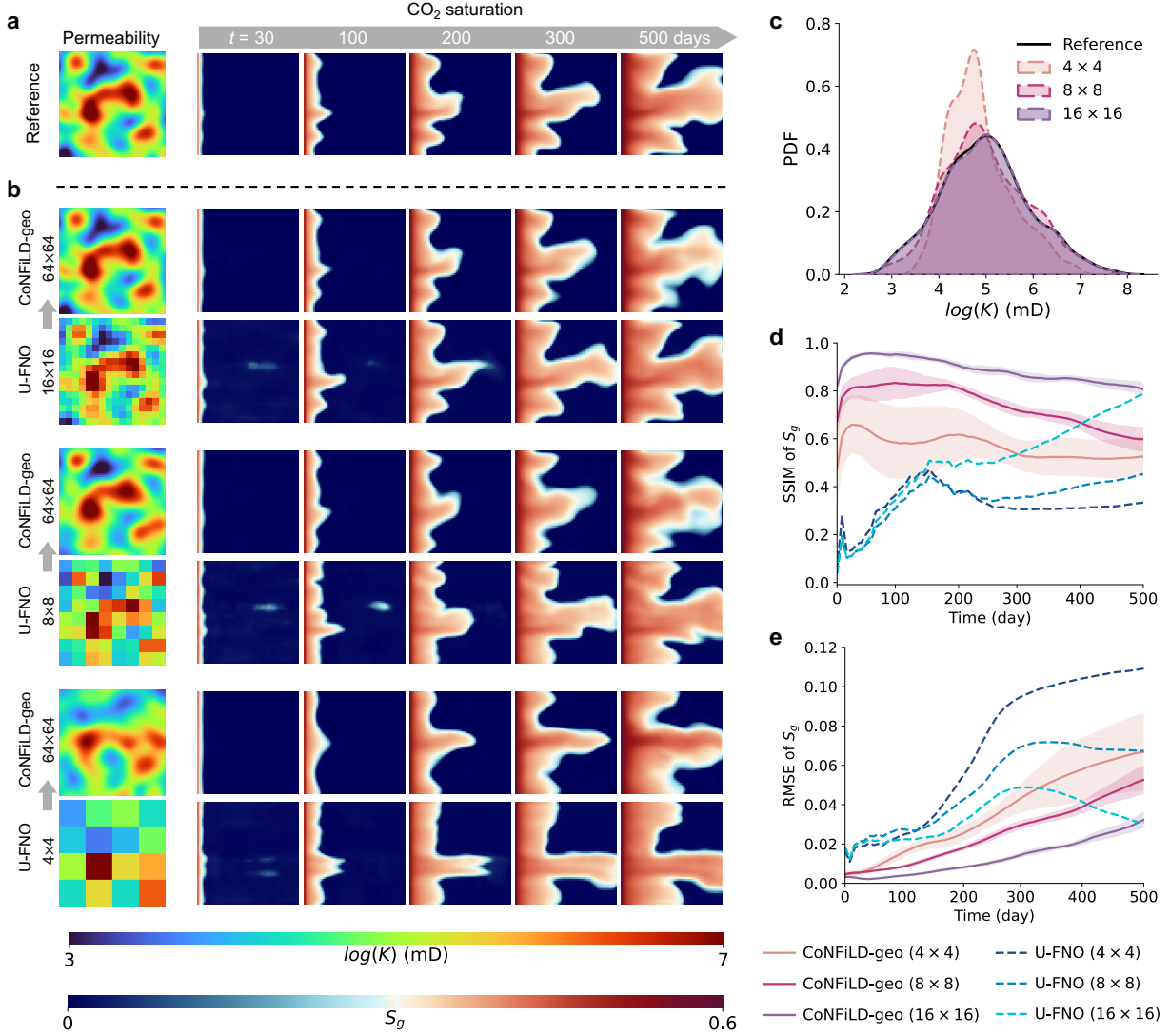

**Fig. S6** Comparison of forward modeling performance between CoNFiLD-geo and U-FNO under coarsely estimated permeability fields. (a) The reference high-resolution permeability field and $CO_2$ saturation trajectory, with snapshots taken at 30, 100, 200, 300, and 500 days. (b) From bottom to top, the resolution of the conditioning low-resolution permeability increases from $4 \times 4$ to $16 \times 16$. For each resolution, the first row shows the CoNFiLD-geo inferred permeability and the corresponding $CO_2$ saturation snapshots; the second row displays the low-resolution permeability input (used as both the input to U-FNO and the conditioning guidance for CoNFiLD-geo), along with the $CO_2$ saturation predicted by U-FNO. (c) PDF plots comparing the reference permeability and the CoNFiLD-geo inferred permeability under varying conditioning resolutions. (d) Temporal variation of SSIM for $CO_2$ saturation predicted by CoNFiLD-geo and U-FNO at varying input resolutions. The shaded regions denote the standard deviation. (e) Temporal variation of RMSE for $CO_2$ saturation predicted by CoNFiLD-geo and U-FNO at varying input resolutions. The shaded regions denote the standard deviation.

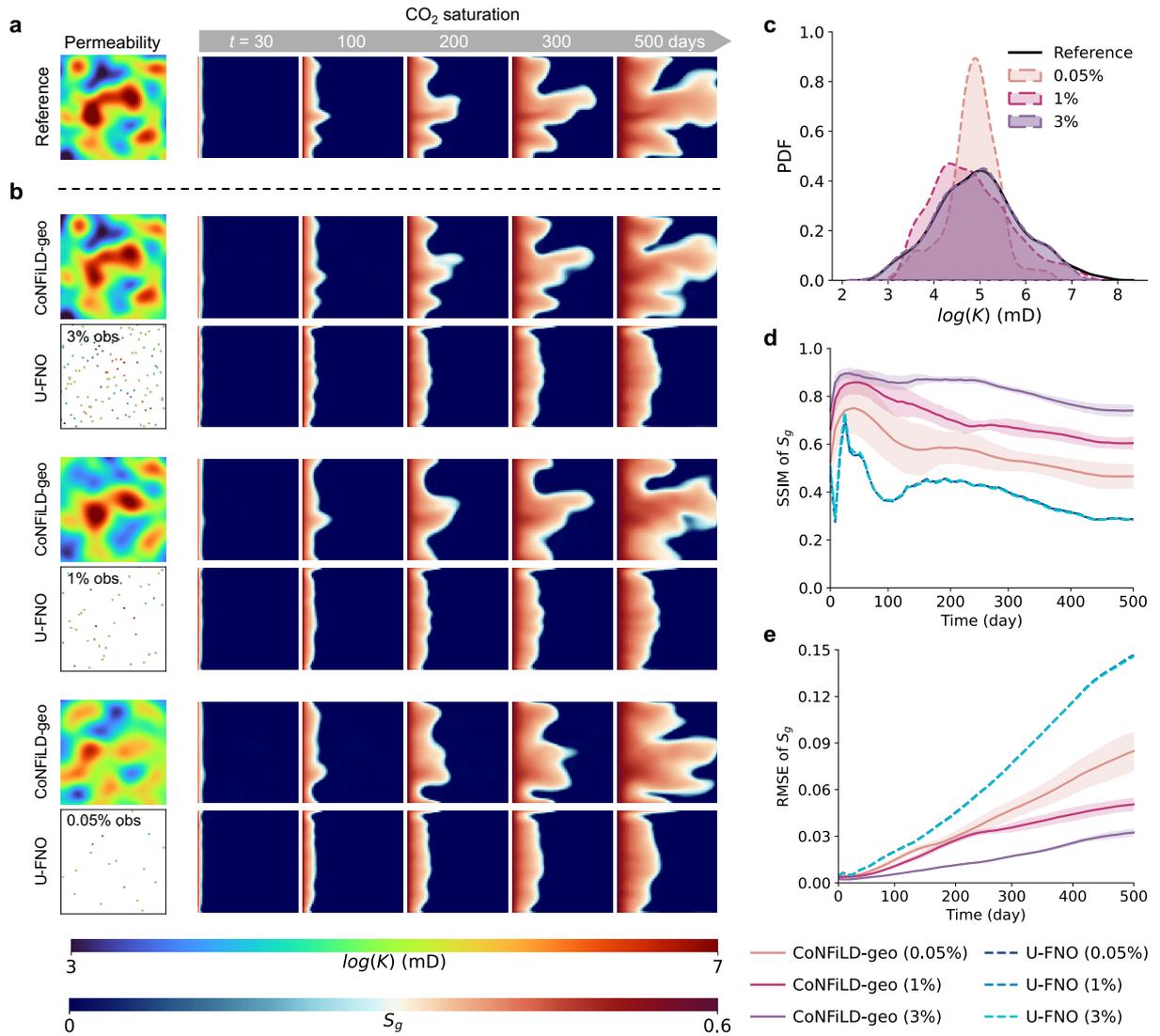

**Fig. S7** Comparison of forward modeling performance between CoNFiLD-geo and U-FNO under sparsely observed permeability fields. (a) The reference full permeability field and $CO_2$ saturation trajectory, with snapshots taken at 30, 100, 200, 300, and 500 days. (b) From bottom to top, the known fraction of the full permeability increases from 0.05% to 3%. For each level of sparsity, the first row shows the CoNFiLD-geo reconstructed permeability and the corresponding $CO_2$ saturation snapshots; the second row displays the sparsely observed permeability input (used as both the input to U-FNO and the conditioning guidance for CoNFiLD-geo), along with the $CO_2$ saturation predicted by U-FNO. (c) PDF plots comparing the reference permeability and the CoNFiLD-geo inferred permeability under different levels of observation sparsity. (d) Temporal variation of SSIM for $CO_2$ saturation predicted by CoNFiLD-geo and U-FNO across varying observation sparsity. The shaded regions denote the standard deviation. (e) Temporal variation of RMSE for $CO_2$ saturation predicted by CoNFiLD-geo and U-FNO across varying observation sparsity. The shaded regions denote the standard deviation.

13

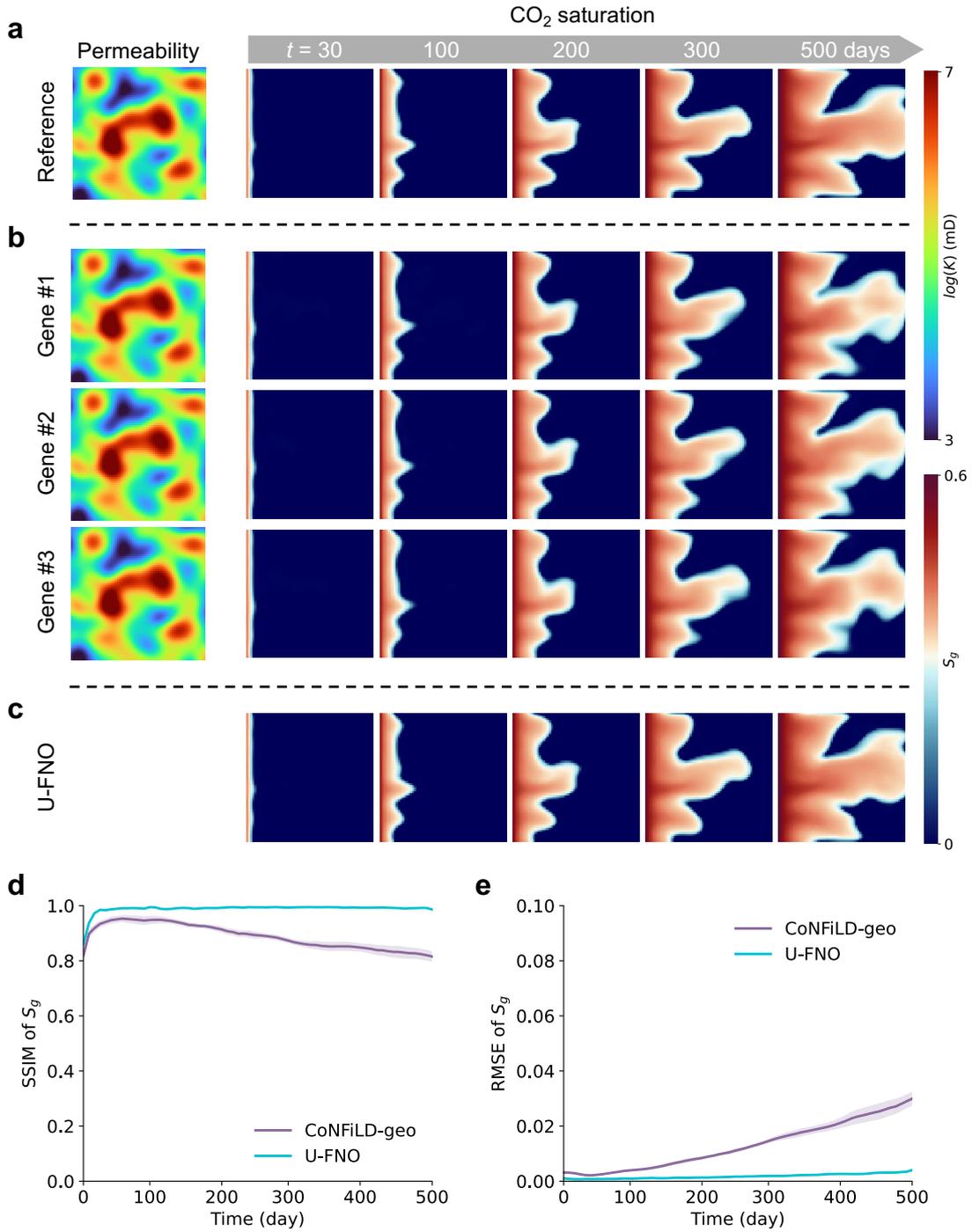

**Fig. S8** Comparison of forward modeling performance between CoNFiLD-geo and U-FNO under fully observed permeability fields. (a) The reference full permeability field and $CO_2$ saturation trajectory, with snapshots taken at 30, 100, 200, 300, and 500 days. (b) Three demonstrating generated permeability fields and their $CO_2$ saturation dynamics of CoNFiLD-geo. (c) U-FNO's predicted $CO_2$ saturation snapshots. (d) Temporal variation of SSIM for $CO_2$ saturation predicted by CoNFiLD-geo and U-FNO. The shaded regions denote the standard deviation. (e) Temporal variation of RMSE for $CO_2$ saturation predicted by CoNFiLD-geo and U-FNO. The shaded regions denote the standard deviation.

14

# 5 Dimension reduction using Proper Orthogonal Decomposition (POD)

The proposed CoNFiLD-geo framework consists of two primary components: a dimensionality reduction module (CNF) and a probabilistic modeling module (LDM). Owing to its modular design, each component can be flexibly replaced with alternative techniques depending on the specific application or domain requirements. In our previous work, we have evaluated the performance of LDM against other generative models such as variational autoencoders (VAEs) and generative adversarial networks (GANs) [18].

For the dimensionality reduction module, a variety of alternatives exist. Classical methods such as Proper Orthogonal Decomposition (POD) [19] approximate system states using a fixed set of orthogonal basis functions weighted by projection coefficients. Learning-based approaches, including convolutional neural networks (CNNs) [20] and graph neural networks (GNNs) [21], have also been widely adopted for surrogate modeling. CNNs are effective on structured grids but are limited to regular spatial domains, while GNNs can operate on unstructured meshes but typically produce node-specific outputs and lack the ability to query continuous spatial locations.

Among these methods, POD is particularly relevant as a comparator due to its structural similarity to CNF — both reconstruct fields as combinations of basis functions and instance-dependent coefficients. In the subsequent analysis, we therefore adopt POD as the primary baseline to assess CNF's performance in representing complex spatiotemporal variability.

## 5.1 Formulation of POD

In CNF, the SIREN can be interpreted as a set of basis functions, while the latent vector $\boldsymbol{L}$ encodes the instance-specific coefficients. The modulation process effectively resembles a functional multiplication between the basis and the coefficients, resulting in the reconstructed field of interest. This formulation is conceptually analogous to Proper Orthogonal Decomposition (POD) [22], where a system state is approximated by a weighted sum of orthogonal spatial modes. Unlike POD, however, the basis functions in CNF are not fixed or orthogonal but are implicitly parameterized by the neural network, allowing for enhanced representational flexibility.

To obtain the basis functions (modes), we first stack the snapshots of $N_s$ samples as,

$$\boldsymbol{\Phi}' = [\boldsymbol{\Phi}_1(\boldsymbol{x}, t_1), \boldsymbol{\Phi}_1(\boldsymbol{x}, t_2), ..., \boldsymbol{\Phi}_{N_s}(\boldsymbol{x}, t_{N_t})] \in \mathbb{R}^{N_h \times N_a}, \qquad (9)$$

where $N_h = N_d \times N_\Phi$ is the full dimension size, $N_a = N_s \times N_t$ is the total number of spatiotemporal snapshots, and $N_h < N_a$ in our case. To reduce the model, a low-rank approximation with rank $N_l \ll N_h$ needs to be sought for the full-order space, $\mathcal{S} = \text{span}\{\boldsymbol{\Phi}_1(\boldsymbol{x}, t_1), \boldsymbol{\Phi}_1(\boldsymbol{x}, t_2), ..., \boldsymbol{\Phi}_{N_s}(\boldsymbol{x}, t_{N_t})\} \subset \mathbb{R}^{N_h}$.

The POD leverages the singular value decomposition (SVD) of $\boldsymbol{\Phi}'$ for dimension reduction,

$$\boldsymbol{\Phi}' = \boldsymbol{P}\boldsymbol{\Sigma}\boldsymbol{Q}^T, \qquad (10)$$

with $\boldsymbol{P} \in \mathbb{R}^{N_h \times N_h}$ and $\boldsymbol{Q} \in \mathbb{R}^{N_a \times N_a}$ being orthogonal matrices, and $\boldsymbol{\Sigma} = \text{diag}\{\sigma_1, \sigma_2, ..., \sigma_{N_h}\}$ containing the singular values. The column space of $\boldsymbol{P}$ spans the full-order space $\mathcal{S}$. Let $\tilde{\boldsymbol{P}} \in \mathbb{R}^{N_h \times N_l}$ denote the first $N_l$ columns of $\boldsymbol{P}$, representing the reduced set of orthogonal bases in the the lower-dimensional subspace. Then, for a new testing snapshot $\boldsymbol{\Phi}_i^{\text{test}}$, the corresponding coefficients used to reconstruct the field are given by $\tilde{\boldsymbol{P}}^T \boldsymbol{\Phi}_i^{\text{test}}$, with the reconstruction error quantified by $\|\boldsymbol{\Phi}_i^{\text{test}} - \tilde{\boldsymbol{P}}\tilde{\boldsymbol{P}}^T \boldsymbol{\Phi}_i^{\text{test}}\|^2$.

## 5.2 Comparison between CNF and POD

We take Case 1 as a benchmark case due to its simplicity and ease of implementation. It is worth noting that the performance gap between CNF and POD is expected to widen in Cases 2 and 3, which involve more complex spatiotemporal patterns. For a fair comparison, both CNF and POD are trained on the same training dataset and evaluated on the same testing dataset.

Fig. S9 presents the results of reconstructing a testing trajectory using CNF and POD. Panels (a) and (b) illustrate the reconstruction outcomes when the reduced dimensionality is set to be the same for both methods, i.e., $N_l^{\text{POD}} = N_l^{\text{CNF}} = 256$. The contour maps indicate that the CNF-reconstructed fields exhibit a closer agreement with the reference solution. In contrast, POD shows noticeable residual errors, particularly in the saturation variable (Fig. S9a). This is further validated in Fig. S9b, where the RMSE of POD is several orders of magnitude higher than that of CNF. In particular, the saturation RMSE increases monotonically over time, indicating cumulative reconstruction error. The poor performance of POD in this case is primarily

attributed to its limited ability to capture transient dynamics. Since POD constructs a fixed set of global basis functions via linear decomposition, it inherently lacks the flexibility to adapt to time-varying features, particularly in systems exhibiting nonlinear or nonstationary behavior. As a result, the method tends to underrepresent evolving patterns, especially during the plume development stages, where strong transients dominate the solution structure. In contrast, CNF leverages learnable, instance-specific representations, enabling it to better accommodate localized and time-dependent variations. The variation of reconstruction error (summed over all three channels) shows that POD requires more than 2048 dimensions to achieve an accuracy comparable to that of CNF (Fig. S9c). However, such high-dimensional latent representations impose a substantial computational burden, rendering them impractical for downstream modeling with LDM. The envelope of POD RMSE variation further indicates that POD remains a suboptimal choice for model reduction when considering both reconstruction fidelity and computational efficiency (Fig. S9c).

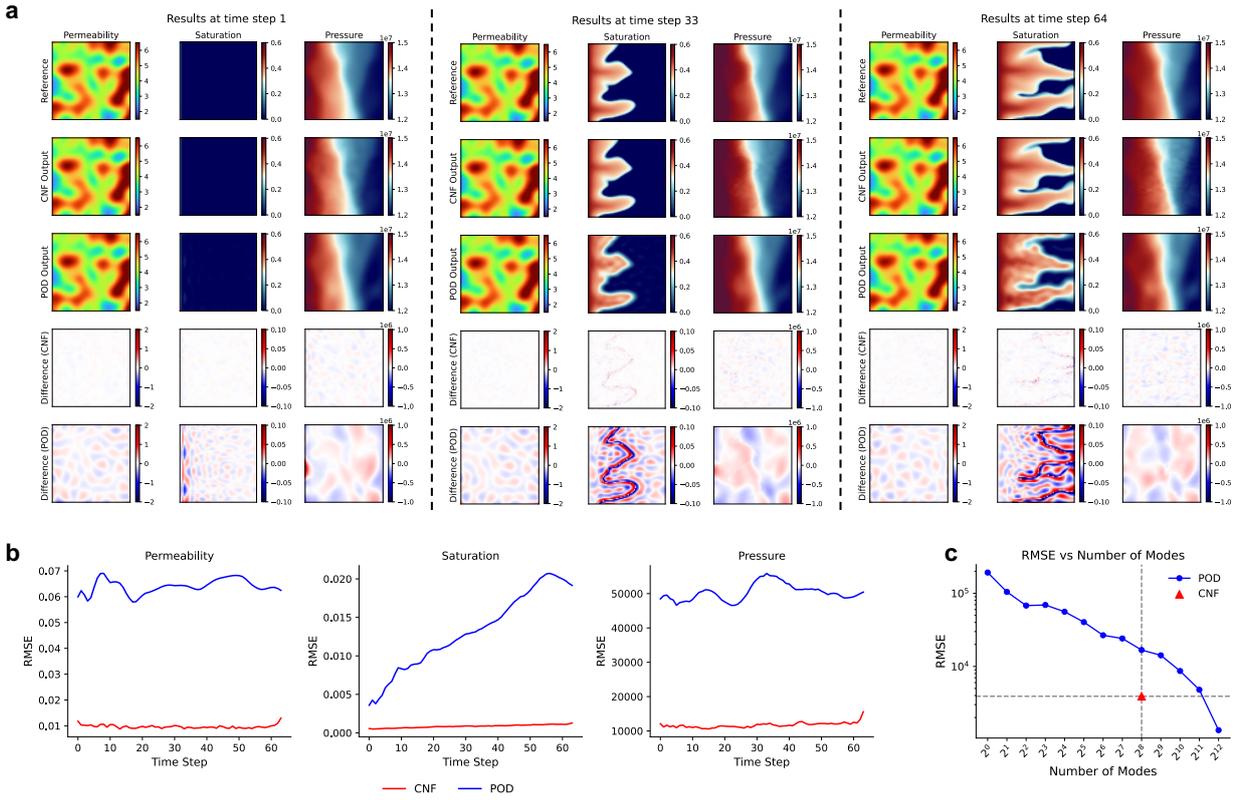

**Fig. S9** Testing trajectory performance for CNF and POD. (a) Contour plots of permeability, saturation, and pressure at three selected time steps. The first row shows the reference solution, followed by reconstructions from CNF and POD in the second and third rows, respectively. The fourth and fifth rows present the corresponding residual errors for CNF and POD. (b) Temporal evolution of RMSE for CNF and POD across the three variables. (c) POD reconstruction RMSE as a function of the number of retained modes.

# 6 Pressure generation results

We have presented the results for pressure in this section. The contours of the generated pressure fields, along with the corresponding evaluating metrics, are elaborated in the following subsections for the three cases. These results serve as a complementary visualization to the main text.

## 6.1 Case 1: $CO_2$ drainage in heterogeneous reservoirs

For the 2D $CO_2$ drainage case, the pressure field exhibits relatively minor temporal variation compared to the transient dynamics of the saturation field. The pressure is elevated near the left-side injection boundary and decreases toward the right-side outflow boundary, forming a transitional gradient that aligns with the dominant direction of $CO_2$ migration (Fig. S10a,b). Notably, the spatial distribution of pressure is influenced by the underlying heterogeneous permeability field. CoNFiLD-geo can basically infer the pressure fields from low-resolution $CO_2$ plume monitoring data, with the $16 \times 16$ resolution scenario yielding the most accurate results (Fig. S10a). As the observation resolution increases, the similarity between the generated and the reference pressure field improves, and the associate uncertainty correspondingly decreases (Fig. S10c). These findings are in agreement with the saturation results discussed in the main text.

The generated pressure field under sparse well observations are illustrated in Fig. S10b and d. As the number of observation well increases, the generated fields manifest improved agreement with the reference (Fig. S10b). Correspondingly, the standard deviation of SSIM decreases (Fig. S10d), reflecting reduced uncertainty in the predictions, which also aligns with the conclusions drawn in the main text.

These results demonstrate that CoNFiLD-geo is capable of accurately generating state variables such as pressure, enabling future extensions of the framework to more complex multiphysics coupling scenarios, including thermal, mechanical and chemical fields. From the contour visualizations, minor ripple-like artifacts can be observed, which stem from the fixed frequency spectrum of the SIREN. Although these artifacts have limited influence on the overall pressure field patterns, they could potentially be mitigated through careful tuning of the activation constant $\omega_0$.

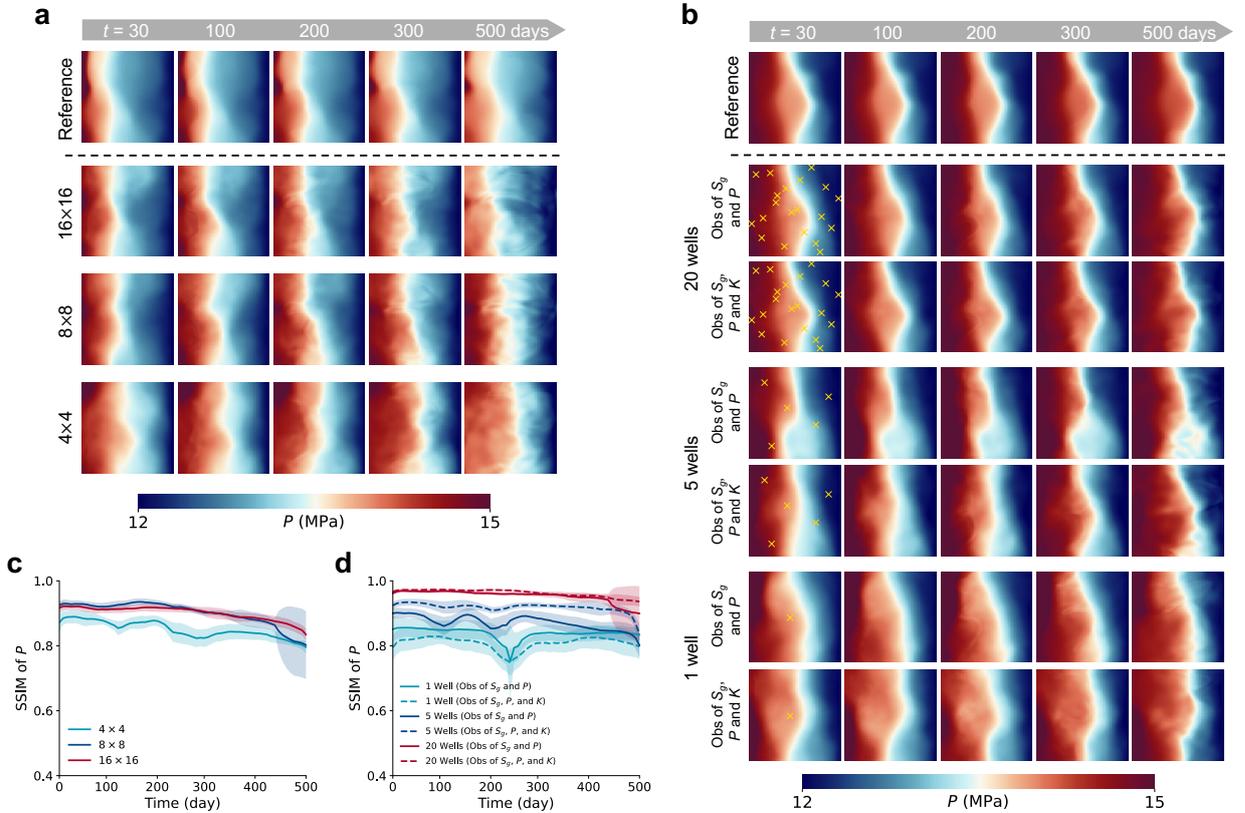

**Fig. S10** Pressure generation results for Case 1. (a) Reference and generated pressure fields at 30, 100, 200, 300, and 500 days under different $CO_2$ resolution conditions. (b) Reference and generated pressure fields with varying numbers of monitoring wells. (c) SSIM of pressure predictions under different $CO_2$ resolution conditions; shaded regions indicate standard deviation. (d) SSIM of pressure predictions under varying numbers of monitoring wells; shaded regions indicate standard deviation.

17

## 6.2 Case 2: field-scale $CO_2$ sequestration at the Sleipner site

For the field-scale Sleipner case, elevated pressure accumulates around the injection well due to injection-induced over-pressurization (see reference contours in Fig. S11a). Over time, this overpressure gradually decreases as the effect spreads through the open boundaries, and the system transitions into a relatively stable diffusive regime. The spatial distribution of pressure is significantly affected by the realistic stratigraphic complexity and the heterogeneity of the permeability field. Among the three monitoring strategies, $CO_2$ plume monitoring yields the most accurate inference results, followed by the scenario with 18 monitoring wells. The case with only 6 monitoring wells exhibits the lowest accuracy and the highest uncertainty, as indicated by the broader shaded region in the RMSE plot (Fig. S11a,b). Despite these differences, CoNFiLD-geo demonstrates reliable capability in generating spatiotemporal pressure dynamics that align well with the reference, highlighting its robustness in modeling state variable evolution under various monitoring conditions. This is further substantiated by the RMSE metric, which remains consistently below $2 \times 10^4$ Pa (Fig. S11b), representing a relative error of less than 0.25% with respect to the reference pressure magnitude of approximately $9 \times 10^6$ Pa.

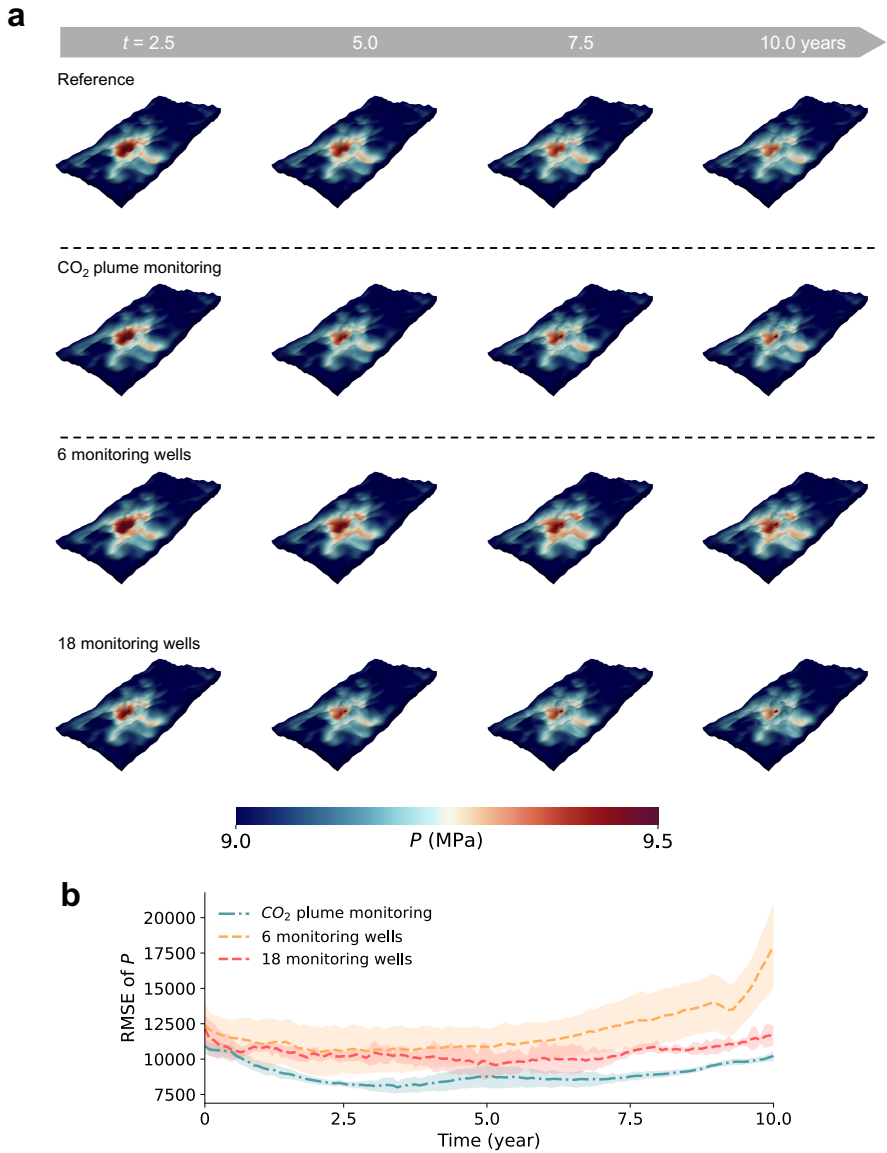

**Fig. S11** Pressure generation results for Case 2. (a) Reference and generated pressure fields at 2.5, 5, 7.5 and 10 years under three monitoring strategies (conditions): $CO_2$ plume monitoring, 6 monitoring wells, and 18 monitoring wells. (b) Temporal variation of RMSE for the three conditions, with shaded areas indicating the standard deviation across generated samples.

## 6.3 Case 3: $CO_2$ injection and brine production in stratigraphically complex reservoirs

The third case simulates the co-existence of $CO_2$ injection and brine production within a closed hydrological system featuring stratigraphically complex reservoir geometry. The pressure is high near the injection well and relatively low near the production well (Fig. S12a). Owing to the closed boundary conditions, the overall level of over-pressurization is higher than that observed in Case 2. The spatial distribution of pressure is primarily governed by the reservoir geometry (i.e., depth and thickness). Again, the spatiotemporal dynamics of pressure is stable compared to that of saturation, as the system has reached a quasi-steady seepage state. Fig. S12b presents the pressure contours generated by CoNFiLD-geo under different monitoring strategies. All monitoring strategies yield results that align well with the reference, except for the $CO_2$ plume monitoring strategy, which shows noticeable deviations. The RMSE evaluation metric plotted in Fig. S12c also shows that this strategy produces the highest error and the largest standard deviation, indicating both lower similarity to the reference and increased uncertainty in the predicted fields. These findings suggest that relying solely on $CO_2$ plume monitoring may be suboptimal for pressure inference in stratigraphically complex reservoirs. Overall, CoNFoLD-geo is able to conditionally generate decent pressure fields, especially when geoseismic survey data and sparse well measurements are available. For all monitoring strategies except $CO_2$ plume monitoring, the RMSE remains consistently below $1 \times 10^5$ Pa (Fig. S12c), which is significantly lower than the reference pressure magnitude of around $11 \times 10^6$ Pa and corresponds to a relative error of less than 1%.

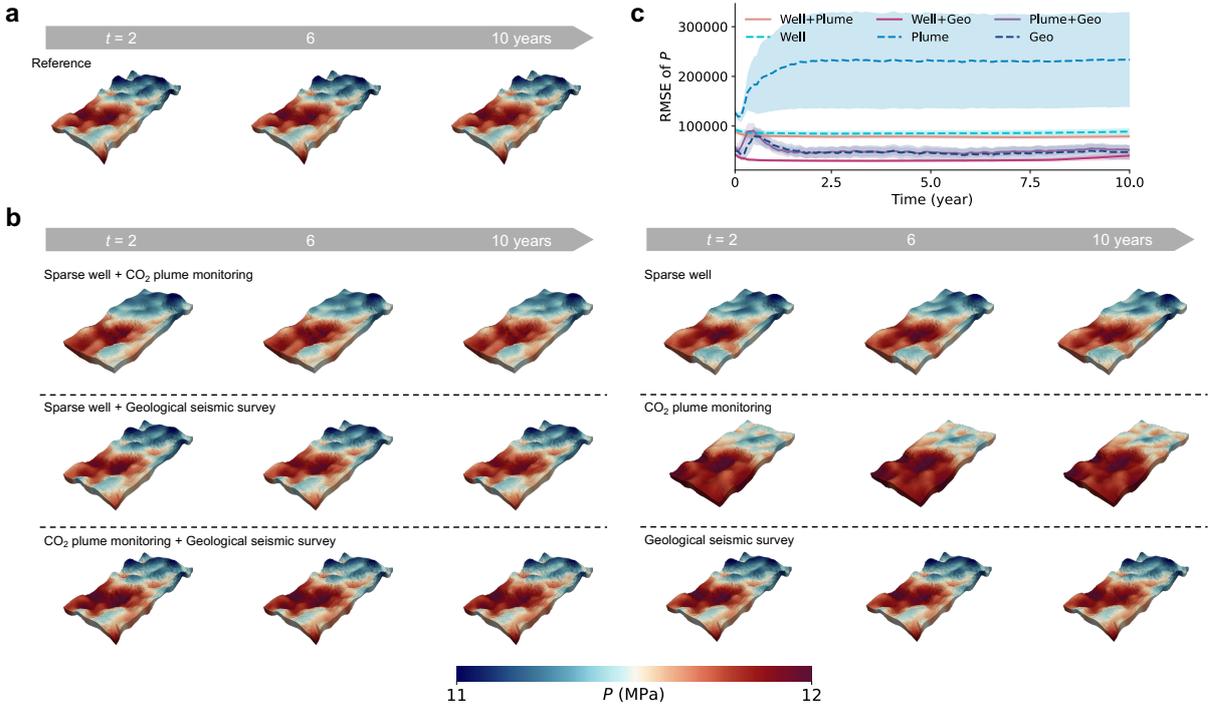

**Fig. S12** Pressure generation results for Case 3. (a) Reference pressure fields at 2, 6 and 10 years. (b) Generated pressure fields under various monitoring strategies. (c) Temporal variation of RMSE for the six monitoring strategies, with shaded areas indicating the standard deviation across generated samples.

# 7 Additional conditional generation results

In the main text, we presented only one randomly selected sample from the generated ensemble of 10 for brevity. In this section, we provide additional visualizations from different generated samples at more time steps to facilitate further evaluation by the reader. We have also included the results of regularly placed monitoring wells for Case 1. Moreover, in the main text, only the multi-source monitoring results are presented for the scenario involving $CO_2$ injection and brine production in stratigraphically complex reservoirs. For completeness, the results under single-source monitoring conditions are provided in the third subsection, while the corresponding pressure results have already been presented in the previous section.

## 7.1 Case 1: $CO_2$ drainage in heterogeneous reservoirs

When monitoring wells are arranged in a regular grid-like pattern, the corresponding conditional generation results are presented in Fig. S13 and S14. The generated permeability and saturation fields show improved agreement with the reference as the number of well measurements increases, accompanied by a reduction in uncertainty. As expected, the inferred permeability fields are more accurate when direct observations of permeability values are available.

Additional conditional generation results for $16 \times 16$-resolution $CO_2$ plume observations and 20 sparse well measurements (with probed permeability) are shown in Fig. S15. When low-resolution $CO_2$ plume data are available, the inferred saturation fields closely match the reference, as the direct observations in the associated functional space effectively guide the generation. The uncertainty is minimal, evidenced by the high similarity among the three generated samples. Although the predicted pressure fields capture the global distribution well, the local details are comparatively less accurate than those of the saturation results. The generated permeability fields basically capture the structural patterns of the reference, effectively distinguishing between high- and low-permeability regions. The variations among the results reflect the uncertainty associated with the inferred permeability fields (Fig. S15a). When sparse well measurements are available, the reconstructed saturation fields generally resemble the global pattern of the reference, with minor discrepancies near the plume front. The pressure fields appear more consistent with the reference, as they exhibit smoother variations that are easier to characterize. The inferred permeability fields align well with the reference, as the direct permeability probes at sparse well locations provide critical information that enables the model to approximate the underlying heterogeneity (Fig. S15b).

## 7.2 Case 2: field-scale $CO_2$ sequestration at the Sleipner site

Fig. S16 presents the additional generation results at finer time steps conditioned on 18 monitoring wells. Both the saturation and pressure fields successfully reproduce the reference data, showcasing the model's excellent capability to reconstruct high-dimensional state variables from extremely sparse observations. The strong similarity among the generated samples indicates a low level of uncertainty in the inferred solutions. The inferred permeability fields also show close match with the reference, with both highly and lowly permeable regions accurately captured by the model. Regions lacking observation wells, such as the peripheral zones of the permeability field, exhibit increased variation/uncertainty, as expected.

## 7.3 Case 3: $CO_2$ injection and brine production in stratigraphically complex reservoirs

Fig. S17 complements Fig. 5 in the main text by providing results under single-source monitoring conditions. From a visual perspective, direct observation of the $CO_2$ plume yields the most accurate saturation results. In contrast, the saturation fields conditioned on "sparse well" and "geological seismic survey" data exhibit some deviations near the plume edge. The inferred reservoir geometry (depth and thickness) closely matches the reference when geological seismic survey data are available, as they serve as informative conditioning data, albeit with inherent noise. The remaining two types of conditional information deliver inferior performance. In particular, the $CO_2$ plume monitoring tends to produce an over-smoothed representation of the reservoir stratigraphy.

Fig. S18 provides additional generation results conditioned on both $CO_2$ plume monitoring and geological seismic survey data, as this combined monitoring strategy yields the most balanced and reliable outputs across different state variables. The generated geomodel and reservoir responses under this conditioning scenario are in good accordance with the reference data, with minor inter-sample variations suggesting low uncertainty. This highlights CoNFiLD-geo's effectiveness in inverse modeling of geological heterogeneity and flow behavior.

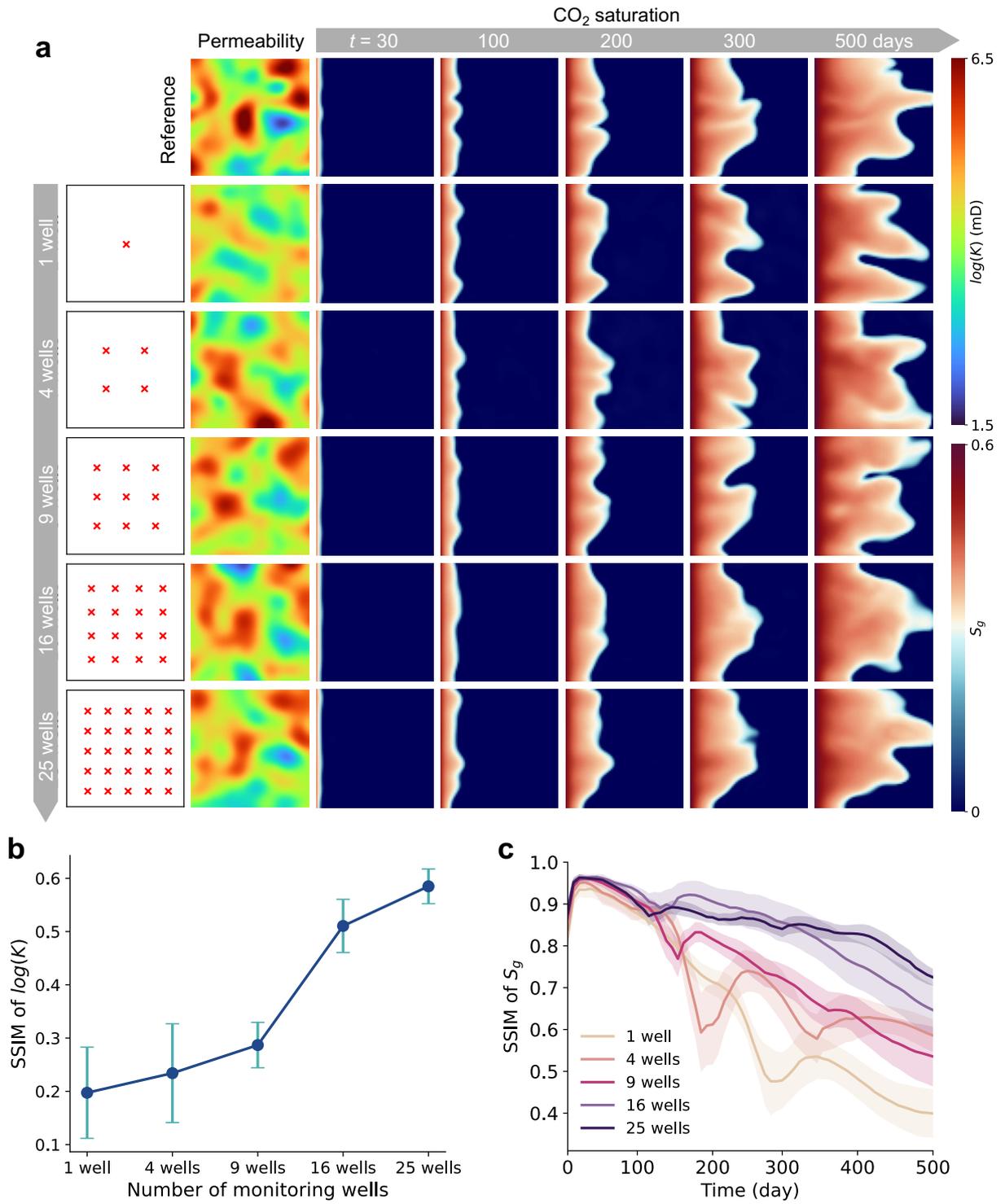

Fig. S13 Conditional generation results under regularly placed monitoring wells for Case 1. Monitoring wells provide observational data on pressure and saturation. (a) Reference and generated permeability fields, along with saturation dynamics at 30, 100, 200, 300, and 500 days. The monitoring well configurations (indicated by red crosses) are displayed in the first column. (b) SSIM of permeability as a function of the number of monitoring wells. Error bars indicate the standard deviation. (c) Temporal variation of SSIM for saturation under different numbers of monitoring wells, with shaded areas denoting the standard deviation.

21

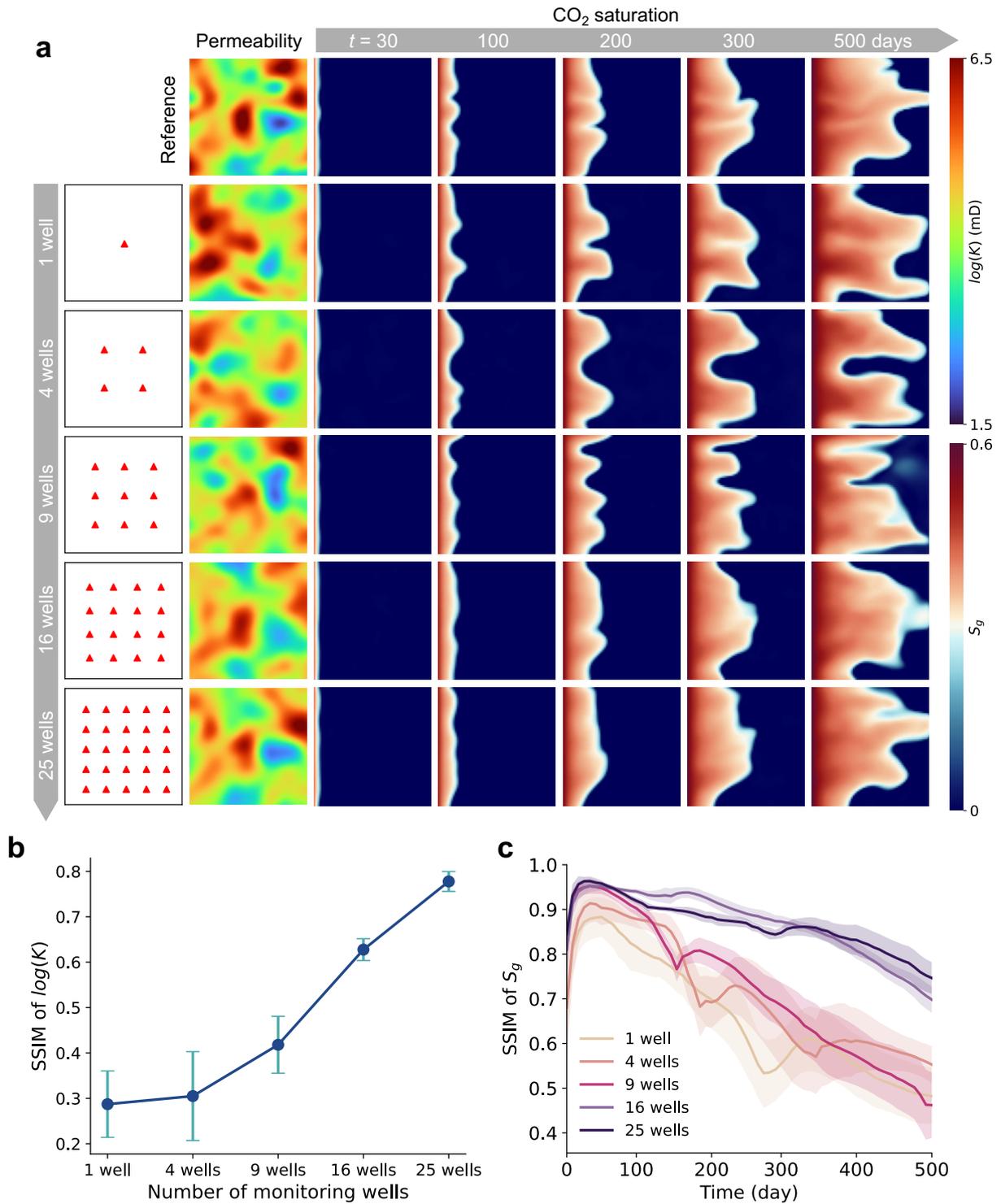

Fig. S14 Conditional generation results under regularly placed monitoring wells for Case 1. Monitoring wells provide observational data on pressure, saturation, and permeability. (a) Reference and generated permeability fields, along with saturation dynamics at 30, 100, 200, 300, and 500 days. The monitoring well configurations (indicated by red triangles) are displayed in the first column. (b) SSIM of permeability as a function of the number of monitoring wells. Error bars indicate the standard deviation. (c) Temporal variation of SSIM for saturation under different numbers of monitoring wells, with shaded areas denoting the standard deviation.

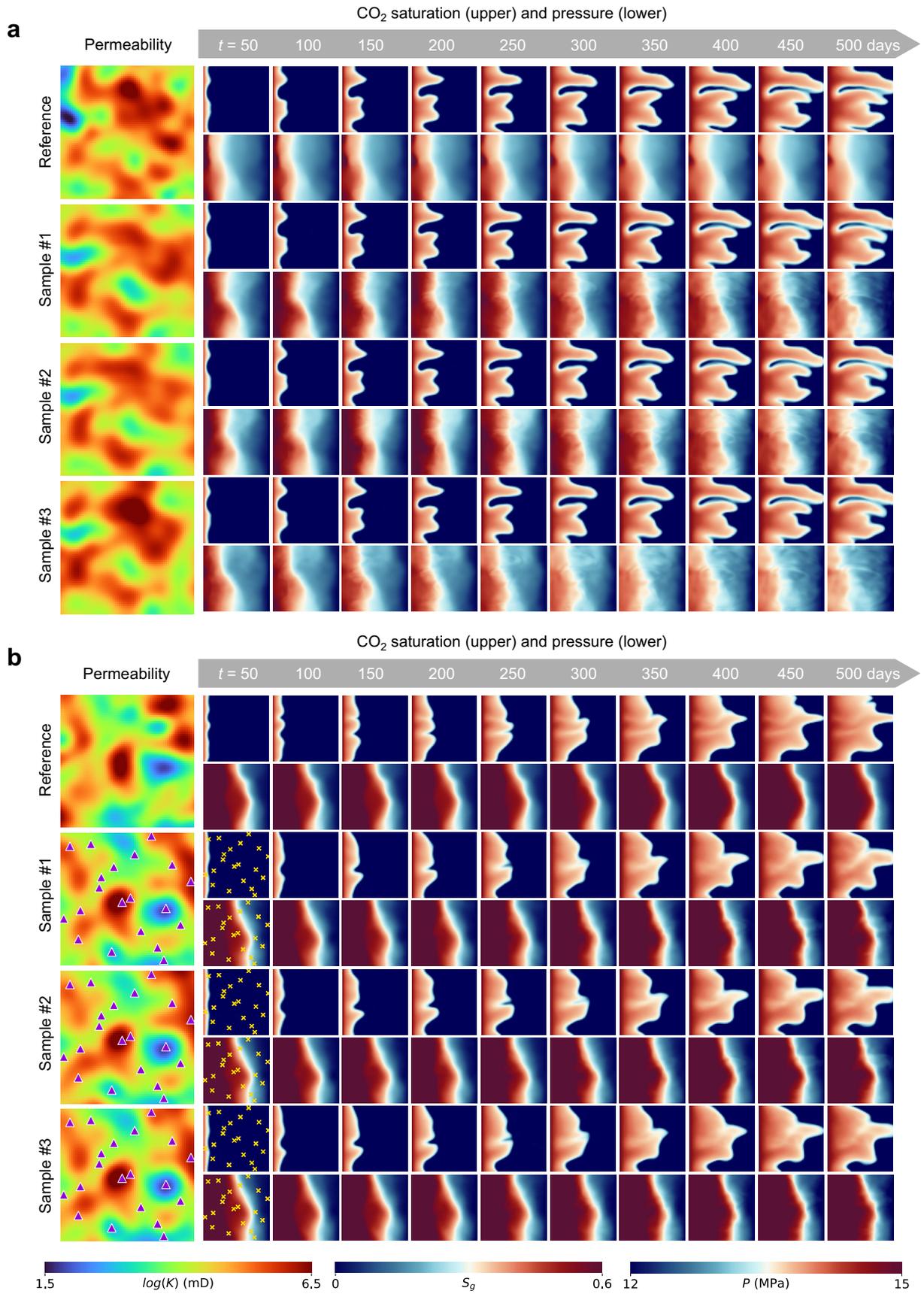

**Fig. S15** Additional generation results at finer time steps for Case 1. The reference and the three randomly selected samples are shown. (a) Condition: 16 × 16 CO$_2$ plume monitoring data. (b) Condition: 20 sparse monitoring wells with permeability, pressure and saturation probed.

23

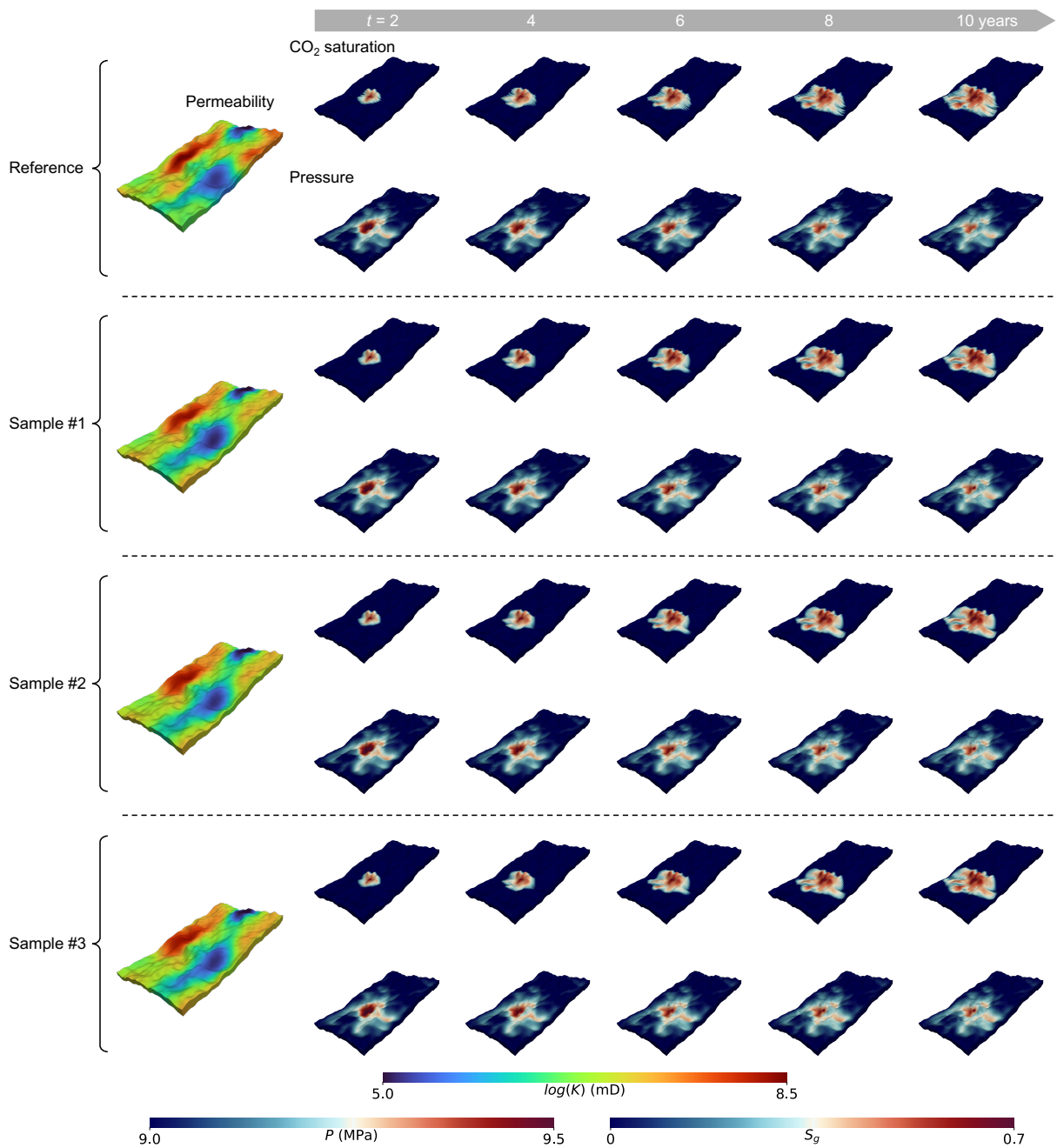

**Fig. S16** Additional generation results at finer time steps for Case 2 under the condition of 18 sparse monitoring wells. From top to bottom, the reference and the three randomly selected samples from the generated ensemble are displayed.

24

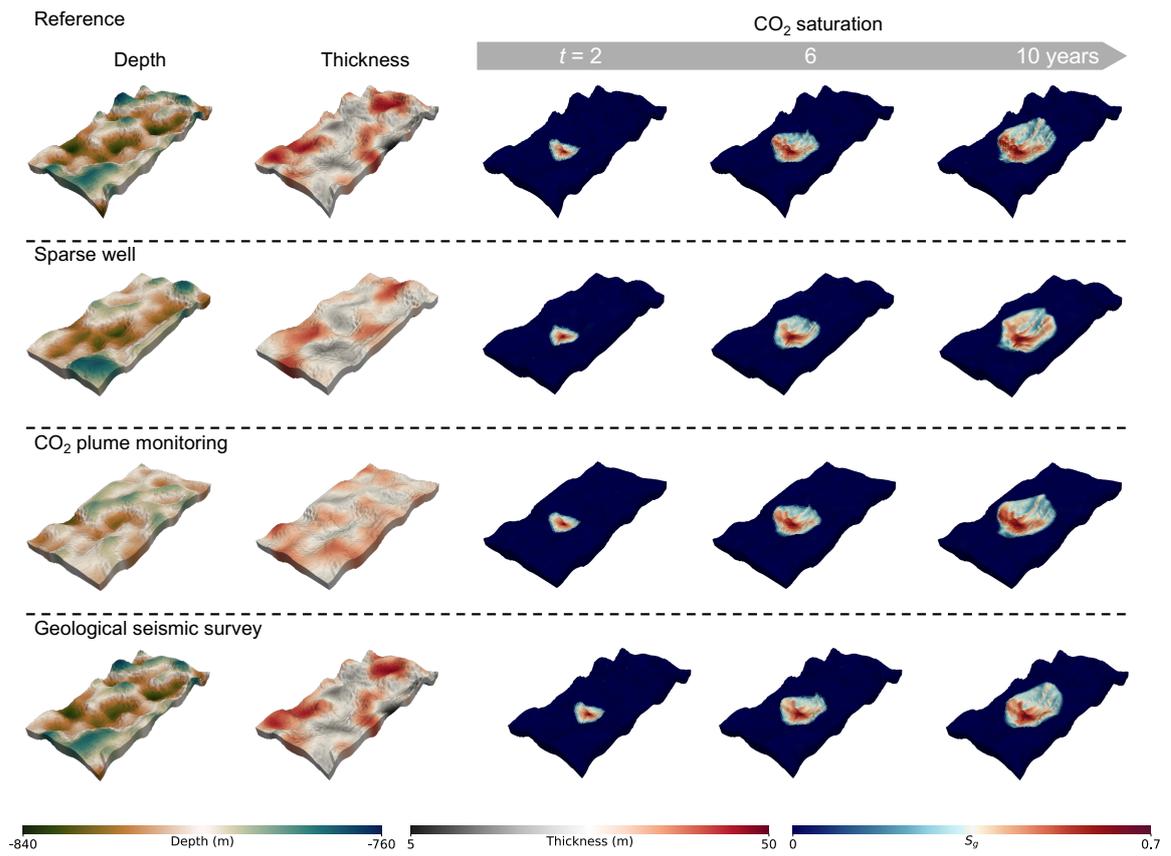

**Fig. S17** Conditional generation results under single-source monitoring data. From top to bottom: the reference, results conditioned on sparse well data, results from $CO_2$ plume monitoring, and results from geological seismic survey.

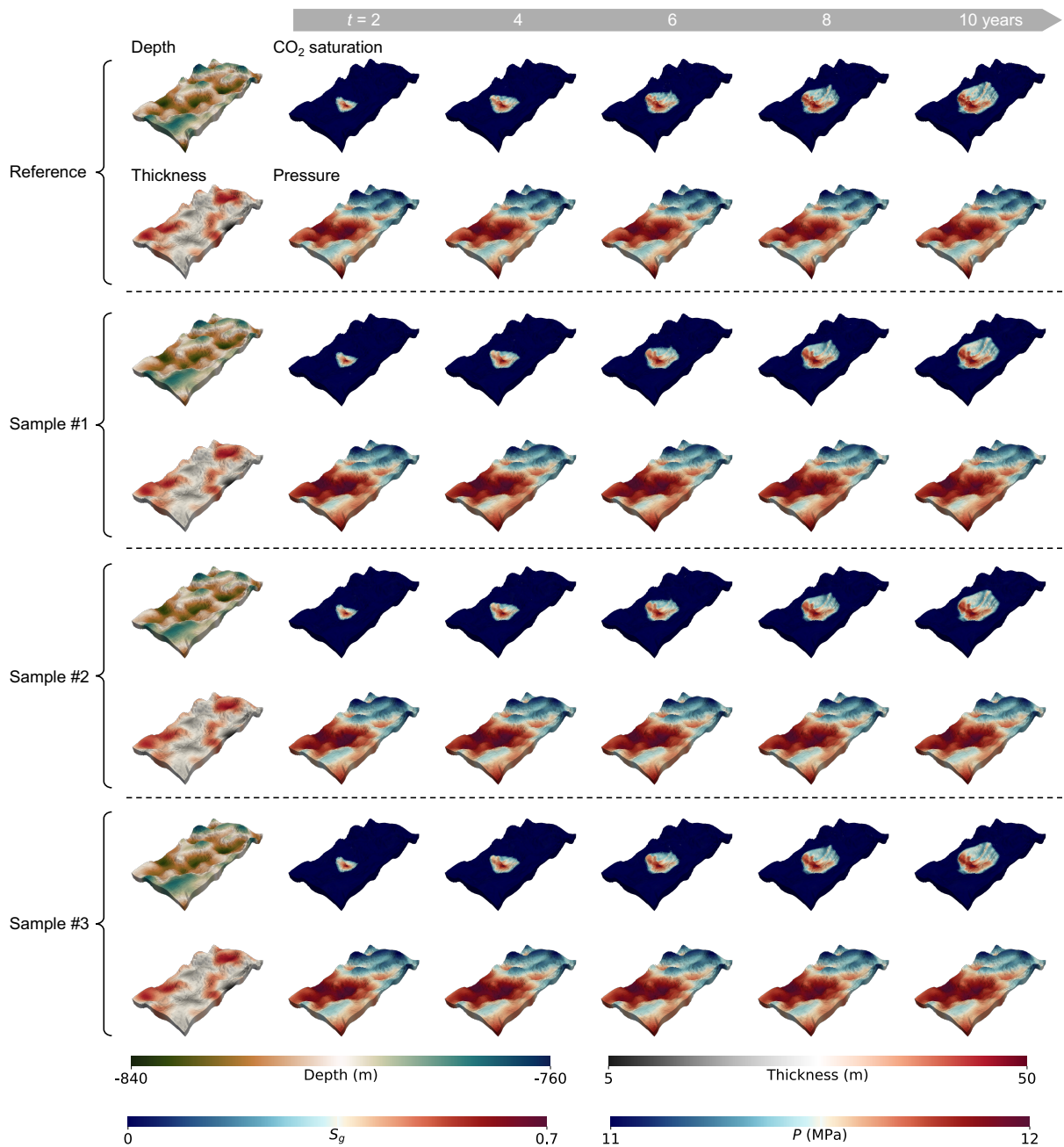

**Fig. S18** Additional generation results at finer time steps for Case 3 conditioned on multi-source monitoring data consisting of 5% noisy $CO_2$ plume observations and geological seismic survey data. From top to bottom, the reference and the three randomly selected samples from the generated ensemble are displayed.

26

# 8 Unconditional generation results

A remarkable feature of CoNFiLD-geo is its ability to perform unconditional generation, in addition to the conditional generation explored so far. Unconditional generation requires no external observations to guide the process, allowing the model to rapidly synthesize physically plausible geomodels and corresponding reservoir responses in a fully data-driven manner. This capability effectively functions as a fast numerical emulator, enabling efficient preliminary uncertainty analysis, prior ensemble construction, and reservoir screening. These applications are also instrumental for pre-operational planning, risk assessment, and strategy development in GCS projects.

## 8.1 Case 1: $CO_2$ drainage in heterogeneous reservoirs

Fig. S19 shows the generated 4 samples starting for different Gaussian white noise. The unconditionally generated samples demonstrate a wide range of geological heterogeneity in the permeability fields, reflecting the model's ability to capture diverse subsurface scenarios. Meanwhile, the resulting saturation and pressure fields consistently exhibit physically coherent patterns, indicating the model's strong generalization capability and adherence to physical processes. Notably, the $CO_2$ plume migration predominantly follows the highly permeable pathways, in accordance with the transport mechanisms characteristic of subsurface multiphase flow systems.

## 8.2 Case 2: field-scale $CO_2$ sequestration at the Sleipner site

As shown in Fig. S20, CoNFiLD-geo explores a wide spectrum of permeability realizations. The generated saturation and pressure fields, while exhibiting some overall similarity, are locally influenced by the spatial heterogeneity of permeability. Their global patterns, however, are primarily governed by the complex stratigraphic structure of the reservoir. The generated parameter-solution pairs reveal physically plausible behaviors. For example, in Sample #2, the presence of a low-permeability region (blue) near the injection well limits the plume's expansion, resulting in significant overpressure buildup.

## 8.3 Case 3: $CO_2$ injection and brine production in stratigraphically complex reservoirs

Unconditional generated reservoirs and the corresponding reservoir responses are presented in Fig. S21. CoNFiLD-geo is capable of synthesizing multifarious reservoirs characterized by stratigraphic complexity and distinct geological features. The generated reservoir pairs also conform to the governing physical principles of subsurface multiphase flow. For instance, supercritical $CO_2$ preferentially accumulates at shallower depths (shown in green), as its buoyancy drives it to rise above denser formation fluids. In addition, thicker reservoirs offer greater storage capacity, which helps confine the plume and results in a more compact spatial distribution. These physical behaviors are effectively captured by CoNFiLD-geo.

In summary, the above unconditional generation results demonstrate that CoNFiLD-geo can efficiently synthesize diverse and physically consistent geomodels along with their corresponding dynamic responses, without relying on external observations. The generated results capture key subsurface flow behaviors, highlighting the model's strong generalization and physical awareness. Future research will explore integrating physics-informed sampling to further enhance the physical consistency of generated outputs.

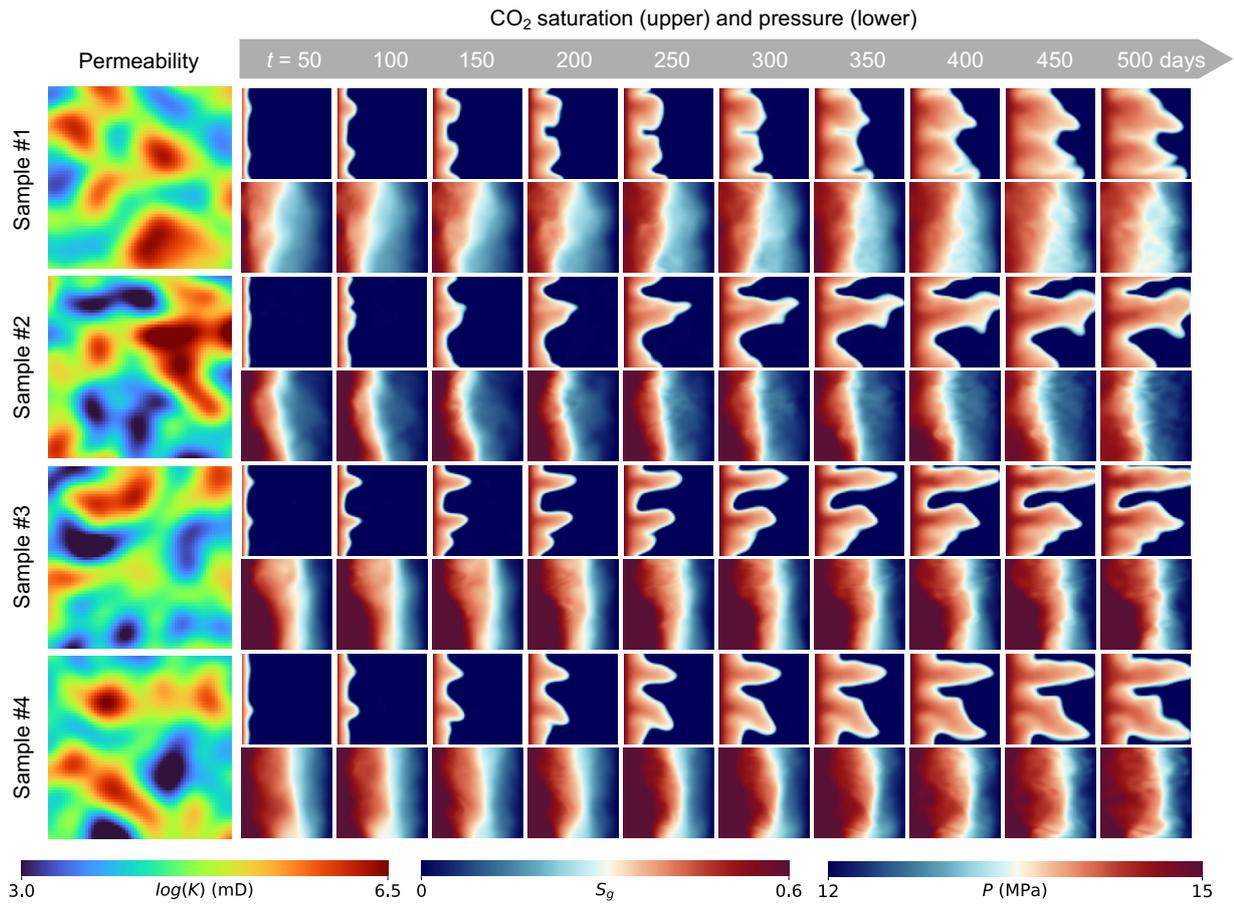

**Fig. S19** Unconditional generation results for Case 1 over a 500-day period. From top to bottom, four generated samples are shown, each initiated from a different Gaussian white noise input.

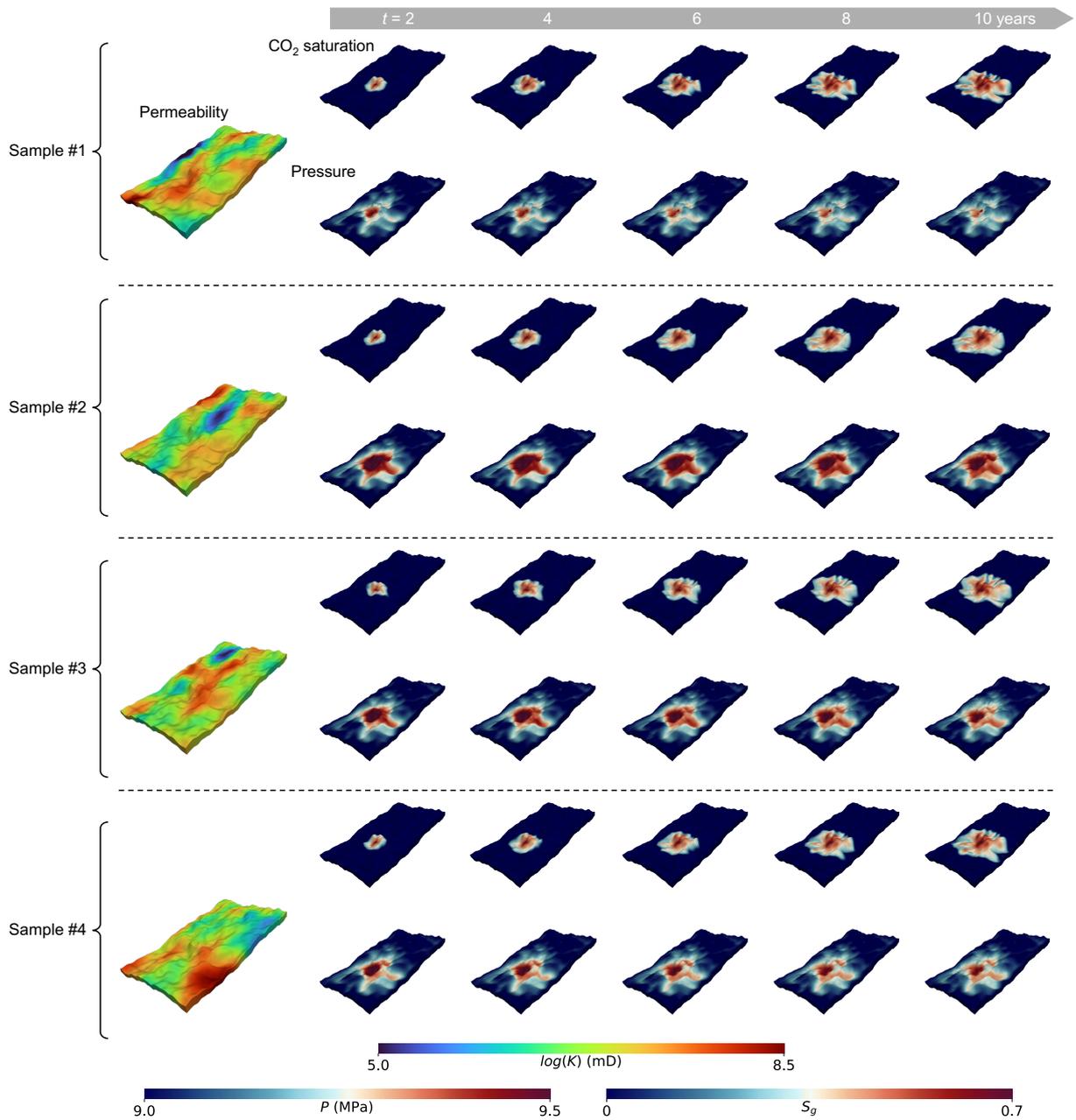

**Fig. S20** Unconditional generation results for Case 2 over a 10-year period. From top to bottom, four generated samples are shown, each initiated from a different Gaussian white noise input.

29

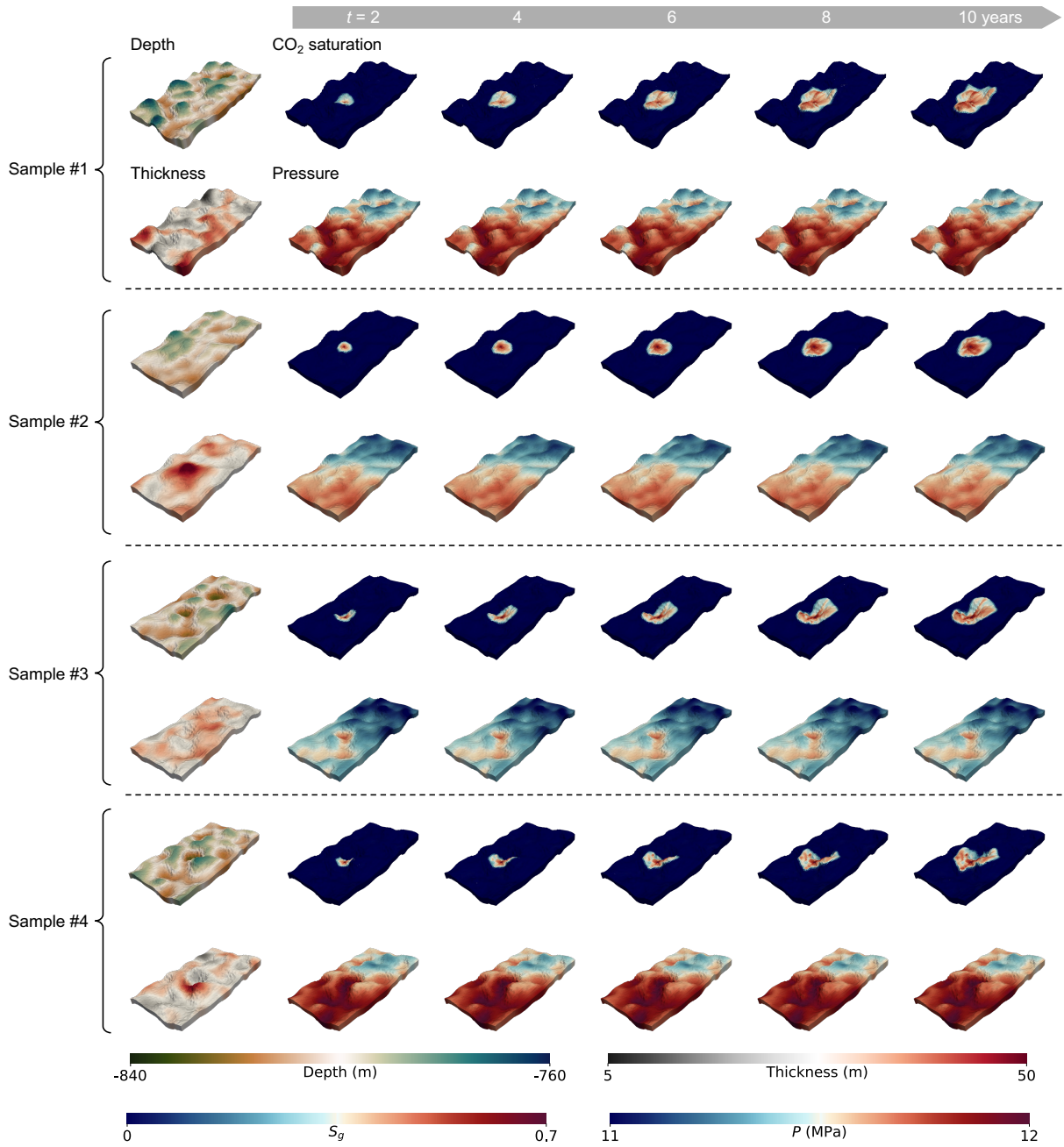

**Fig. S21** Unconditional generation results for Case 3 over a 10-year period. From top to bottom, four generated samples are shown, each initiated from a different Gaussian white noise input.

# 9 Computational time

Table S7 presents the computational cost of CoNFiLD-geo under conditional and unconditional settings, alongside the corresponding cost of traditional numerical simulation. As a fast numerical emulator, CoNFiLD-geo is able to unconditionally generate a geomodel and its corresponding reservoir responses in approximately 20 seconds, which is substantially less than the time required by conventional numerical simulation. For conditional generation, the time cost varies depending on observational information and model complexity, ranging from 48 to 192 seconds for the three representative cases, which remains within an acceptable and efficient computational range. It should be noted that CoNFiLD-geo directly generates posterior samples via zero-shot generation, in contrast to traditional gradient-based or gradient-free methods that progressively approximate the posterior through iterative calls to a surrogate model or a numerical simulator. In addition, CoNFiLD-geo can generate samples in batch mode, facilitating the fast evaluation of uncertainties. The computational cost of numerical simulation is significantly higher, ranging from 5 to 30 minutes depending on the complexity of the case. Hence, CoNFiLD-geo opens a new pathway toward the real-time uncertainty quantification of GCS using generative AI.

Table S7 Computational time of CoNFiLD-geo for unconditional generation, conditional generation and numerical simulation. The reported conditional generation time corresponds to the following scenarios: (i) 20 sparse well measurements for Case 1; (ii) 18 sparse well measurements for Case 2; (iii) 18 sparse well measurements for Case 3.

|  | Case 1 | Case 2 | Case 3 |
| --- | --- | --- | --- |
| Unconditional generation (s) | 22 | 19 | 20 |
| Conditional generation (s) | 62 | 48 | 192 |
| Numerical simulation (min) | 5 | 18 | 30 |

[18] Du, P., Parikh, M. H., Fan, X., Liu, X.-Y. & Wang, J.-X. Conditional neural field latent diffusion model for generating spatiotemporal turbulence. *Nature Communications* **15**, 10416 (2024). URL https://www.nature.com/articles/s41467-024-54712-1.

[19] Guo, M. & Hesthaven, J. S. Data-driven reduced order modeling for time-dependent problems. *Computer Methods in Applied Mechanics and Engineering* **345**, 75–99 (2019). URL https://linkinghub.elsevier.com/retrieve/pii/S0045782518305334.

[20] Vlachas, P. R., Arampatzis, G., Uhler, C. & Koumoutsakos, P. Multiscale simulations of complex systems by learning their effective dynamics. *Nature Machine Intelligence* **4**, 359–366 (2022).

[21] Lam, R. *et al.* Learning skillful medium-range global weather forecasting. *Science* **382**, 1416–1421 (2023).

[22] Berkooz, G., Holmes, P. & Lumley, J. The Proper Orthogonal Decomposition in the Analysis of Turbulent Flows. *Annual Review of Fluid Mechanics* **25**, 539–575 (2003).



\end{document}